\RequirePackage{fix-cm}
\documentclass[smallextended]{svjour3wide} 
\smartqed  
\usepackage[
colorlinks=true, 
pdfstartview=FitV, 
linkcolor=blue, 
citecolor=magenta, 
urlcolor=blue, 
bookmarks=true,
bookmarksnumbered=true,
pdftitle={},
pdfauthor={Masaru Hongo (RIKEN)}
]{hyperref}

\usepackage{amsmath,amssymb,bm}
\usepackage{graphicx}
\usepackage{bbold}
\usepackage{color}
\usepackage[normalem]{ulem}  
\usepackage[option]{colortbl}
\usepackage{cite}

\usepackage{feynmp}

\newcommand{\ptc}[1]{{\bar{#1}}}

\newcommand{\pt}{\bar{t}}
\newcommand{\pzero}{\bar{0}}
\newcommand{\ps}{\bar{s}}
\newcommand{\px}{\bar{x}}
\newcommand{\pmu}{\bar{\mu}}
\newcommand{\pnu}{\bar{\nu}}
\newcommand{\prho}{\bar{\rho}}
\newcommand{\psigma}{\bar{\sigma}}

\newcommand{\bzero}{\bm{0}}

\newcommand{\pbx}{\bar{\bm{x}}}

\newcommand{\tilt}{\tilde{t}}

\newcommand{\tilx}{\tilde{x}}

\newcommand{\tilh}{\tilde{h}}
\newcommand{\tilj}{\tilde{j}}
\newcommand{\tilv}{\tilde{v}}
\newcommand{\tiln}{\tilde{n}}
\newcommand{\tils}{\tilde{s}}
\newcommand{\tilell}{\tilde{\ell}}
\newcommand{\tilgamma}{\tilde{\gamma}}
\newcommand{\tildelta}{\tilde{\delta}}
\newcommand{\tildel}{\tilde{\partial}}

\newcommand{\tilLam}{\tilde{\Lambda}}
\newcommand{\tilLcal}{\tilde{\mathcal{L}}}

\newcommand{\tilP}{\tilde{P}}
\newcommand{\tilD}{\tilde{D}}

\newcommand{\tilA}{\tilde{A}}
\newcommand{\tila}{\tilde{a}}
\newcommand{\tilUcal}{\widetilde{\mathcal{U}}}

\newcommand{\current}{\mathcal{J}}

\newcommand{\bra}[1]{\langle {#1} |}
\newcommand{\braket}[1]{\langle {#1} \rangle }
\newcommand{\ket}[1]{| {#1} \rangle}

\newcommand{\ip}[1]{\bm{(} {#1}\bm{)} }
\newcommand{\overlr}[1]{ \overleftrightarrow{#1}}
\newcommand{\lie}{\mathsterling}

\newcommand{\hK}{\hat{K}}
\newcommand{\hT}{\hat{T}}
\newcommand{\hU}{\hat{U}}
\newcommand{\hA}{\hat{A}}
\newcommand{\hB}{\hat{B}}

\newcommand{\hJ}{\hat{J}}
\newcommand{\hS}{\hat{S}}

\newcommand{\hrho}{\hat{\rho}}

\newcommand{\hrhoLG}{\hat{\rho}_{\mathrm{LG}}}

\newcommand{\hpi}{\hat{\pi}}
\newcommand{\hlambda}{\hat{\lambda}}
\newcommand{\hkappa}{\hat{\kappa}}
\newcommand{\hp}{\hat{p}}
\newcommand{\hphi}{\hat{\phi}}
\newcommand{\hSigma}{\hat{\Sigma}}
\newcommand{\hn}{\hat{n}}
\newcommand{\hcurrent}{\hat{\mathcal{J}}}
\newcommand{\hOcal}{\hat{\mathcal{O}}}

\newcommand{\hTcal}{\hat{\mathcal{T}}}
\newcommand{\hPcal}{\hat{\mathcal{P}}}
\newcommand{\hEcal}{\hat{\mathcal{E}}}
\newcommand{\hScal}{\hat{\mathcal{S}}}
\newcommand{\hUcal}{\hat{\mathcal{U}}}
\newcommand{\hc}{\hat{c}}

\newcommand{\PT}{\mathcal{P}\mathcal{T}}

\newcommand{\Lcal}{\mathcal{L}}

\newcommand{\Ecal}{\mathcal{E}}
\newcommand{\Gcal}{\mathcal{G}}

\newcommand{\Tcal}{\mathcal{T}}
\newcommand{\Pcal}{\mathcal{P}}
\newcommand{\Dcal}{\mathcal{D}}
\newcommand{\Ocal}{\mathcal{O}}
\newcommand{\Scal}{\mathcal{S}}

\newcommand{\average}[1]{\langle#1\rangle}
\newcommand{\averageLG}[1]{\langle#1\rangle^{\mathrm{LG}}}
\newcommand{\baverageLG}[1]{\Big\langle#1\Big\rangle^{\mathrm{LG}}}
\newcommand{\Tr}{\mathop{\mathrm{Tr}}}

\newcommand{\with}{\quad\mathrm{with}\quad}

\journalname{Journal of Statistical Physics}

\begin{document}

\title{Nonrelativistic hydrodynamics from quantum field theory:}
\subtitle{
(I) Normal fluid composed of spinless Schr\"odinger fields}
\titlerunning{Nonrelativistic hydrodynamics from quantum field theory} 

\author{Masaru Hongo}

\institute{
M. Hongo \at
iTHES Research Group, RIKEN, Wako 351-0198, Japan \\
\email{masaru.hongo@riken.jp}
}

\date{\today}

\maketitle

 \begin{abstract}
  We provide a complete derivation of hydrodynamic equations 
  for nonrelativistic systems 
  based on quantum field theories of spinless Schr\"odeinger fields, 
  assuming that an initial density operator takes 
  a special form of the local Gibbs distribution.
  The constructed optimized/renormalized perturbation theory 
  for real-time evolution
  enables us to separately evaluate dissipative and nondissipative 
  parts of constitutive relations.
  It is shown that the path-integral formula for local thermal equilibrium 
  together with the symmetry properties of the resulting action--- 
  the nonrelativistic diffeomorphism and gauge symmetry in the 
  thermally emergent Newton-Cartan geometry---provides 
  a systematic way to derive the nondissipative part of
  constitutive relations. 
  We further show that 
  dissipative parts are accompanied with the entropy production operator 
  together with two kinds of fluctuation theorems by the use of which 
  we derive the dissipative part of constitutive relations and 
  the second law of thermodynamics. 
  After obtaining the exact expression for constitutive relations, 
  we perform the derivative expansion and derive the first-order 
  hydrodynamic (Navier-Stokes) equation with the Green-Kubo formula for 
  transport coefficients. 
 \end{abstract}

\tableofcontents

\section{Introduction and Summary}
\label{sec:Intro}
\subsection{Introduction}
Hydrodynamics is one of the most established 
theoretical framework by the use of which we can describe 
real-time dynamics of many-body systems.
To be more precise, hydrodynamics captures the 
macroscopic spacetime evolution of the conserved charge densities 
such as the energy and momentum densities~\cite{Landau:Fluid}. 
One most important feature of hydrodynamics is 
that it provides universal description of \textit{any} many-body system:
We can apply it to air and water in our daily life, 
strongly correlated electron systems in condensed matter physics, 
nuclear matter inside the neutron stars in astrophysics,
the quark-gluon plasma in high-energy physics, 
and also active matters in biological systems.
Hydrodynamic analysis including precise numerical simulations 
is now indispensable tool
to investigate real-time dynamics in many fields of science.

Nevertheless, compared to its successful applications, 
foundation of hydrodynamics 
based on underlying microscopic theories---in particular, quantum field 
theories---remains unclear.
In fact, although this problem has been pursued for a long time in 
development of nonequilibrium (classical) statistical mechanics~\cite{Nakajima,Mori1,McLennan,McLennan1,Zubarev:1979,Zubarev1,Zubarev2,PhysRevA.8.2048}, 
it was just recent that a much greater understanding of
hydrodynamics has been promoted  from the modern viewpoint of nonequilibrium 
statistical mechanics and quantum field theory.
On the one hand, a considerable technique has been developed in the 
quantum field-theoretical side, which enables us to construct 
the generating functional of relativistic hydrodynamics 
for the hydrostatic situations~\cite{Banerjee:2012iz,Jensen:2012jh}.
In these works, the curved spacetime technique together 
with symmetry consideration are fully utilized to clarify the possible form of 
the hydrostatic generating functional for relativistic hydrodynamics%
\footnote{
There are further notable advances to include the non-hydrostatic effects 
and construct the Wilsonian effective action of relativistic (fluctuating) hydrodynamics on the basis of 
the Schwinger-Keldysh formalism (See Refs.~\cite{Haehl:2015pja,Crossley:2015evo,Haehl:2015uoc,Haehl:2016pec,Haehl:2016uah,Jensen:2017kzi,Glorioso:2017fpd,Haehl:2017zac}). 
}.
On the other hand, based on the recent development of nonequilibrium statistical mechanics, 
one simplest derivation of the nonrelativistic hydrodynamic 
equations from the classical Hamiltonian description of systems is given in Ref.~\cite{sasa:2013}.
The key idea in that work is an efficient use of the nonequilibrium identity, 
or the variant of the so-called fluctuation theorem \cite{YamadaKawasaki,Jarzynski1,Evans,Gallavotti,Kurchan,Maes,Lebowitz,Crooks,Jarzynski2,Seifert}. 
This treatment is generalized to systems composed of relativistic 
quantum fields, which brings about relativistic hydrodynamic 
equations~\cite{Hayata:2015lga}. 
In addition, it is also clarified that 
the same curved spacetime structure as the hydrostatic 
situations naturally emerges when we consider 
a partition functional of systems in local thermal equilibrium~\cite{Hongo:2016mqm}.

In spite of these interesting development, there remain two 
unsatisfactory points for the field theoretical derivation of 
nonrelativistic hydrodynamics.
Firstly,
the above quantum field-theoretical technique has been mainly applied to 
not \textit{nonrelativistic} hydrodynamics but 
\textit{relativistic} hydrodynamics except for a few works~\cite{Gromov:2014vla,Jensen:2014aia,Jensen:2014ama}. 
This is partially because the geometric structure is rather complicated in the 
nonrelativistic situation compared to relativistic one.
In fact, instead of using a familiar pseudo-Riemannian (relativistic) geometry, 
we have to use an unfamiliar Newton-Cartan (nonrelativistic) geometry 
in order to discuss the nonrelativistic hydrodynamics~\cite{Son:2013rqa,Geracie:2014nka,Bergshoeff:2014uea,Geracie:2015dea}.
This exotic geometric language looks unnecessarily complicated at first glance, 
but it provides a considerably efficient way to clearly specify 
spacetime symmetries for nonrelativistic systems 
inescapably tied to hydrodynamic equations.
Secondly, compared to the clear formulation of the nondissipative part, 
dissipative part of transport phenomena are often handled 
in a phenomenological way \cite{Jensen:2014ama,Geracie:2015xfa}, 
in which they assume the local version of the second law of thermodynamics, 
or the existence of the entropy current $s^\mu$ satisfying 
$\partial_\mu s^\mu \geq 0$.
However, this local second law of thermodynamics should not be assumed but be 
derived based on a certain assumption from the viewpoint of 
nonequilibrium statistical mechanics. 
Indeed, due to the phenomenological aspect of the entropy current analysis,
transport coefficients are not calculable within their formalism, 
and considered as physical constants taking positive values dependent on 
microscopic constituents of the fluid.

Then, it's time to clarify a complete derivation of 
nonrelativistic hydrodynamic equations based on recent developments 
of both nonequilibrium statistical mechanics and quantum field theory with 
the help of the geometric language for nonrelativistic systems.
The purpose of this paper is thus to derive 
nonrelativistic hydrodynamic equations 
based on quantum field-theoretical description of 
nonrelativistic systems 
composed of interacting Bosonic or Fermionic spinless Schr\"odinger fields%
\footnote{
Spinful cases will be discussed in the subsequent paper \cite{Hongo}.
}.
To accomplish this purpose, we first develop a canonical generalization of 
imaginary-time (Matsubara) formalism \cite{Matsubara,AGD} for local thermal equilibrium
based on the Newton-Cartan geometry, which enables us to describe
the nondissipative transports, e.g. given by convective and hall transport 
terms. 
Furthermore, we construct the optimized/renormalized perturbation theory 
(See e.g. \cite{Kleinert2009path,Jakovac2015resummation} 
and references therein for references of optimized perturbation theory)
for real-time dynamics and derive the dissipative transport together with 
the Green-Kubo formula for the transport coefficients \cite{Green,Nakano,Kubo}. 
This completes the derivation of conventional 
normal nonrelativistic hydrodynamic equations without thermal fluctuations.

\subsection{Summary of result}
We here briefly summarize our result along with 
the introductory remarks on the basic structure of 
conventional hydrodynamics.
Hydrodynamic equations are based on (covariant) conservation laws, 
whose microscopic parents are given by the following operator identities: 
$(\nabla_\mu - \Gcal_\mu ) \hcurrent^\mu_{~a}(x) = \hScal_a(x)$. 
Here $\nabla_\mu$ and $\Gcal_\mu$ denote the covariant derivative and 
the possible torsional contribution to the conservation laws, and 
$\hcurrent^\mu_{~a}(x) \equiv \{\hTcal^\mu_{~\nu}(x),\hJ_M^\mu(x),\hJ_Q^\mu \}$ 
conserved current operators like the mass current, 
$\hScal_a (x)$ source terms like the Lorentz force, respectively. 
We note that the source terms $\hScal_a(x)$ are written in terms of 
the conserved currents $\hcurrent^\mu_{~a}(x)$ and the external fields 
$j(x)$ under consideration.
We will provide the detailed description of them in the subsequent section.
In order to derive the macroscopic hydrodynamic equation, 
we need to take the average of conservation laws over some density operator:
\begin{equation}
 (\nabla_\mu - \Gcal_\mu ) \average{\hcurrent^\mu_{~a}(x)} 
  = \average{\hScal_a(x)},
\end{equation}
where we employed the Heisenberg picture and 
introduced the average $\average{\hOcal} \equiv \Tr (\hrho_0 \hOcal )$ 
with the initial density operator $\hrho_0$.
Although these macroscopic continuity equations serve as 
basic equations of hydrodynamics, 
we cannot solve them unless we express the spatial part 
of $\average{\hcurrent^\mu_{~a}(x)}$ in terms of the time component:
\begin{equation}
 \average{\hcurrent^\mu_{~a}(x)} 
  = \current^\mu_{~a} [\average{\hcurrent^{\ptc{0}}_{~a}}].
\end{equation}
These relations are called \textit{constitutive relations}.
While we do not know the existence of these relations in a 
general nonequilibrium situation, 
we empirically know that they do exist around local thermal equilibrium. 
Furthermore, the form of the constitutive relation is universal and 
information on the microscopic ingredients is reflected in only a handful of 
physical properties, which split up into two groups: 
The first group is static properties of systems, 
among which a representative is the equation of state 
$p = p (\average{\hcurrent^{\ptc{0}}_{~a}})$ with a fluid pressure $p$. 
The second group is dynamic properties which, for example, expresses the ease 
with which energy current can pass through a medium.
These transport properties are installed to the so-called 
transport coefficients.
Introducing the pressure $p$ and a set of transport coefficients 
$L_i = \{\eta, \zeta, \kappa,\cdots\}$ with a shear (bulk) viscosity 
$\eta\,(\zeta)$, and the heat conductivity $\kappa$, 
we thus need a way to determine them for any given value of 
$\average{\hcurrent^{\ptc{0}}_{~a}}$,
\begin{equation}
 p = p [\average{\hcurrent^{\ptc{0}}_{~a}}], \quad 
  L_i = L_i [\average{\hcurrent^{\ptc{0}}_{~a}}],
\end{equation}
on the basis of microscopic models.
We then formulate our problems to derive hydrodynamics as follows: 
\textit{Can we derive the universal form of the constitutive relation
together with a way to calculate necessary physical properties
based on the underlying quantum theory?}

The foremost point in the present setting is to specify the 
correct form of the density operator which captures 
the full non-linear dynamics of the conserved charge densities.
Recalling that hydrodynamics is applied to systems near local thermal 
equilibrium, we introduce the local Gibbs distribution for the 
density operator, which describes systems in local thermal equilibrium. 
The local Gibbs distribution at time $\pt$ is given by
\begin{equation}
 \hrhoLG [\pt;\lambda] 
  \equiv \exp \left( - \hS[\pt;\lambda] \right), \with
  \hS [\pt;\lambda] \equiv \hK [\pt;\lambda] + \Psi [\pt;\lambda], 
\end{equation}
where $\hK[\pt;\lambda]$ is defined as 
\begin{equation}
 \begin{split}
  \hK[\pt;\lambda] 
  &\equiv 
  - \int d \Sigma_{\pt \mu} \hcurrent^\mu_{~a} \lambda^a
  = - \int d \Sigma_{\pt \mu}
  \left( \hTcal^\mu_{~\nu}(x) \beta^\nu(x)  
  + \hJ_M^\mu (x) \nu_M (x) + \hJ_Q^\mu (x) \nu_Q (x) \right).
 \end{split}
\end{equation}
Here $\hTcal^\mu_{~\nu}$, $\hJ_M^\mu$ and $\hJ_Q^\mu$ 
denote the nonrelativistic 
energy-momentum tensor, mass current and electric current operator, 
respectively, and $\Psi[\pt;\lambda,j]$ gives 
the normalization of the density operator:
\begin{equation}
 \Psi[\pt;\lambda,j] \equiv \log \Tr \exp \left( - \hK[\pt;\lambda]  \right),
\end{equation}
This funcional $\Psi[\pt;\lambda,j]$ is identified 
as the (local) thermodynamic functional known as the Massieu-Planck functional.
We also introduced the hypersurface vector $d\Sigma_{\pt\mu}$ whose 
direction is perpendicular to constant-time hypersurface, 
and magnitude is given by its spatial volume element.
This form of the density operator describes local thermal equilibrium 
with the help of the conjugate local thermodynamic parameter 
$\lambda^a(x) \equiv \{\beta^\mu(x), \nu_M(x), \nu_Q(x)\}$ 
for the conserved charge densities at time $\pt$. 
As is clarified in the subsequent section, these parameters 
correspond to the local inverse temperature, fluid-velocity, and 
local chemical potential.
This form of density operator is a local generalization of the 
familiar Gibbs distribution used in the grand canonical ensemble.

We then put a critical assumption that the density operator takes a form of 
the local Gibbs distribution at initial time $\pt_0$: 
$\hrho_0 = \hrhoLG[\pt_0;\lambda]$, 
and consider the subsequent time evolution.
Here $\lambda^a (x) \big|_{\pt_0}$ 
denotes a set of local thermodynamic parameters at initial time $\pt_0$. 
Since we fix our initial condition, the problem seems to be simple: 
we only need to evaluate $\average{\hcurrent^\mu_{~a}(x)} 
\equiv \Tr \big( \hrho_0 \hcurrent^\mu_{~a}(x))$.
Nevertheless, contrary to our optimistic expectation, 
even if we accept the above critical assumption, 
the derivation of hydrodynamic equations requires further work.
In fact, when we calculate the expectation values of the conserved current 
operator $\average{\hcurrent^\mu_{~a}(x)}$ at later time $\pt~ (>\pt_0)$, 
we have to approximate it unless we are able to obtain exact results.
To make meaningful approximations at later $\pt~ (>\pt_0)$, 
we have to reconstruct the perturbarive expansion in a similar manner with 
the optimized/renormalized perturbation theory. 
For that purpose, introducing a new set of parameters 
$\lambda^a (x)\big|_{\pt}$ 
at later time, we decompose our density operator as 
\begin{equation}
 \begin{split}
  \hrho_0 
  &= \exp \left( - \hS[\pt_0;\lambda] \right)
  = \exp \left( - \hS[\pt;\lambda] 
  + \hS[\pt;\lambda] - \hS[\pt_0;\lambda]\right) \\
  &= \exp \left( - \hS[\pt;\lambda] \right)
  T_\tau \exp \left( \int_0^1 d\tau \hSigma_\tau [\pt,\pt_0;\lambda] \right),
 \end{split}
\end{equation}
where we defined the entorpy production operator
$\hSigma[\pt,\pt_0;\lambda] \equiv \hS[\pt,\lambda] - \hS[\pt_0;\lambda]$, 
and 
$\hOcal_\tau \equiv e^{\tau\hS[\pt,\lambda]} \hOcal e^{-\tau\hS[\pt,\lambda]}$.
By the use of this decomposition, we obtain 
\begin{equation}
 \average{\hcurrent^\mu_{~a}(x)} \big|_{\pt}
  = \averageLG{\hcurrent^\mu_{~a}(x)}_{\pt}
  + \averageLG{\hU \delta \hcurrent^\mu_{~a}(x)}_{\pt}, \with 
  \hU \equiv 
  T_\tau \exp \left( \int_0^1 d\tau \hSigma_\tau [\pt,\pt_0;\lambda] \right),
  \label{eq:Decomp}
\end{equation}
where we introduced 
$\averageLG{\hOcal}_{\pt} \equiv \Tr \big( \hrhoLG[\pt;\lambda] \hOcal \big)$ 
and $\delta \hOcal \equiv \hOcal - \averageLG{\hOcal}_{\pt}$.
At this stage, this is just an identity, and if we can exactly evaluate e.g. 
$\average{\hcurrent^\mu_{~a}(x)}$, it does not depend on 
$\lambda^a (x) \big|_{\pt}$.
We, however, cannot accomplish such exact calculations in almost every 
situations, and rather
cut the perturbative expansion at some order 
on the top of a local Gibbs distribution with new parameters 
$\lambda^a (x) \big|_{\pt}$. 
Then, our result will depend on a way to define the new parameters 
$\lambda^a(x) \big|_{\pt}$.
Recalling that we are interested in the spacetime evolution of 
the conserved charge 
densities $\average{\hcurrent^\ptc{0}_{~a}(x)}$, 
we employ a condition like the fastest apparent convergence (FAC) 
in the optimized perturbation theory~\cite{Stevenson:1981vj}
for $\average{\hcurrent^\ptc{0}_{~a}(x)}$.
In other words, 
we put a condition that the deviation of conserved charge density will 
vanishes: $\averageLG{\hU \delta \hcurrent^{\ptc{0}}_{~a}(x)}_{\pt} = 0 $.
This condition is equivalent to 
\begin{equation}
 \average{ \hcurrent^{\ptc{0}}_{~a}(x)} \big|_{\pt}
  =  \averageLG{\hcurrent^{\ptc{0}}_{~a}(x)}_{\pt} , \label{eq:DefParameter1}
\end{equation}
which means that we defined the new parameters $\lambda^a (x)\big|_{\pt}$ 
so as to match with local thermodynamics for a given value of 
$\average{ \hcurrent^{\ptc{0}}_{~a}(x)} \big|_{\pt}$.
Through this procedure, Eq.~\eqref{eq:Decomp} provides us a meaningful 
decomposition to isolate the problem containing different types of 
physical properties:
One is to evaluate the average values of conserved current operators  
in local thermal equilibrium, and the other is a deviation from it. 
The nondissipative first part is shown to be
fully captured by the single functional $\Psi[\pt;\lambda,j]$
which is given by the path integral of quantum field theories: 
\begin{equation}
 \Psi [\pt;\lambda,j] 
  = \log \int \Dcal \phi \Dcal \phi^\dag \, e^{\Scal[\phi,\phi^\dag;\tilj]}.
\end{equation}
Here $\phi$ and $\Scal[\phi,\phi^\dag;\tilj]$ denotes 
a matter field (Schr\"odinger field) and its action under consideration.
The notable point is that the resulting action have full diffeomorphism 
invariance and gauge invariance in thermally emergent curved spacetime
with imaginary-time independent background fields $\tilj$. 
Since we are considering nonrelativistic systems, 
the emergent spacetime structure is given by the Newton-Cartan geometry.
The dissipative second part is associated with the entropy production 
operator $\hSigma[\pt,\pt_0;\lambda]$ accompanied with time evolution.
After writing down the exact formulae for both of them, 
we eventually perform the derivative expansion and derive 
e.g. the first-order constitutive relation as
\begin{align}
 \average{\hTcal^\mu_{~\nu}(x)} &= 
  - \big( \Ecal \cdot n \big) u^\mu  n_\nu 
  + n_M u^\mu u_\nu
  + p (\delta^\mu_\nu - u^\mu n_\nu)  \nonumber
  \\ &\quad 
  + \frac{\kappa}{\beta} h^{\mu\lambda}
  (\nabla_{\perp\lambda} \beta
  + \beta^\sigma P^\rho_\lambda F^n_{\sigma\rho}  ) n_\nu 
  - P^\mu_\nu
  \frac{\zeta}{\beta} h_{\rho\sigma} \nabla_\perp^{\rho} \beta^\sigma
  - \frac{2\eta}{\beta} P^\mu_\rho h_{\nu\sigma} 
  \nabla_{\perp}^{\langle\rho} \beta^{\sigma\rangle} ,  \\
 \average{\hJ_M^\mu(x)} &= n_M u^\mu  ,
\end{align}
which correctly reproduces the Navier-Stokes equation. 
We, moreover, show that information on static properties, 
e.g.~the equation of state,
is extracted from the local thermodynamic functional $\Psi[\pt;\lambda,j]$, 
and dynamic properties from the Green-Kubo formula:
\begin{align}
 \zeta 
 &= \beta(x) \int_{-\infty}^{\pt} d\pt' \int d\Sigma_{\pt}' N' 
 \ip{\tildelta \hp(x), \tildelta \hp(x')}_{\pt}, \\
 \eta 
 &= \frac{\beta(x)}{(d+1)(d-2)} 
 \int_{-\infty}^{\pt} d\pt' \int d\Sigma_{\pt}' N' 
 \ip{\tildelta \hpi_{\mu\nu}(x), \tildelta \hpi_{\rho\sigma}(x')}_{\pt} 
 h^{\mu\rho} h^{\nu\sigma} ,  \\
 \kappa
 &= \frac{\beta(x)}{d-1} 
 \int_{-\infty}^{\pt} d\pt' \int d\Sigma_{\pt}' N' 
 \ip{\tildelta \hEcal^\mu(x), \tildelta \hEcal^\nu(x')}_{\pt} h_{\mu\nu},
\end{align}
where we introduced the local version of the Kubo-Mori-Bogoliubov inner product
in Eq.~\eqref{eq:ip}. 
Note that all of the above quantities are, in principle, calculable 
for any given value of
$\lambda^a(x) \big|_{\pt}$ which has one-to-one correspondence to 
the conserved charge densities 
$\average{\hcurrent^{\ptc{0}}_{~a}(x)} \big|_{\pt}$ 
through Eq.~\eqref{eq:DefParameter1}.
From these results, we can say that we have derived the hydrodynamic equations 
based on the underlying quantum theories%
\footnote{
We note that the conventional hydrodynamics without thermal 
fluctuation may break down in the low-dimensional systems due to 
the existence of the so-called long-time tail 
\cite{YamadaKawasaki,PhysRevLett.25.1254,PhysRevLett.25.1257,Pomeau}.
In order to consider the possible long-time tail effect, 
we need to construct the nonlinearly fluctuating hydrodynamics 
and take into account their mode-coupling.
We however do not discuss the effect of hydrodynamic fluctuations 
in this paper.
}.

This paper is organized as follows: 
In Sec.~\ref{sec:Pre}, we put preliminaries for 
the nonrelativistic geometry, symmetries, and local Gibbs ensemble, 
all of which serves as a basis for our discussion.
In Sec.~\ref{sec:PathIntegral}, we provide the path-integral formula 
for the local thermodynamic functional $\Psi[\pt;\lambda,j]$ which enables us 
to evaluate nondissipative part of constitutive relations: 
$\averageLG{\hcurrent^\mu_{~a}(x)}_{\pt}$. 
In Sec.~\ref{sec:TimeEvo}, we construct the optimized/renormalized 
perturbation theory for time evolution and obtain the exact formula for the 
dissipative part of constitutive relation: 
$\averageLG{\hU \delta \hcurrent^\mu_{~a}(x)}_{\pt}$. 
In Sec.~\ref{sec:Derivation}, we eventually perform the derivative expansion,
and derive the constitutive relations 
together with the Green-Kubo formula for the transport coefficients.
Section.~\ref{sec:Discussion} is devoted to a discussion.

 \section{Preliminaries for nonrelativistic geometry, symmetry, 
and local Gibbs distribution} 
\label{sec:Pre}
In this section, we provide preliminaries for nonrelativistic geometry, 
symmetries, and local Gibbs distribution, which gives a solid basis 
to develop our discussion on the derivation of hydrodynamics.
Our consideration is on the nonrelativistic quantum system, whose Lagrangian 
is e.g. given by the nonlinear Schr\"odinger field:
\begin{equation}
 \Scal [\varphi] 
  = \int d^dx 
  \left[i \phi^\dag \partial_0 \phi
   - \frac{1}{2m} \delta^{ij} 
   \partial_i \phi^\dag \partial_j \phi  
  - \lambda |\phi|^4  \right]  .
  \label{eq:NLSchro}
\end{equation}
Starting from this kind of Lagrangian, we would like to put this system under 
the external $U(1)$ gauge field and background curved geometry. 
The former is not so difficult except for the fact that 
we have a mass current as a conserved current 
in addition to the electric current in the nonrelativistic setup.
Of course, in the simple setup where 
we only have a single component charged field, 
they give the same current except for its unit, and thus, 
they are indeed not independent. 
However, if we have two or more components of charged and/or uncharged fields 
we have to distinguish them. 
We therefore introduce two $U(1)$ gauge fields: the former 
is one for the mass current and the latter one is for the electric current. 
Introducing two background $U(1)$ gauge fields ($a_\mu$ for 
the $U(1)_M$ mass gauge field, and $A_\mu$ 
for the $U(1)_Q$ electromagnetic gauge field), 
we can put the system under the external $U(1)$ gauge field 
by replacing the partial derivative 
$\partial_\mu$ with the covariant derivative 
$\partial_\mu - i m a_\mu - i e A_\mu$. 
Compared with this, putting systems in the background curved geometry 
is somewhat complicated since we have to clarify a basic structure 
of the nonrelativistic geometry, which is known as the Newton-Cartan geometry. 
Although a part of contents in this section looks too elaborate, 
it provides a solid basis to derive the nondissipative part 
of constitutive relations based on quantum field theories. 
In fact, as will be shown in Sec.~\ref{sec:PathIntegral}, 
the nonrelativistic curved geometry, 
or the so-called Newton-Cartan geometry 
will naturally emerge when we construct the imaginary-time formalism 
for local thermal equilibrium.

In Sec.~\ref{sec:ADM}, we clarify the basic structure of the 
nonrelativistic curved geometry, or 
the twistless torsional Newton-Cartan (TTNC) geometry. 
In Sec.~\ref{sec:Matter}, we derive the covariant (non-)conservation laws 
and relation between some currents 
based on gauge and diffeomorphism invariance. 
In Sec.~\ref{sec:LG}, we introduce the local Gibbs distribution 
which describes systems in local thermal equilibrium.

\subsection{Spacetime decomposition of Newton-Cartan geometry}
\label{sec:ADM}
\begin{figure}[b]
 \centering
 \includegraphics[width=1.0\linewidth]{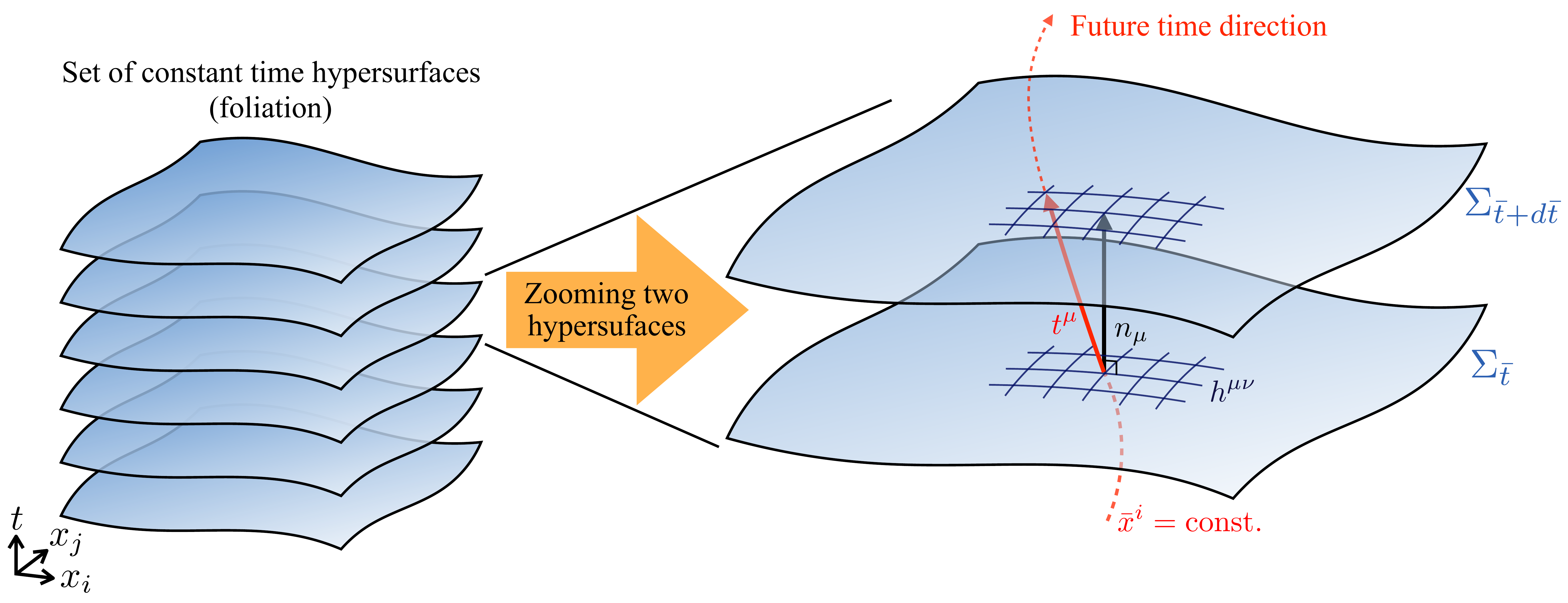}
 \caption{Illustration of the spacetime decomposition of the 
 twistless torsional Newton-Cartan (TTNC) geometry. 
 $\Sigma_{\pt}$ denotes a hypersurface parametrized by $\pt (x)=\ $const., 
 and $n_{\mu}$ is a normal vector perpendicular to the hypersurface.
 On each constant hypersurface, we introduce a curved spatial coordinate
 parametrized by $h^{\mu\nu}$ (or $h_{\mu\nu}$). 
 }
 \label{Fig:hypersurface}
\end{figure}

We first review the basic structure of the Newton-Cartan geometry 
which gives a way to describe the nonrelativistic curved spacetime geometry. 
As the metric $g_{\mu\nu}$ or the vielbein $e_\mu^{~a}$ 
plays a central role to describe the spacetime structure 
in the relativistic theory, 
we have similar geometric objects in the Newton-Cartan geometry. 
The Newton-Cartan data is given by $\{n_\mu,v^\mu,h_{\mu\nu},h^{\mu\nu} \}$ 
which satisfies 
\begin{equation}
 n_\mu v^\mu = 1, \quad n_\mu h^{\mu\nu}=0, \quad v^\mu h_{\mu\nu} = 0,
  \quad h^{\mu\rho} h_{\rho\nu} = \delta^\mu_\nu -v^\mu n_\nu
  \equiv P^\mu_{\nu}. 
  \label{eq:NCdata}
\end{equation}
Note that $h^{\mu\nu}$ is not the inverse of $h_{\mu\nu}$ as is 
demonstrated in the last relation.
Throughout this paper, 
we assume that $n_\mu$ does not necessarily satisfies $dn = 0$ 
but satisfies $n \wedge dn = 0$. 
In other words, we consider the so-called twistless torsional 
Newton-Cartan (TTNC) geometry ($dn \neq 0$ but $n \wedge dn = 0 $).
When we finally take flat limit, we will take
$n_\mu \big|_{\mathrm{flat}} = (1,\bzero)$ satisfying 
$dn \big|_{\mathrm{flat}} = 0$.
Thanks to this hypersurface orthogonality ($n \wedge dn = 0$), we can 
introduce a set of constant time hypersurfaces, or foliations which parametrize 
time coordinate (See Fig.~\ref{Fig:hypersurface}).
We then introduce a time coordinate function $\pt(x)$ which defines 
the constant time hypersurface $\Sigma_{\pt}$ and take $n_\mu$ 
as the normal vector perpendicular to $\Sigma_{\pt}$: 
$n_\mu = N(x) \partial_\mu \pt(x)$. 
In addition, introducing the spatial coordinate systems $\pbx = \pbx (x)$ on 
$\Sigma_{\pt}$, we define the time-direction vector $t^\mu$ by 
\begin{equation}
 t^\mu \equiv \partial_{\pt} x^\mu (\pt,\pbx). 
  \label{eq:Timevector}
\end{equation}
While the time vector $t^\mu$, which determines the local time direction, 
does not play a central role in the Newton-Cartan geometry, 
it is used to define a useful gauge in Sec.~\ref{sec:PathIntegral}.
When we employ the new coordinate system $\{\pt,\pbx\}$, 
we only have the zeroth component for the normal vector in the 
new coordinate systems: $n_{\pmu} = N \delta^{\pzero}_{\pmu}$.
Therefore, the second relation in Eq.~\eqref{eq:NCdata} reveals 
that $h^{\mu\nu}$ is the degenerate rank-$(d-1)$ tensor which only contains
spatial components. 
Noting that the vector $v^\mu$ has a non-vanishing time component 
due to the first relation in Eq.~\eqref{eq:NCdata}, 
we can introduce the non-degenerate rank-$d$ tensor $\gamma^{\mu\nu}$ 
and its inverse $\gamma_{\mu\nu}$ by 
\begin{equation}
 \gamma^{\mu\nu} \equiv h^{\mu\nu} + v^\mu v^\nu, \quad 
  \gamma_{\mu\nu} \equiv h_{\mu\nu} + n_\mu n_\nu,
\end{equation}
where we can show they are indeed inverse by the use of 
all relations in Eq.~\eqref{eq:NCdata}. 
Leaving unfixed spatial components $v^{\ptc{i}}$, 
we can explicitly write down the Newton-Cartan data in 
the coordinate system $\{\pt,\pbx\}$ as 
\begin{equation}
 n_{\pmu} \equiv 
  (N, \bzero) , \quad
 v^{\pmu} \equiv 
 \begin{pmatrix}
  N^{-1} \\
  v^{\ptc{i}}
 \end{pmatrix}, \quad
 h_{\pmu\pnu} 
  \equiv 
  \begin{pmatrix}
   N^2 v^2 & - N v_{\ptc{i}} \\
   - N v_{\ptc{j}} & h_{\ptc{i}\ptc{j}}
  \end{pmatrix} ,\quad
  h^{\pmu\pnu} \equiv 
  \begin{pmatrix}
   0 & 0 \\
   0 & h^{\ptc{i}\ptc{j}}
  \end{pmatrix}, 
\end{equation}
where we defined $v_{\ptc{i}} \equiv h_{\ptc{i}\ptc{j}}v^{\ptc{j}}$, 
and $v^2 \equiv v^\ptc{i} v_{\ptc{i}} = h_{\ptc{i}\ptc{j}} v^\ptc{i} v^\ptc{j}$.
Note that $h_{\ptc{i}\ptc{j}}$ is the inverse of $h^{\ptc{i}\ptc{j}}$, 
so that they satisfy 
$h_{\ptc{i}\ptc{k}}h^{\ptc{k}\ptc{j}} = \delta_{\ptc{i}}^{\ptc{j}}$.
In this coordinate system, the non-degenerate metric $\gamma_{\pmu\pnu}$ 
and its inverse take forms as
\begin{equation}
 \gamma_{\pmu\pnu} 
  = \begin{pmatrix}  
     N^2 (v^2 + 1) & - N v_{\ptc{i}}  \\
     - N v_{\ptc{j}} & h_{\ptc{i}\ptc{j}}
    \end{pmatrix}, \quad
  \gamma^{\pmu\pnu} 
  = \begin{pmatrix}  
     N^{-2} &  N^{-1} v^{\ptc{i}} \\
     N^{-1} v^{\ptc{j}}  & h^{\ptc{i}\ptc{j}} + v^{\ptc{i}} v^{\ptc{j}}
    \end{pmatrix}.
\end{equation}
Based on the above, we introduce the $d$-dimensional spacetime volume element 
for nonrelativistic geometry as 
\begin{equation}
 \int d^d x \sqrt{\gamma} = \int d^d x N \sqrt{h} , 
  \with 
  \gamma \equiv \det \gamma_{\mu\nu}, \quad
  h \equiv \det h_{\ptc{i}\ptc{j}}.
\end{equation}
Furthermore, with the help of the normal vector $n_\mu$
and spatial metric $h_{\mu\nu}$, 
we also introduce the hypersurface vector $d\Sigma_{\pt\mu}$ as 
\begin{equation}
 d\Sigma_{\pt \mu} 
  \equiv d\Sigma_{\pt} n_\mu , \with 
  d\Sigma_{\pt} \equiv d^{d-1} \px\sqrt{h} ,
\end{equation}
where $d\Sigma_{\pt}$ denotes the $(d-1)$-dimensional spatial volume element 
on the constant time hypersurface as is the case for relativistic geometry.
The integral of the spatial volume element can be rewritten 
in terms of spacetime integral as follows:
\begin{equation}
 \int d\Sigma_{\pt}
  = \int d^{d-1} \px \sqrt{h}
  = \int d^d x \sqrt{\gamma}  N^{-1} \delta( \pt -\pt(x) ),
  \label{eq:SpaceInt}
\end{equation}
which will be used in the subsequent discussion.

In addition to the above geometric data, 
we introduce the covariant derivative $\nabla_\mu$ acting on 
e.g. $(1,1)$-tensor $A^\nu_{~\rho}$ as 
\begin{equation}
 \nabla_\mu A^\nu_{~\rho} 
  \equiv  \partial_\mu A^\nu_{~\rho} 
  + \Gamma^\nu_{~\mu\sigma} A^\sigma_{~\rho} 
  - \Gamma^\sigma_{~\mu\rho} A^\nu_{~\sigma},
\end{equation}
where $\Gamma^\mu_{~\nu\rho}$ denotes a connection for 
the nonrelativistic curved geometry.
As the metric compatibility condition
$\nabla_\mu g_{\nu\rho} = \nabla_\mu g^{\nu\rho} = 0$ 
(together with the torsion-free assumption)
determines the Levi-Civita connection 
$\Gamma^{\mu}_{~\nu\rho} $, 
we put the compatibility condition for the Newton-Cartan data: 
\begin{equation}
 \nabla_\mu n_\nu = 0 , \quad \nabla_\mu h^{\nu\rho} = 0 ,
\end{equation}
in order to determine our nonrelativistic connection.
Here note that $\nabla_\mu h_{\nu\rho}$ does not vanish 
since $h^{\mu\nu}$ is not the inverse of $h_{\mu\nu}$.
Nevertheless, these compatibility conditions are not enough to determine the 
complete form of the connection $\Gamma^{\mu}_{~\nu\rho}$, 
and we have to put another condition to fix it. 
There are two competing options for the additional condition 
\cite{Jensen:2014aia,Bergshoeff:2014uea}: 
One is invariance under the $U(1)_M$ mass gauge transformation 
$a_\mu \to a_\mu + \partial_\mu \alpha $, and 
another is invariance under the so-called Milne boost transformation 
defined in Eq.~\eqref{eq:finiteMilne} in the next subsection.
These conditions are incompatible with each other and 
we cannot respect both of them.
Following the Ref.~\cite{Jensen:2014aia}, 
we employ mass gauge invariance and 
use the following expressions for the connection
\begin{equation}
 \Gamma^\mu_{~\nu\rho} 
  \equiv v^\mu \partial_{\nu} n_{\rho}
  + \frac{1}{2} h^{\mu\sigma} 
  \big( \partial_\nu  h_{\rho\sigma} + \partial_\rho  h_{\nu\sigma} 
   - \partial_\sigma  h_{\nu\rho} \big) 
  + \frac{1}{2} h^{\mu\sigma} 
  ( n_{\nu} F^a_{\rho\sigma} + n_{\rho} F^a_{\nu\sigma} ),
  \label{eq:Gamma}
\end{equation}
where we defined the field strength tensor for the $U(1)_M$ 
mass gauge field $a_\mu$ as
\begin{equation}
   F^a_{\mu\nu} \equiv \partial_\mu a_\nu - \partial_\nu a_\mu.
    \label{eq:massStrength}
\end{equation}
This connection is manifestly invariant under the $U(1)_M$ 
gauge transformation but not invariant under the Milne boost transformation%
\footnote{
Alternatively, 
we can employ the Milne boost invariant connection given by 
\begin{equation}
 \Gamma^\mu_{~\nu\rho} 
  \equiv \bar{v}^\mu \partial_{\nu} n_{\rho}
  + \frac{1}{2} h^{\mu\sigma} 
  \left( \partial_\nu  \bar{h}_{\rho\sigma} + \partial_\rho \bar{h}_{\nu\sigma} 
   - \partial_\sigma  \bar{h}_{\nu\rho} \right),
  \with
 \bar{v}^\mu \equiv v^\mu - h^{\mu\nu} a_\nu, \quad 
  \bar{h}_{\mu\nu} 
  \equiv h_{\mu\nu} + n_\mu a_\nu + n_\nu a_\mu.
\end{equation}
Since $\bar{v}^\mu$ and $\bar{h}_{\mu\nu}$ are Milne boost invariant, 
this connection is manifestly Milne boost invariant
(See e.g. Refs.~\cite{Jensen:2014aia,Bergshoeff:2014uea} in more detail). 
}.
The first term in Eq.~\eqref{eq:Gamma} brings about
the anti-symmetric (torsional) part of the connection: 
\begin{equation}
 \Gamma^{\mu}_{~\nu\rho} - \Gamma^\mu_{~\rho\nu} = v^\mu F^n_{\nu\rho} , 
  \with F^n_{\mu\nu} \equiv \partial_\mu n_\nu - \partial_\nu n_\mu,
  \label{eq:TorsionalPart}
\end{equation}
which does not vanish if $dn = F^n_{\mu\nu} dx^\mu \wedge dx^\nu \neq 0$.

We then show a useful formula for a later discussion:
\begin{equation}
 \partial_{\pt} \int d\Sigma_{\pt\mu} f^\mu 
  = \int d\Sigma_{\pt} N(x) (\nabla_\mu - \Gcal_\mu) f^\mu ,
  \label{eq:Stokes}
\end{equation}
where $f^\mu(x)$ denotes an arbitrary smooth vector, and 
$\Gcal_\mu \equiv 
 \Gamma^{\nu}_{~\nu\mu} - \Gamma^{\nu}_{~\mu\nu} = v^\nu F^n_{\nu\mu}$
results from the torsional contribution for the 
covariant derivative.
This formula can be derived as follows:
Using Eq.~\eqref{eq:SpaceInt} and the definition of the normal vector 
$n_\mu \equiv N \partial_\mu \pt(x)$ and assuming that 
$f^\mu(x)$ vanishes at the spacetime boundary, we obtain 
\begin{equation}
 \begin{split}
  \int d\Sigma_{\pt\mu} f^\mu 
  &= \int d^d \px \sqrt{h} \delta(\pt - \pt(x)) N 
  \partial_\mu \pt(x) f^\mu \\
  &= - \int d^d \px \sqrt{\gamma} \partial_\mu \theta (\pt - \pt(x)) f^\mu \\
  &= \int d^d \px \sqrt{\gamma} \theta (\pt - \pt(x)) 
  (\nabla_\mu -  \Gcal_\mu) f^\mu ,
 \end{split} \label{eq:Step}
\end{equation}
where we performed the integration by parts to proceed the second line and 
used $\partial_\mu \sqrt{\gamma} = \sqrt{\gamma}\, \Gamma^\nu_{~\mu\nu} 
= \sqrt{\gamma} (\Gamma^\nu_{~\nu\mu} - \Gcal_\mu)$. 
We then differentiate Eq.~\eqref{eq:Step} with respect to $\pt$
\begin{equation}
 \begin{split}
  \partial_{\pt} \int d\Sigma_{\pt\mu} f^\mu 
  &= \partial_{\pt} \int d^d \px \sqrt{\gamma} \theta (\pt - \pt(x)) 
  (\nabla_\mu -  \Gcal_\mu) f^\mu  \\
  &= \int d\Sigma_{\pt} N(x) (\nabla_\mu - \Gcal_\mu) f^\mu .
 \end{split}
\end{equation}
This is just Eq.~\eqref{eq:Stokes} which we want to prove. 
This formula, which belongs to a family of the Stokes theorem, 
will be often used in the subsequent discussion.

\subsection{Nonrelativistic symmetry and conservation law}
\label{sec:Matter}
We next discuss the relation between symmetries for nonrelativistic systems
and its consequences like conservation laws 
based on gauge and coordinate reparametrization symmetry.
With the help of the geometric preliminary developed in the previous section, 
our system is now put in the presence of the external 
$U(1)_M$ and $U(1)_Q$ gauge fields and background curved (TTNC) geometry. 
Writing all the external fields together by 
$j(x) =\{n_\mu(x),v^\mu(x),h_{\mu\nu}(x),h^{\mu\nu}(x),a_\mu(x),A_\mu(x) \}$, 
we can express our action as 
\begin{equation}
 \Scal [\varphi; j] 
  = \int d^d x \sqrt{\gamma} 
  \Lcal (\varphi_i(x),\partial_\mu \varphi_i(x);
  j(x)),
\end{equation}
where $\varphi_i$ denotes a set of dynamical matter fields under consideration,
and spacetime integral covers all region where matter fields exist.
For example, the explicit form of the Lagrangian for the 
charged and/or uncharged spinless nonlinear Schr\"odinger fields 
$\phi_n$ in the Newton-Cartan geometry is given by 
\begin{equation}
 \begin{split}
  \Lcal (\phi, D_\mu \phi) =  
  \sum_{n}  \Bigg[ 
  \frac{i}{2} v^\mu \phi_n^\dag \overlr{D_\mu} \phi_n
   - \frac{1}{2m_n} h^{\mu\nu} D_\mu \phi_n^\dag D_\nu \phi_n
  - \sum_{m}\lambda_{n,m} |\phi_n|^2  |\phi_m|^2
  \Bigg],
  \label{eq:GaugedNLSchro}
 \end{split}
\end{equation}
where we defined 
$\phi_n^\dag \overlr{D_\mu} \phi_n 
\equiv \phi_n^\dag D_\mu\phi_n -\phi_n D_\mu \phi_n^\dag$
with the covariant derivative $D_\mu$:
\begin{equation}
 D_\mu \phi_n 
  \equiv (\partial_\mu + i m_n a_\mu + i q_n A_\mu) \phi_n, \quad 
  D_\mu \phi_n^\dag \equiv 
  (\partial_\mu - i m_n a_\mu - i q_n A_\mu) \phi_n^\dag,
\end{equation}
where we take a mass (electric charge) of the matter field $\phi_n$ 
as $m_n~(q_n)$.

Let us then consider a set of infinitesimal transformation with 
the general coordinate transformation, 
$U(1)_M$ and $U(1)_Q$ gauge transformation, 
and the so-called \textit{Milne boost transformation}
driven by parameters 
$\chi(x) \equiv \{ \xi^\mu(x), \alpha_M(x), \alpha_Q (x),  \Lambda_\mu(x) \}$:
\begin{equation}
 \begin{cases}
  \delta_\chi n_\mu 
  = \xi^\nu \partial_\nu n_\mu + n_\nu \partial_\mu \xi^\nu ,\\
  \delta_\chi h^{\mu\nu} 
  = \xi^\rho \partial_\rho h^{\mu\nu} 
  - h^{\mu\rho} \partial_\rho \xi^\nu - h^{\nu\rho} \partial_\rho \xi^\mu ,\\
  \delta_\chi v^\mu 
  = \xi^\nu \partial_\nu v^\mu - v^\nu \partial_\nu \xi^\mu
  +  h^{\mu\nu} \Lambda_\nu ,  \\ 
  \delta_\chi h_{\mu\nu} 
  = \xi^\rho \partial_\rho h_{\mu\nu} 
  + h_{\mu\rho} \partial_\nu \xi^\rho + h_{\nu\rho} \partial_\mu \xi^\rho 
  - (n_\mu P^\rho_\nu + n_\nu P^\rho_\mu ) \Lambda_\rho, \\
  \delta_\chi a_\mu 
  = \xi^\nu \partial_\nu a_\mu + a_\nu \partial_\mu \xi^\nu 
  + \partial_\mu \alpha_M - P^\nu_\mu \Lambda_\nu  , \\ 
  \delta_\chi A_\mu 
  = \xi^\nu \partial_\nu A_\mu + A_\nu \partial_\mu \xi^\nu 
  + \partial_\mu \alpha_Q , \\ 
  \delta_\chi \varphi_i = \lie_\xi \varphi_i
  - i \alpha_M  m_i \varphi_i
  - i \alpha_Q q_i \varphi_i ,
 \end{cases}  \label{eq:SymmetryTr}
\end{equation}
where $\chi(x) = \{\xi^\mu(x)$, $\alpha_M(x), \alpha_Q(x), \Lambda_\mu(x) \}$ 
denotes a set of arbitrary infinitesimal vector and scalar functions 
which vanish at the boundary of the spacetime region,
and $m_i~ (q_i)$ does the mass ($U(1)_Q$ charge) 
of the matter field $\varphi_i(x)$, 
respectively. 
We then consider systems whose action is invariant under these transformations%
\footnote{
When we will take into account the magnetic moment $g_s$ 
in $d=2+1$ dimensions, 
we have to slightly modify the Milne boost transformation for the gauge field 
$a_\mu$ to include the effect of $g_s$. 
In this paper, we do not take into account the magnetic moment. 
}
: 
$\delta_\chi \Scal = 0 $.
On the other hand, we can express $\delta_\chi \Scal$ 
in terms of variations of the action with respect to external fields.
We, however, have to pay attention to the fact that all variation 
of the Newton-Cartan data $\{n_\mu, h^{\mu\nu},v^\mu,h_{\mu\nu}\}$ 
is not independent due to the condition \eqref{eq:NCdata}.
For example, if we take the variation of the first relation 
of Eqs.~\eqref{eq:NCdata}, we obtain
\begin{equation}
 \delta (n_\mu v^\mu) = v^\mu \delta n_\mu + n_\mu \delta v^\mu = 0,
\end{equation}
which shows that the variations of $n_\mu$ and $v^\mu$ are
related with each other.
Following Ref.~\cite{Jensen:2014aia}, 
we make the variation of $n_\mu$ to be arbitrary, 
and, as a consequence, the variation of $\{h^{\mu\nu},v^\mu,h_{\mu\nu}\}$ 
is constrained as follows:
\begin{equation}
 \begin{cases}
  \delta v^\mu = - v^\mu v^\nu \delta n_\nu + P^\mu_\nu \delta \ptc{v}^\nu, \\
  \delta h^{\mu\nu} 
  = - ( v^\mu h^{\nu\rho} + v^\nu h^{\mu\rho} )\delta n_\rho 
  + P^\mu_\rho P^\nu_\sigma \delta \ptc{h}^{\rho\sigma}  , \\
  \delta h_{\mu\nu} = 
  - ( n_\mu h_{\nu\rho} + n_\nu h_{\mu\rho} )\delta \ptc{v}^\rho 
  - h_{\mu\rho} h_{\nu\sigma} \delta \ptc{h}^{\rho\sigma}, 
 \end{cases}
\end{equation}
where $\delta \ptc{v}^\mu$ and $\delta \ptc{h}^{\mu\nu}$ are unconstrained.
Taking into account this, 
we only have the independent variation
$\{\delta n_\mu, \delta \ptc{h}^{\mu\nu}, \delta \ptc{v}^\mu, 
\delta a_\mu, \delta A_\mu \}$ as  
\begin{equation}
 \begin{cases}
  \delta_\chi n_\mu 
  = \nabla_\mu (\xi^\nu n_\nu) - F_{\mu\nu}^n \xi^\nu, \\
  \delta_\chi \ptc{h}^{\mu\nu} 
  = P^\mu_\alpha P^\nu_\beta \delta_\chi h^{\alpha\beta} 
  = P^\mu_\alpha P^\nu_\beta 
  ( \xi^\rho \nabla_\rho h^{\alpha\beta}  - h^{\alpha\rho} \nabla_\rho \xi^\beta 
  - h^{\beta\rho} \nabla_\rho \xi^\alpha ), \\
  \delta_\chi \ptc{v}^\mu 
  = P^\mu_\nu \delta_\chi v^\nu 
  = P^\mu_\nu (\xi^\rho \nabla_\rho v^\nu - v^\rho \nabla_\rho \xi^\nu)
  + h^{\mu\nu} \Lambda_\nu,
  \\ 
  \delta_\chi a_\mu 
  = \nabla_\mu (\xi^\nu a_\nu) - F^a_{\mu\nu}\xi^\nu
  + \partial_\mu \alpha_M - P^\nu_\mu \Lambda_\nu ,  
  \\
  \delta_\chi A_\mu 
  = \nabla_\mu (\xi^\nu A_\nu) - F_{\mu\nu}\xi^\nu
  + \partial_\mu \alpha_Q  ,  
 \end{cases}
\end{equation}
where, in addition to the field strength tensors for $a_\mu$ and $n_\mu$ 
defined in Eqs.~\eqref{eq:massStrength} and \eqref{eq:TorsionalPart}, 
we further introduced the field strength tensor for $A_\mu$:
\begin{equation}
  F_{\mu\nu} \equiv \partial_\mu A_\nu - \partial_\nu A_\mu.
\end{equation}
Using these expressions together with the fact that variations of 
dynamical fields $\varphi_i$ does not contribute by the use of 
the equation of motion ($\delta \Scal/\delta \varphi_i = 0 $), 
we can express $\delta_\chi \Scal$ as
\begin{equation}
 \begin{split}
  \delta_\chi \Scal 
  &= \int d^d x 
  \left[
  \frac{\delta \Scal}{\delta n_\mu} \delta_\chi n_\mu
  + \frac{\delta \Scal}{\delta \ptc{h}^{\mu\nu}} \delta_\chi \ptc{h}^{\mu\nu} 
  + \frac{\delta \Scal}{\delta \ptc{v}^\mu } \delta_\chi \ptc{v}^\mu
  + \frac{\delta \Scal}{\delta a_\mu} \delta_\chi a_\mu
  + \frac{\delta \Scal}{\delta A_\mu} \delta_\chi A_\mu\right] \\
  &= \int d^d x \sqrt{\gamma}
  \left[
  - \Ecal^\mu\delta_\chi n_\mu
  - \frac{1}{2} T_{\mu\nu} \delta_\chi \ptc{h}^{\mu\nu} 
  - \Pcal_\mu \delta_\chi \ptc{v}^\mu
  - J_M^\mu\delta_\chi a_\mu 
  - J_Q^\mu\delta_\chi A_\mu \right] 
  \\
  &= \int d^d x \sqrt{\gamma}
  \xi^\nu \Big[
  - (\nabla_\mu - \Gcal_\mu) 
  \Tcal^\mu_{~\nu} 
  + J_M^\mu F^a_{\mu\nu} 
  + J_Q^\mu F_{\mu\nu} 
  - \Tcal^{\mu}_{~\rho} (F^n_{\mu\nu} v^\rho + n_\mu \nabla_\nu v^\rho)
  \\
  &\hspace{70pt}  
  + a_\nu (\nabla_\mu - \Gcal_\mu) J_M^\mu 
  + A_\nu (\nabla_\mu - \Gcal_\mu) J_Q^\mu \Big]  \\
  &\quad
  + \int d^d x \sqrt{\gamma}
  \Big[   
  \alpha_M (\nabla_\mu - \Gcal_\mu) J_M^\mu  
  + \alpha_Q (\nabla_\mu - \Gcal_\mu) J_Q^\mu 
  - h^{\mu\nu} \Lambda_\nu
  ( \Pcal_\mu - h_{\mu\rho} J_M^\rho ) \Big]
  + (\mathrm{surface\ terms}),
  \label{eq:deltaS}
 \end{split}
\end{equation}
where we defined a set of conserved currents---energy current $\Ecal^\mu$, 
stress tensor $T_{\mu\nu}$, 
momentum density $\Pcal_\mu$, mass current $J_M^\mu$ 
and electric current $J_Q^\mu$---by the 
variation of the action with respect to external fields:
\begin{equation}
 \Ecal^\mu \equiv - \frac{1}{\sqrt{\gamma}} \frac{\delta \Scal}{\delta n_\mu} 
 ,\quad
 T_{\mu\nu} \equiv 
 - \frac{2}{\sqrt{\gamma}} \frac{\delta \Scal}{\delta \ptc{h}^{\mu\nu}}  
 ,\quad
 \Pcal_\mu \equiv 
 - \frac{1}{\sqrt{\gamma}} \frac{\delta \Scal}{\delta \ptc{v}^\mu} 
 ,\quad
 J_M^\mu 
 \equiv - \frac{1}{\sqrt{\gamma}} \frac{\delta \Scal}{\delta a_\mu} ,\quad
 J_Q^\mu 
 \equiv - \frac{1}{\sqrt{\gamma}} \frac{\delta \Scal}{\delta A_\mu}.
 \label{eq:Jset}
\end{equation}
To obtain the last line in Eq.~\eqref{eq:deltaS}, 
we performed the integral by parts and introduced 
the nonrelativistic energy-momentum tensor $\Tcal^\mu_{~\nu}$ 
defined by a combination of conserved currents,
\begin{equation}
 \Tcal^\mu_{~\nu} \equiv - \Ecal^\mu n_\nu + v^\mu \Pcal_\nu  + T^\mu_{~\nu} ,
  \label{eq:NonrelaEM}
\end{equation}
where we raised the index of stress tensor by $h^{\mu\nu}$:
$T^\mu_{~\nu} \equiv h^{\mu\rho} T_{\rho\nu}$. 
Recalling that $\xi^\nu(x)$, $\alpha_M(x)$, $\alpha_Q(x)$ and $\Lambda_\nu(x)$ 
are arbitrary functions, 
invariance of the action ($\delta_\chi \Scal = 0$) implies 
\begin{align}
 (\nabla_\mu - \Gcal_\mu) \Tcal^\mu_{~\nu}
 &= J_M^\mu F^a_{\mu\nu} 
 + J_Q^\mu F_{\mu\nu} 
 - \Tcal^\mu_{~\rho} ( F^n_{\mu\nu}  v^\rho + n_\mu \nabla_\nu v^\rho ), 
 \label{eq:EMcons}\\
 (\nabla_\mu - \Gcal_\mu ) J_M^\mu &= 0 \label{eq:JMcons}, \\
 (\nabla_\mu - \Gcal_\mu ) J_Q^\mu &= 0 \label{eq:JQcons}, \\
 \Pcal_\mu &= h_{\mu\nu} J_M^\nu \label{eq:Milne} ,
\end{align}
the operator version of which will be used as a basic building block to derive
hydrodynamic equations. 
The first two equations provides a complete set of the 
nonrelativistic covariant conservation laws---Eq.~\eqref{eq:EMcons} gives 
the energy-momentum conservation law, 
Eq.~\eqref{eq:JMcons} the mass conservation law,
and Eq.~\eqref{eq:JQcons} the electric charge conservation law%
\footnote{
As is clearly seen from the right-hand-side of Eq.~\eqref{eq:EMcons}, 
they are associated with the source terms like the Lorentz force, and indeed, 
\textit{non}-conservation laws.
Nevertheless, we hereafter refer to Eqs.~\eqref{eq:EMcons}-\eqref{eq:JQcons} 
as the covariant conservation laws for the sake of simplicity.
}.
In fact, if we see the conserved charge density 
associated with $\Tcal^\mu_{~\nu}$, they are given by 
$n_\mu \Tcal^\mu_{~\nu} 
= - n\cdot \Ecal n_\nu + \Pcal_\nu$.
The first term of this represents the energy density, and the second term 
the momentum density. 
We can thus regard Eq.~\eqref{eq:EMcons} as 
the nonrelativistic energy-momentum conservation laws. 
We, however, note that if we defined 
$\Tcal^{\mu\nu} \equiv h^{\nu\rho}\Tcal^\mu_{~\rho}$, 
this is not necessarily symmetric with respect to their indices: 
$\Tcal^{\mu\nu} \neq \Tcal^{\nu\mu}$.
Contrary to the first three equations, 
the fourth equation $\eqref{eq:Milne}$ does not 
provide the conservation law but provides the relation between 
the momentum density $\Pcal_\mu$ and the mass current $J_M^\mu$, 
which is well known for nonrelativistic systems. 
This results from the fact that 
the Milne boost invariance is not the dynamical symmetry of our action; 
in other words, as is demonstrated in Eq.~\eqref{eq:SymmetryTr}, 
we do not have the transformation of the dynamical fields $\varphi_i$ unlike 
the gauge and coordinate transformations.
The Milne boost invariance thus represents the redundancy how we describe 
the external field applied to the systems 
by the use of $\{v^\mu, h_{\mu\nu}, a_\mu\}$.
Although this relation does not look so important compared to the 
conservation laws, 
it will lead to the worthwhile result that the electrical conductivity 
vanishes for single-component  nonrelativistic systems as discussed in 
Sec.~\ref{sec:TimeEvo}-\ref{sec:Derivation}.
We also note that our covariant action under consideration%
---e.g. one constructed by \eqref{eq:GaugedNLSchro}---%
enjoys symmetry under the \textit{finite} Milne boost transformation given by
\begin{equation}
 \begin{cases}
  \delta_\Lambda v^\mu
  = h^{\mu\nu} \Lambda_\nu , \\
  \delta_\Lambda h_{\mu\nu} 
  =  - ( n_\mu P_{\nu}^{\rho} + n_{\nu} P_{\mu}^{\rho}  ) \Lambda_\rho
  + n_{\mu} n_{\nu} h^{\rho\sigma} \Lambda_\rho \Lambda_\sigma, \\
  \delta_\Lambda a_\mu
   = - P^{\nu}_{\mu} \Lambda_{\nu}  
  + \dfrac{1}{2} n_{\mu} h^{\rho\sigma} \Lambda_\rho \Lambda_\sigma ,
 \end{cases}
 \label{eq:finiteMilne}
\end{equation}
where $\Lambda_\mu (x)$ in this equation denotes 
not the infinitesimal vector but the finite vector.
This finite Milne boost invariance becomes important when we will consider 
the path-integral formula associated with local thermal equilibrium in 
Sec.~\ref{sec:PathIntegral}.

\subsection{Local Gibbs distribution}
\label{sec:LG}
As is the case for relativistic quantum field theories~\cite{Hayata:2015lga,Hongo:2016mqm}, 
we introduce the \textit{local} Gibbs distribution for the density operator
to describe the nonrelativistic system in local thermal equilibrium.
The local Gibbs distribution is a generalization of 
the (global) Gibbs distribution employed in the grandcanonical ensemble.
It provides a way to describe locally thermalized systems 
in terms of a set of the (intensive) local thermodynamic parameters 
$\lambda^a(x) \equiv \{\beta^\mu(x),\nu'_M(x), \nu'_Q(x) \}$\,---the conjugate 
local thermodynamic variables, or the Lagrange multipliers 
to determine the average values of 
the conserved charge densities $\hc_a(x) \equiv \{\hp_{\nu}(x), \hn_M'(x), 
\hn_Q'(x) \}$ 
with $\hp_\nu(x) \equiv n_\mu(x) \hTcal^\mu_{~\nu}(x)$, 
$\hn_M'(x) \equiv n_\mu (x)\hJ_M^\mu(x)$, and 
$\hn_Q'(x) \equiv n_\mu (x)\hJ_Q^\mu(x)$.
The explicit form of the local Gibbs distribution%
\footnote{
The local Gibbs distribution is uniquely determined by solving 
the maximization problem with constraints on the conserved charge densities
(See e.g. Ref.~\cite{Hongo:2016mqm} in detail). 
} is given by
\begin{equation}
 \hrhoLG [\pt;\lambda] 
  \equiv \exp \left( - \hS[\pt;\lambda] \right), \with
  \hS [\pt;\lambda] \equiv \hK [\pt;\lambda] + \Psi [\pt;\lambda], 
  \label{eq:DefLG}
\end{equation}
where $\hK[\pt;\lambda]$ is defined as 
\begin{equation}
 \begin{split}
  \hK[\pt;\lambda] 
  &\equiv 
  - \int d \Sigma_{\pt \mu} \hcurrent^\mu_{~a} \lambda^a
  = - \int d \Sigma_{\pt \mu}
  \left( \hTcal^\mu_{~\nu}(x) \beta^\nu(x)  
  + \hJ_M^\mu (x) \nu'_M(x) + \hJ_Q^\mu (x) \nu'_Q(x)  \right).
  \label{eq:DefK}
 \end{split}
\end{equation}
Here we used the hypersurface vector $d\Sigma_{\pt\mu} = d\Sigma_{\pt} n_\mu$ 
introduced in the previous section, and 
$\hcurrent^\mu_{~a}(x) \equiv \{\hTcal^\mu_{~\nu}(x), \hJ_M^\mu(x), 
\hJ_Q^\mu(x) \}$ denotes a set of conserved current operators.
We also introduced the normalization factor $\Psi[\pt;\lambda,j]$ which 
is identified as the local thermodynamic functional, or the so-called 
Massieu-Planck functional:
\begin{equation}
 \Psi [\pt;\lambda] = \log \Tr \exp ( - \hK [\pt;\lambda]).
\end{equation}
This functional is one of the most important quantities in our formulation 
since it is shown to contain complete information on transport properties 
of systems in local thermal equilibrium;
namely, as will be shown in the next section, 
we can extract the average value of all the conserved current operators
$\averageLG{\hcurrent^\mu_{~a}(x)}_{\pt}$ 
from this single functional $\Psi[\pt;\lambda,j]$.
Here we defined 
$\averageLG{\hOcal}_{\pt} \equiv \Tr (\hrhoLG[\pt;\lambda] \hOcal)$ 
representing the average value of an arbitrary operator $\hOcal$
in local thermal equilibrium at time $\pt$.

Before moving to the discussion on average values of conserved currents 
in local thermal equilibrium, we here quickly review some basic properties 
of thermodynamic functionals for later use.
As is clear from the definition of the Massieu-Planck functional, 
taking variation of $\Psi[\pt;\lambda,j]$ with respect to $\lambda^a(x)$ 
provides the average value of the conserved charge densities over 
local Gibbs distribution:
\begin{equation}
 c_a(x) \big|_{\pt} \equiv \averageLG{\hc_a(x)}_{\pt} 
  = \frac{\delta \Psi [\pt;\lambda]}{\delta \lambda^a(x)}. 
  \label{eq:delPsi}
\end{equation}
This relation shows that the Massieu-Planck functional is properly 
regarded as the local thermodynamic functional.
Furthermore, we introduce the entropy functional $S[\pt;c]$
for local thermal equilibrium as 
\begin{equation}
 \begin{split}
  S[\pt;c] 
  &\equiv \averageLG{\hS [\pt;\lambda]}_{\pt} 
  = - \Tr 
  \big( \log \hrhoLG[\pt;\lambda] \log \hrhoLG[\pt;\lambda] \big) \\
  &= - \int d\Sigma_{\pt} \lambda^a (x) c_a (x) + \Psi[\pt;\lambda,j],
 \end{split}
\end{equation}
which is nothing but the Legendre transformation of $\Psi[\pt;\lambda]$%
\footnote{
To be precise, based on the convexity of 
the Massieu-Planck functional $\Psi[\pt;\lambda,j]$, 
we define the entropy functional $S[\pt;c]$ 
in accordance with the proper manner of the Legendre transformation:
\begin{equation}
 S[\pt;c] \equiv \inf_{\lambda} 
  \left[ \int d \Sigma_{\pt} \lambda^a c_a + \Psi [\pt;\lambda] \right].
\end{equation}
}. 
We note that the natural variables for the entropy functional is 
not $\lambda^a(x)$ but conserved charge densities $c_a(x)$ as usual.
This is clarified by taking a variation of the entropy functional:
\begin{equation}
 \begin{split}
  \delta S 
  &= - \int d \Sigma_{\pt}
  \big( \lambda^a \delta c_a  + c_a \delta \lambda^a \big)
  + \delta \Psi[\pt;\lambda,j] \\ 
  &= - \int d \Sigma_{\pt}
  \big( \lambda^a(x) \delta c_a (x)  + c_a (x)\delta \lambda^a(x) \big)
  + \int d\Sigma_{\pt} 
  \delta \lambda^a (x) \frac{\delta \Psi [\pt;\lambda]}{\delta \lambda^a(x)} \\ 
  &= - \int d \Sigma_{\pt}  \lambda^a(x) \delta c_a (x)  
 \end{split}
\end{equation}
where we used Eq.~\eqref{eq:delPsi} to proceed the last line.
We therefore obtain the following formula for local thermodynamics 
parameters $\lambda^a(x)$:
\begin{equation}
 \lambda^a (x) \big|_{\pt} = - \frac{\delta S[\pt;c]}{\delta c_a(x)} .
  \label{eq:delS}
\end{equation}
It is worthwhile to emphasize that we have the one-to-one correspondence 
between the local thermodynamic parameters $\lambda^a(x)$ and 
the averaged conserved charge densities $c_a(x)$ owing to 
the convexity of the Massieu-Planck functional $\Psi[\pt;\lambda,j]$.

\section{Path-integral formula for local thermal equilibrium}
\label{sec:PathIntegral}
In this section, we develop the imaginary-time path-integral 
formalism for spinless Schr\"odinger fields
in local thermal equilibrium, 
which enables us to evaluate nondissipative part of constitutive relations 
(See Ref.\,\cite{Hongo:2016mqm} in the relativistic case).
In Sec.~\ref{sec:Variation}, we first show that 
the Massieu-Planck functional is regarded as a generating functional 
for the average value of conserved current operators over 
the local Gibbs distribution $\averageLG{\hcurrent^\mu_{~a}(x)}_{\pt}$. 
We provide the exact variational formula 
based on invariance of the functional operator $\hK[\pt;\lambda]$.
We also introduce a useful gauge choice, 
a \textit{hydrostatic gauge}, which simplify our calculation. 
In Sec.~\ref{sec:PathIntegral1}, 
explicitly dealing with spinless Bosonic and Fermionic Schr\"odinger fields, 
we obtain the path-integral formula for 
the Massieu-Planck functional, and show that it is written in terms of 
quantum field theories in the thermally emergent Newton-Cartan geometry. 
In Sec.~\ref{sec:Symmetry}, we summarize the symmetry properties 
of the Massieu-Planck functional.

\subsection{Variational formula for nondissipative constitutive relation}
\label{sec:Variation}
Here we provide the variational formula for the Massieu-Planck functional 
and show that it is regarded as a generating functional for 
$\averageLG{\hcurrent^\mu_{~a}(x)}_{\pt}$. 
After deriving the variational formula without a gauge fixing in 
Sec.~\ref{sec:Variation1}, we introduce the useful gauge 
which we call hydrostatic gauge in Sec.~\ref{sec:Variation2}. 

\subsubsection{Derivation of variational formula in general setup}
\label{sec:Variation1}
We here show the derivation of the variational formula in a general setup 
without gauge fixing. 
First of all, we rewrite $\hK[\pt;\lambda]$ as 
\begin{equation}
 \begin{split}
  \hK[\pt,\lambda,j] 
  = \int d^dx  \delta (\pt - \pt (x) )
  \hkappa (\lambda^a,j) , \with 
  \hkappa(\lambda^a,j) 
  \equiv  \sqrt{\gamma} N^{-1}  n_\mu \hcurrent^\mu_{~a} \lambda^a,
 \end{split}
\end{equation}
from which we can see that this functional operator is 
manifestly invariant under the coordinate reparametrization.
In order to emphasize arguments of functional operator $\hK[\pt;\lambda]$, 
we explicitly write down them. 
We also note that this functional operator is $U(1)_M$ and $U(1)_Q$
gauge invariant. 
Let us then consider the combination of 
the infinitesimal coordinate reparametrization 
and $U(1)_{M,Q}$ gauge transformation to the specific direction 
whose parameters are given by  
$\xi^\mu(x) = \epsilon \beta^\mu(x)$, 
$\alpha_M (x) = - \epsilon(\nu'_M + \beta \cdot a)$, and 
$\alpha_Q (x) = - \epsilon(\nu'_Q + \beta \cdot A)$. 
Invariance of $\hK[\pt,\lambda,j]$ under this transformation 
leads to the following operator identity%
\footnote{
We can also derive this operator identity in the following simple way.
For that purpose, we note that the functional operator 
$\epsilon \hK[\pt;\lambda]$ 
generates a set of the coordinate reparametrization and 
gauge transformation with parameters 
$\xi^\mu(x) = \epsilon \beta^\mu(x)$,
$\alpha_M (x) = \epsilon(\nu'_M - \beta \cdot a)$, 
and $\alpha_Q (x) = \epsilon(\nu'_Q - \beta \cdot A)$.
Then, we can immediately see that the above identity is nothing but 
the commutativity of themselves:
$\big[ i\epsilon \hK[\pt;\lambda,j], \hK[\pt,\lambda,j] \big] 
= \delta_\lambda \hK [\pt,\lambda,j] = 0$
}
\begin{equation}
 \delta_\lambda \hK[\pt,\lambda,j] 
  \equiv 
  \hK[\pt + \epsilon \lie_\beta \pt, \lambda^a + \lie_\beta \lambda^a,
  j + \epsilon \delta_\lambda  j]
  - \hK[\pt, \lambda^a, j]  = 0 ,
\end{equation}
where $\lie_\beta$ denotes the Lie derivative along $\beta^\mu$. 
We write the variation of external fields as $\delta_\lambda j$ to reveal that 
our variation of the $U(1)_{M,Q}$ gauge field is not 
a simple Lie derivative but given by 
\begin{align}
 \delta_\lambda a_\mu 
  &\equiv \lie_\beta a_\mu - \nabla_\mu (\nu'_M + \beta \cdot a)
  = - \beta^\nu F^a_{\mu\nu} - \nabla_\mu \nu'_M , 
 \label{eq:dela}\\
 \delta_\lambda A_\mu 
  &\equiv \lie_\beta A_\mu - \nabla_\mu (\nu'_Q + \beta \cdot A)
  = - \beta^\nu F_{\mu\nu} - \nabla_\mu \nu'_Q , 
  \label{eq:delA}
\end{align}
while the variations of the other external fields, or 
the Newton-Cartan data is simply given by their Lie derivatives:
e.g. $\delta_\lambda n_\mu = \lie_\beta n_\mu$.
We note that the lie derivatives of $\nu_{M,Q}$ are expressed as
$\lie_\beta \nu'_M = - \beta^\mu \delta_\lambda a_\mu$ 
and $\lie_\beta \nu'_Q = - \beta^\mu \delta_\lambda A_\mu$ 
due to Eqs.~\eqref{eq:dela}-\eqref{eq:delA}.
Based on these, we can express $\delta_\lambda \hK[\pt,\lambda,j]$ 
in terms of the variations as
\begin{equation}
 \begin{split}
  \delta_\lambda \hK
  &=  \int d^d x
  \epsilon \left[ \frac{\delta \hK}{\delta \pt(x)} \lie_\beta \pt(x)
  + \frac{\delta \hK}{\delta \lambda^a(x)} \lie_\beta \lambda^a(x)
  + \frac{\delta \hK}{\delta j_a (x)} \delta_\lambda j_a (x) \right]  
  \\
  &= \int d^d x
  \epsilon \Bigg[
  \sqrt{\gamma} \delta (\pt - \pt(x)) 
  \left(
  \hEcal^\mu \lie_\beta n_\mu  
  + \hPcal_\mu \lie_\beta \ptc{v}^\mu
  + \frac{1}{2} \hT_{\mu\nu} \lie_\beta \ptc{h}^{\mu\nu}
  + \hJ_M^\mu \delta_\lambda a_\mu 
  + \hJ_Q^\mu \delta_\lambda A_\mu \right)
  \beta^\mu \partial_\mu \pt
  \\ &\hspace{24pt}
  + \frac{\delta \hK}{\delta n_\mu} \lie_\beta n_\mu
  + \frac{\delta \hK}{\delta \ptc{v}^\mu} \lie_\beta \ptc{v}^\mu
  + \frac{\delta \hK}{\delta \ptc{h}^{\mu\nu}} \lie_\beta \ptc{h}^{\mu\nu}
  + \left(  \frac{\delta \hK}{\delta a_\mu} 
  - \frac{\delta \hK}{\delta \nu'_M} \beta^\mu  \right) \delta_\lambda a_\mu 
  + \left( \frac{\delta \hK}{\delta A_\mu} 
  - \frac{\delta \hK}{\delta \nu'_Q} \beta^\mu \right) \delta_\lambda A_\mu 
  \Bigg]
  \\
  &= \int d^d x
  \epsilon 
  \Bigg[
  \left( \beta \sqrt{h}\, \hEcal^\mu \big|_{\pt} 
  + \frac{\delta \hK}{\delta n_\mu} \right) \lie_\beta n_\mu
  + \left( \beta \sqrt{h} \hPcal_\mu \big|_{\pt} 
  + \frac{\delta \hK}{\delta \ptc{v}^\mu} \right) \lie_\beta \ptc{v}^\mu 
  + \left( \frac{\beta \sqrt{h}}{2} \hT_{\mu\nu} \big|_{\pt} 
  + \frac{\delta \hK}{\delta \ptc{h}^{\mu\nu}} \right) 
  \lie_\beta \ptc{h}^{\mu\nu} 
  \\ &\hspace{42pt}
  + \left( \beta \sqrt{h} \hJ_M^\mu \big|_{\pt}
  - \frac{\delta \hK}{\delta \nu'_M} \beta^\mu 
  + \frac{\delta \hK}{\delta a_\mu}
  \right) \delta_\lambda a_\mu
  + \left( \beta \sqrt{h} \hJ_Q^\mu \big|_{\pt}
  - \frac{\delta \hK}{\delta \nu'_Q} \beta^\mu 
  + \frac{\delta \hK}{\delta A_\mu}
  \right) \delta_\lambda A_\mu \Bigg],
 \end{split}
 \label{eq:deltaK}
\end{equation}
where we defined 
$\beta (x) \equiv n_\mu (x) \beta^\mu (x)= N (x) \beta^{\ptc{0}} (x)$ 
and used the following expression for the first term in the first line:
\begin{equation}
 \begin{split}
  \frac{\delta \hK}{\delta \pt(x)}
  &= \frac{\delta}{\delta \pt (x)}
  \int  d^d x \sqrt{\gamma} \theta (\pt - \pt(x)) 
  (\nabla_\mu - \Gcal_\mu) (\hcurrent^\mu_{~a} \lambda^a) \\ 
  &= -  \int  d^d x \sqrt{\gamma} \delta (\pt - \pt(x)) 
  (\nabla_\mu - \Gcal_\mu) (\hcurrent^\mu_{~a} \lambda^a) \\ 
  &= \int  d^d x \sqrt{\gamma} \delta (\pt - \pt(x)) 
  \left( 
  \hEcal^\mu \lie_\beta n_\mu
  + \hPcal_\mu \lie_\beta v^\mu
  + \frac{1}{2} \hT_{\mu\nu} \lie_\beta h^{\mu\nu}
  + \hJ_M^\mu \delta_\lambda a_\mu 
  + \hJ_Q^\mu \delta_\lambda A_\mu  \right).
 \end{split}
\end{equation}
Here we used Eq.~\eqref{eq:Step} to rewrite $\hK[\pt,\lambda,j]$ by the use of
the step function $\theta (\pt - \pt(x))$, and substitute
the following results for the covariant divergence of 
$\hcurrent^\mu_{~a} \lambda^a$ with the help of 
the conservation laws \eqref{eq:EMcons}-\eqref{eq:JQcons}:
\begin{equation}
 \begin{split}
  (\nabla_\mu - \Gcal_\mu) \big(\hcurrent^\mu_{~a} \lambda^a\big) 
  &= \hTcal^\mu_{~\nu} 
  \Big( 
  \nabla_\mu \beta^\nu
  - \beta^\rho ( F^n_{\mu\rho} v^\nu + n_\mu \nabla_\rho v^\nu ) 
  \Big) 
  + \hJ_M ^\mu (\beta^\nu F^a_{\mu\nu} + \nabla_\mu \nu'_M) 
  + \hJ_Q^\mu (\beta^\nu F_{\mu\nu} + \nabla_\mu \nu'_Q) \\
  &= ( - \hEcal^\mu n_\nu + v^\mu \hPcal_\nu + \hT^\mu_{~\nu}) 
  \nabla_\mu \beta^\nu 
  + \hEcal^\mu F^n_{\mu\nu} \beta^\nu
  - \hPcal_\mu \beta^\nu \nabla_\nu v^\mu
  - \hJ_M^\mu \delta_\lambda a_\mu
  - \hJ_Q^\mu \delta_\lambda A_\mu \\
  &=  - \hEcal^\mu \lie_\beta n_\mu
  - \hPcal_\mu \lie_\beta \ptc{v}^\mu 
  - \frac{1}{2} \hT_{\mu\nu} \lie_\beta \ptc{h}^{\mu\nu}
  - \hJ_M^\mu \delta_\lambda a_\mu 
  - \hJ_Q^\mu \delta_\lambda A_\mu .
  \label{eq:Divergence}
 \end{split}
\end{equation}
We then take average of Eq.~\eqref{eq:deltaK} 
over the local Gibbs distribution at that time $\pt$, which results in 
\begin{equation}
 \begin{split}
  \averageLG{\delta_\lambda \hK}_{\pt} 
  &= \int d^d x
  \epsilon \Bigg[
  \left( \beta\sqrt{h} \averageLG{\hEcal^\mu}_{\pt} 
  - \frac{\delta \Psi}{\delta n_\mu} \right) \lie_\beta n_\mu 
  \\
  &\hspace{42pt}
  + \left( \beta\sqrt{h} \averageLG{\hPcal_\mu}_{\pt}
  - \frac{\delta \Psi}{\delta \ptc{v}^\mu} \right) \lie_\beta \ptc{v}_\mu
  + \left( \frac{\beta\sqrt{h}}{2} \averageLG{\hT_{\mu\nu}}_{\pt} 
  - \frac{\delta \Psi}{\delta \ptc{h}^{\mu\nu}} \right) 
  \lie_\beta \ptc{h}^{\mu\nu}
  \\
  &\hspace{42pt}
  + \left( \beta \sqrt{h} \averageLG{\hJ_M^\mu}_{\pt}
  + \frac{\delta \Psi}{\delta \nu'_M} \beta^\mu 
  - \frac{\delta \Psi}{\delta a_\mu}
  \right) \delta_\lambda a_\mu
  + \left( \beta \sqrt{h} \averageLG{\hJ_Q^\mu}_{\pt}
  + \frac{\delta \Psi}{\delta \nu'_Q} \beta^\mu 
  - \frac{\delta \Psi}{\delta A_\mu}
  \right) \delta_\lambda A_\mu \Bigg] ,
 \end{split}
\end{equation}
where we replaced the average value of the variation of $\hK[\pt,\lambda,j]$ 
with respect to external fields 
as the variation of the Massieu-Planck functional.
Therefore, using the identity $\averageLG{\delta_\lambda \hK}_{\pt} = 0 $, 
we have eventually obtained a set of the variational 
formulae for the Massieu-Planck functional:
\begin{equation}
 \begin{cases}
  \averageLG{\hEcal^\mu(x)}_{\pt}  
  = \dfrac{1}{\beta \sqrt{h}}
  \dfrac{\delta \Psi[\pt;\lambda,j]}{\delta n_\mu(x)}, \vspace{5pt} \\
  \averageLG{\hPcal_\mu(x)}_{\pt}  
  =  \dfrac{1}{\beta \sqrt{h}}
  \dfrac{\delta \Psi[\pt;\lambda,j]}{\delta \ptc{v}^\mu(x)}, \vspace{5pt} \\
  \averageLG{\hT_{\mu\nu}(x)}_{\pt}  
  = \dfrac{2}{\beta \sqrt{h}}
  \dfrac{\delta \Psi[\pt;\lambda,j]}{\delta \ptc{h}^{\mu\nu}(x)}, \\   
 \end{cases}
 \quad
 \begin{cases}
  \averageLG{\hJ_M^\mu(x)}_{\pt}  
  = - \dfrac{1}{\beta \sqrt{h}}
  \left(
  \dfrac{\delta \Psi[\pt;\lambda,j]}{\delta \nu'_M (x)} \beta^\mu
  - \dfrac{\delta \Psi[\pt;\lambda,j]}{\delta a_\mu(x)}
  \right), \vspace{5pt} \\
  \averageLG{\hJ_Q^\mu(x)}_{\pt}  
  = - \dfrac{1}{\beta \sqrt{h}}
  \left(
  \dfrac{\delta \Psi[\pt;\lambda,j]}{\delta \nu'_Q(x)} \beta^\mu
  - \dfrac{\delta \Psi[\pt;\lambda,j]}{\delta A_\mu(x)}
  \right), 
 \end{cases} 
\label{eq:VarGeneral}
\end{equation}
Combining the first three equations 
we can construct the variational formula 
for nonrelativistic energy-momentum tensor as 
\begin{align}
 \averageLG{\hTcal^\mu_{~\nu}(x)}_{\pt} 
  &= \frac{1}{\beta \sqrt{h}}
  \left(
  - \frac{\delta \Psi[\pt;\lambda,j]}{\delta n_\mu(x)} n_\nu(x)
  + v^\mu (x) \frac{\delta \Psi[\pt;\lambda,j]}{\delta \ptc{v}^\nu(x)} 
  + 2 h^{\mu\rho} (x)
 \frac{\delta \Psi[\pt;\lambda,j]}{\delta \ptc{h}^{\rho\nu}(x)}
  \right). \label{eq:Tleq}
\end{align}
These variational formulae relate the expectation values of 
all the conserved charge currents in local thermal equilibrium 
to the single functional $\Psi[\pt;\lambda,j]$.
Therefore, once we evaluate the Massieu-Planck functional $\Psi[\pt;\lambda,j]$,
we can immediately calculate the average values of the conserved current 
operators over the local Gibbs distribution by the use of these formulae. 
We thus only need to evaluate the Massieu-Planck functional 
$\Psi[\pt;\lambda,j]$ to obtain $\averageLG{\hcurrent^\mu_{~a}(x)}_{\pt}$. 
Note that these variational formulae are exact in the sense that 
we do not perform the derivative expansion at this stage. 
Of course, in order to derive the usual hydrodynamic equations 
in Sec.~\ref{sec:Derivation}, we will perform the derivative expansion 
of $\Psi[\pt;\lambda,j]$. 
Nevertheless, this formula can be applied in more general situations 
where some components of derivative expansion breaks down%
\footnote{Such an interesting situation is often realized in condensed matter system 
under a strong magnetic field, 
which will be discussed in our subsequent paper \cite{Hongo}.}.

\subsubsection{Variational formula in the hydrostatic gauge}
\label{sec:Variation2}
In the previous subsection, we derived the variational formulae 
\eqref{eq:VarGeneral} in the general setup without 
choosing any special coordinate system or special gauge. 
Although variational formulae \eqref{eq:VarGeneral}
tell us complete information on nondissipative transports taking place 
in local thermal equilibrium, it is more useful to reexpress them 
in a special gauge which we call the \textit{hydrostatic gauge}. 
Furthermore, it is also notable that we can considerably simplify 
the derivation of variation formulae in that gauge. 
Thus, we here rederive and reexpress the variational formulae for 
the Massieu-Planck functional in the hydrostatic gauge. 

\begin{figure}[t]
 \centering
 \includegraphics[width=0.95\linewidth]{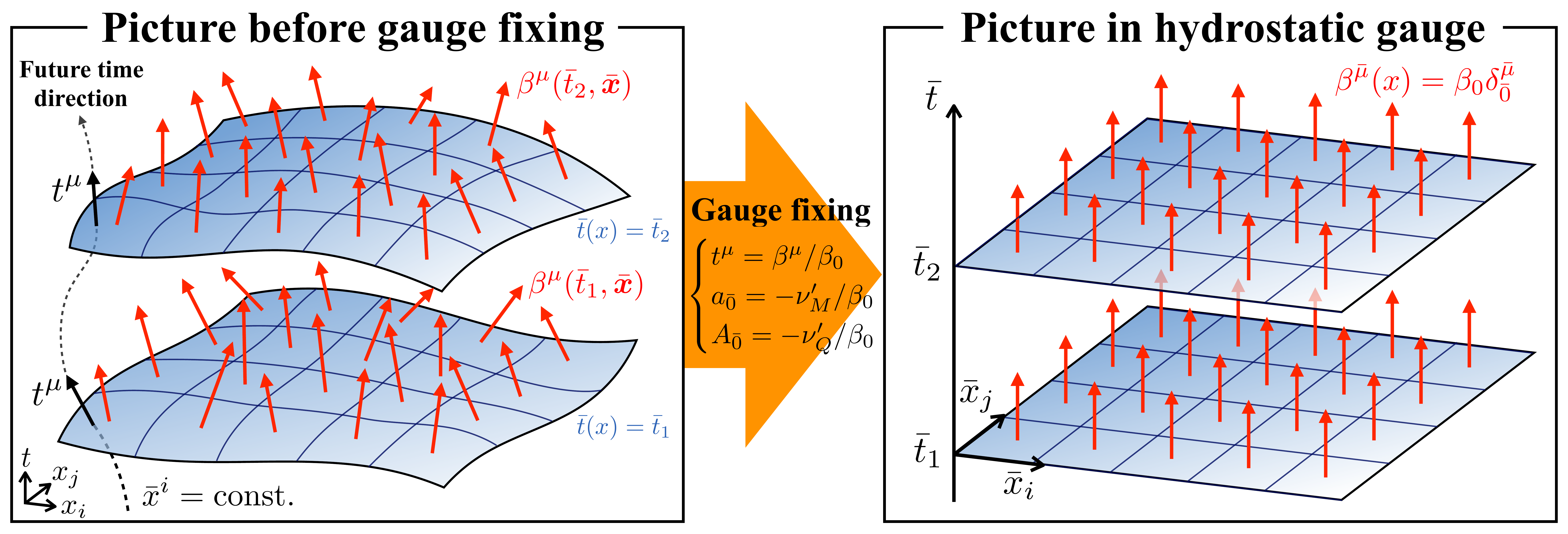}
 \caption{
 Schematic picture of the hydrostatic gauge fixing. 
 While we have an arbitrary inhomogeneous fluid vector configuration 
 $\beta^\mu (x)$ before the hydrostatic gauge fixing (left figure), 
 we have homogeneously ordered fluid vector 
 $\beta^{\ptc{\mu}}(x) \big|_{\mathrm{hs}} 
 = \beta_0 \delta^{\ptc{\mu}}_{\pzero}$ in the hydrostatic gauge (right figure).
  }
 \label{Fig:hydrostatic-gauge}
\end{figure}

The key identity in our discussion is the time derivative of 
the Massieu-Planck functional:
\begin{equation}
 \begin{split}
  \partial_{\pt} \Psi[\pt;\lambda,j]
  &= - \averageLG{\partial_{\pt} \hK [\pt;\lambda,j]}_{\pt}  
  = \baverageLG{ \partial_{\pt} 
  \int d\Sigma_{\pt\mu} \hcurrent^\mu_{~a} \lambda^a }_{\pt} \\
  &= \baverageLG{\int d\Sigma_{\pt} N 
  ( \nabla_\mu - \Gcal_\mu ) \big( \hcurrent^\mu_{~a} \lambda^a \big) }_{\pt} \\
  &= - \baverageLG{\int d\Sigma_{\pt} N 
  \Big( \hEcal^\mu \lie_\beta n_\mu
  + \hPcal_\mu \lie_\beta \ptc{v}^\mu 
  + \frac{1}{2} \hT_{\mu\nu} \lie_\beta \ptc{h}^{\mu\nu}
  + \hJ_M^\mu \delta_\lambda a_\mu 
  + \hJ_Q^\mu \delta_\lambda A_\mu  \Big)}_{\pt} ,
 \end{split}
 \label{eq:DtPsi}
\end{equation}
where we used Stokes theorem~\eqref{eq:Stokes} and 
Eq.~\eqref{eq:Divergence} to proceed the second and third line, respectively. 
This is a general identity without gauge fixing.
Then, let us introduce the hydrostatic gauge by matching
the time-direction vector $t^\mu(x)$ in Eq.~\eqref{eq:Timevector}, 
and zeroth component of gauge fields 
$a_{\pzero}(x) = t^\mu (x) a_\mu (x)$ and $A_{\pzero} = t^\mu(x) A_\mu (x)$ 
with local thermodynamic parameters $\lambda^a(x)$: 
\begin{equation}
 t^\mu (x) = \beta^\mu (x)/\beta_0, \quad
  a_{\pzero} (x) = - \nu'_M (x) / \beta_0, \quad 
  A_{\pzero} (x) = - \nu'_Q (x) / \beta_0,
  \label{eq:HydrostaticGauge}
\end{equation}
where $\beta_0$ denotes an arbitrary constant, 
or reference inverse temperature.
This equation is a hydrostatic gauge fixing condition. 
The second and third conditions in Eq.~\eqref{eq:HydrostaticGauge} are common;
it means that we can regard the chemical potential as the zeroth component of 
corresponding gauge fields in the hydrostatic gauge. 
The first equation enables us to see our inhomogeneous fluid configurations 
in an extremely simplified way. 
Indeed, as is shown in Fig.~\ref{Fig:hydrostatic-gauge}, 
our fluid vector $\beta^\mu (x)$ becomes a homogeneous constant vector 
directed to the time direction in the hydrostatic $(\pt,\pbx)$-coordinate.
Thus, in this coordinate system, the fluid looks entirely at rest, 
which is the origin of the name \textit{hydrostatic} gauge. 
Nevertheless, it is important to emphasize that our system is, in general, not 
stationary at all contrary to its hydrostatic appearance. 
This results from the fact that our fluid vector $\beta^\mu$, or 
the time-direction vector $t^\mu$ in the hydrostatic gauge, is not 
a killing vector: $\lie_t g_{\mu\nu} \neq 0$. 
Therefore, to choose the hydrostatic gauge in the whole spacetime, 
we need to track the time-dependent fluid vector $\beta^\mu$ to match it with
the time-direction vector at every moment. 

Then, let us derive the variational formula in the hydrostatic gauge. 
Thanks to the hydrostatic gauge condition \eqref{eq:HydrostaticGauge},
we can replace $\delta_\lambda a_\mu$ and $\delta_\lambda A_\mu$ 
in the hydrostatic gauge by the simple lie derivatives:
\begin{equation}
 \begin{split}
  \delta_\lambda a_\mu \big|_{\mathrm{hs}} 
  &= \lie_\beta a_\mu - \nabla_\mu ( \nu_M' + \beta \cdot a)
  \big|_{\nu_M' = - \beta \cdot a}
  = \lie_\beta a_\mu,
  \\
  \delta_\lambda A_\mu \big|_{\mathrm{hs}} 
  &= \lie_\beta A_\mu - \nabla_\mu ( \nu_Q' + \beta \cdot a)
  \big|_{\nu_Q' = - \beta \cdot A}
  = \lie_\beta A_\mu.
 \end{split}
\end{equation}
Using these relations together with 
$\lie_t = (\beta_0)^{-1} \lie_\beta \beta^\mu$ which also results from the 
gauge fixing condition \eqref{eq:HydrostaticGauge}, 
we obtain the following expression for $\partial_{\pt} \Psi [\pt;\lambda,j]$:
\begin{equation}
 \begin{split}
  \partial_{\pt} \Psi[\pt;j] \big|_{\mathrm{hs}}
  &= - \int d^{d-1} \px \sqrt{\gamma} \beta_0 
  \Big( \averageLG{\hEcal^\mu}_{\pt} \lie_t n_\mu
  + \averageLG{\hPcal_\mu}_{\pt} \lie_t \ptc{v}^\mu 
  + \frac{1}{2} \averageLG{\hT_{\mu\nu}}_{\pt} \lie_t \ptc{h}^{\mu\nu} 
  + \averageLG{\hJ_M^\mu}_{\pt} \lie_t a_\mu 
  + \averageLG{\hJ_Q^\mu}_{\pt} \lie_t A_\mu \Big) .
 \end{split}
\end{equation}
On the other hand, the left-hand-side of this equation is nothing but 
the lie derivative of the Massieu-Planck functional 
$\Psi[\pt;j]\big|_{\mathrm{hs}}$ along the time-direction vector $t^\mu$, 
and thus, we can simply express it in terms of variations as 
\begin{equation}
 \begin{split}
  \partial_{\pt} \Psi[\pt;j] \big|_{\mathrm{hs}}
  &= \int d^{d-1} \px 
  \left( \frac{\delta \Psi}{\delta n_\mu} \lie_t n_\mu 
  + \frac{\delta \Psi}{\delta \ptc{v}^\mu} \lie_t \ptc{v}^\mu 
  + \frac{\delta \Psi}{\delta \ptc{h}^{\mu\nu}} \lie_t \ptc{h}^{\mu\nu}
  + \frac{\delta \Psi}{\delta a_\mu} \lie_t a_\mu 
  + \frac{\delta \Psi}{\delta A_\mu} \lie_t A_\mu  
  \right) .
 \end{split}
\end{equation}
Therefore, matching these equation immediately provides 
the following simple variational formulae in the hydrostatic gauge: 
\begin{equation}
 \begin{cases}
  \averageLG{\hEcal^\mu(x)}_{\pt}  
  = - \dfrac{1}{\beta_0 \sqrt{\gamma}}
  \dfrac{\delta \Psi[\pt;j]}{\delta n_\mu(x)} \Bigg|_{\mathrm{hs}},
  \vspace{5pt} \\
  \averageLG{\hPcal_\mu(x)}_{\pt}  
  = - \dfrac{1}{\beta_0 \sqrt{\gamma}}
  \dfrac{\delta \Psi[\pt;j]}{\delta \ptc{v}^\mu(x)} \Bigg|_{\mathrm{hs}} ,
   \vspace{5pt} \\
  \averageLG{\hT_{\mu\nu}(x)}_{\pt}  
  = - \dfrac{2}{\beta_0 \sqrt{\gamma}}
  \dfrac{\delta \Psi[\pt;j]}{\delta \ptc{h}^{\mu\nu}(x)} \Bigg|_{\mathrm{hs}},
 \end{cases} 
 \quad
 \begin{cases}
  \averageLG{\hJ_M^\mu(x)}_{\pt}  
  = - \dfrac{1}{\beta_0 \sqrt{\gamma}}
  \dfrac{\delta \Psi[\pt;j]}{\delta a_\mu(x)} \Bigg|_{\mathrm{hs}}, 
   \vspace{5pt}\\
  \averageLG{\hJ_Q^\mu(x)}_{\pt}  
  = - \dfrac{1}{\beta_0 \sqrt{\gamma}}
  \dfrac{\delta \Psi[\pt;j]}{\delta A_\mu(x)} \Bigg|_{\mathrm{hs}} .
 \end{cases}
 \label{eq:VarHS}
\end{equation}
From these variational formulae, we see that the Massieu-Planck functional 
in the hydrostatic gauge $\Psi[\pt;j] \big|_{\mathrm{hs}}$ simply serves as 
a generating functional (or a kind of the action) for the conserved current 
operator averaged over the local Gibbs distribution 
(See Eq.~\eqref{eq:Jset} for the definition of the microscopic 
current operators). 
Again, combining the first three equation enables us to evaluate 
the nonrelativistic energy-momentum tensor in local thermal equilibrium 
as follows: 
\begin{equation}
 \averageLG{\hTcal^\mu_{~\nu}(x)}_{\pt} 
 = - \frac{1}{\beta_0 \sqrt{\gamma}}
 \left(
 - \frac{\delta \Psi[\pt;j]}{\delta n_\mu(x)} \Bigg|_{\mathrm{hs}} n_\nu(x)
 + v^\mu (x) \frac{\delta \Psi[\pt;j]}{\delta \ptc{v}^\nu(x)} 
 \Bigg|_{\mathrm{hs}}
 + 2 h^{\mu\rho} (x)
 \frac{\delta \Psi[\pt;j]}{\delta \ptc{h}^{\rho\nu}(x)} \Bigg|_{\mathrm{hs}} 
 \right). \label{eq:TleqHS}
\end{equation}
When we will perform the derivative expansion and derive the hydrodynamic 
equation in Sec.~\ref{sec:Derivation}, we will use these variational formulae
instead of the general ones \eqref{eq:VarGeneral}.

\subsection{Path-integral formula and thermally emergent Newton-Cartan geometry}
\label{sec:PathIntegral1}
As demonstrated in the previous section, 
we can extract information on all the conserved currents 
in local thermal equilibrium from the single functional $\Psi[\pt;\lambda,j]$.
Then, the problem is reduced to evaluating this functional 
based on underlying quantum theories. 
In this section, dealing with the spinless Bosonic or Fermionic 
Schr\"odinger field as a concrete example, 
we write down the path-integral formula 
for $\Psi[\pt;\lambda,j]$, which brings about the emergence of 
thermally induced curved spacetime.
Since we are considering nonrelativistic systems, 
emergent thermal spacetime is not the (pseudo) Riemannian geometry
but the Newton-Cartan geometry. 

The Lagrangian for the single-component 
interacting spinless Bosonic or Schr\"odinger field
$\phi$ in the nonrelativistic curved spacetime reads  
\begin{equation}
 \begin{split}
 \Scal [\phi;j] 
  &= \int d^dx \sqrt{\gamma} 
  \left[\frac{i}{2} v^\mu \phi^\dag \overlr{D_\mu} \phi
  - \frac{1}{2m} h^{\mu\nu} D_\mu \phi^\dag D_\nu \phi  
  \right] 
  + \Scal_{\mathrm{int}}[\phi;j]
 \end{split}
\end{equation}
where we defined 
$\phi^\dag \overlr{D_\mu} \phi \equiv \phi^\dag D_\mu\phi -\phi D_\mu \phi^\dag$
with the covariant derivative $D_\mu$ acting on $\phi$ and $\phi^\dag$ as 
\begin{equation}
 D_\mu \phi \equiv (\partial_\mu + i m a_\mu + ie A_\mu) \phi, \quad 
  D_\mu \phi^\dag \equiv (\partial_\mu - i m a_\mu - i e A_\mu) \phi^\dag. 
\end{equation}
Here $\Scal_{\mathrm{int}}[\phi;j]$ contains interaction terms of our system:
For example, while
$\Scal_{\mathrm{int}} = - \dfrac{1}{2} \displaystyle{\int} d^dx \sqrt{\gamma} \lambda |\phi|^4$
describes the nonlinear Schr\"odinger field realized in cold atom systems, 
we can also describe systems with long range interaction in the covariant 
manner---e.g. systems interacting through the Lennard-Jones potential---by 
the use of auxiliary massless field living in extra dimensions 
(See e.g. \cite{Hoyos:2011ez,Fujii:2016mbc} for such a treatment).
The essential point here is that we assume that $\Scal_{\mathrm{int}}[\phi;j]$ 
respects diffeomorphism, gauge, and Milne boost invariance as is the case of 
for the above two examples.
We then have the corresponding covariant conservation laws, 
or the operator identities
\eqref{eq:EMcons}-\eqref{eq:Milne} with a set of conserved current 
defined in Eq.~$\eqref{eq:Jset}$.
In the following discussion, we consider the nonlinear Schr\"odinger 
fields as a concrete example: 
\begin{equation}
 \Scal [\phi;j] = \int d^d x \sqrt{\gamma} \Lcal ,\with 
  \Lcal \equiv \frac{i}{2} v^\mu \phi^\dag \overlr{D_\mu} \phi
  - \frac{1}{2m} h^{\mu\nu} D_\mu \phi^\dag D_\nu \phi   
  - \frac{1}{2} \lambda |\phi|^4.
 \label{eq:Action}
\end{equation} 
Then, taking variations of this action
brings about a following set of conserved current operators:
\begin{align}
 \hEcal^\mu 
 &\equiv - \frac{1}{\sqrt{\gamma}} \frac{\delta \Scal}{\delta n_\mu} 
 =  v^\mu  \left[ 
 \frac{1}{2m} h^{\mu\nu} D_\mu \hphi^\dag D_\nu \hphi 
 + \frac{1}{2} \lambda |\hphi|^4 \right]
 - \frac{1}{2m} \big( h^{\mu\sigma} v^\rho + h^{\mu\sigma} v^\sigma \big)
 D_\rho \hphi^\dag D_\sigma \hphi  
 , \\
 \hPcal_\mu 
 &\equiv - \frac{1}{\sqrt{\gamma}} \frac{\delta \Scal}{\delta \bar{v}^\mu} 
 = - \frac{i}{2} \hphi^\dag \overlr{D_\perp}_\mu \hphi, \\
 \hT_{\mu\nu}
 &\equiv - \frac{2}{\sqrt{\gamma}} \frac{\delta \Scal}{\delta \bar{h}^{\mu\nu}} 
 = \frac{1}{2m} 
 \left[ D_{\perp\mu} \hphi^\dag D_{\perp\nu} \hphi 
 + D_{\perp\nu} \hphi^\dag D_{\perp\mu} \hphi \right]
 + h_{\mu\nu} \Lcal(\hphi,D_\rho \hphi)  , \\ 
 \hJ_M^\mu 
 &\equiv - \frac{1}{\sqrt{\gamma}} \frac{\delta \Scal}{\delta a_\mu} 
 = m \left[ v^\mu \hphi^\dag \hphi 
  - \frac{i}{2m} h^{\mu\nu} \hphi^\dag \overlr{D_\nu}\hphi \right] , \\
 \hJ_Q^\mu 
 &\equiv - \frac{1}{\sqrt{\gamma}} \frac{\delta \Scal}{\delta A_\mu} 
 = e \left[ v^\mu \hphi^\dag \hphi 
  - \frac{i}{2m} h^{\mu\nu} \hphi^\dag \overlr{D_\nu} \hphi \right] ,
\end{align}
where we defined a spatial projection of the covariant derivative as
$D_{\perp\mu} \equiv P^\nu_\mu D_\nu$. 
Since we are considering the single-component charged matter here,
the mass current and electric current are connected with the trivial 
relation: $e\hJ_M^\mu = m \hJ_Q^\mu$. 
We thus consider the mass densities as the independent conserved 
quantities and do not include the electric charge densities 
in the local Gibbs distribution. 
We also note that the relation $\hPcal_\mu = h_{\mu\nu} \hJ_M^\nu$ 
due to the Milne boost invariance is certainly satisfied. 
By using these conserved current operators 
together with the canonical commutation relation, 
we can explicitly write down the path-integral formula as follows:
\begin{equation}
  \begin{split}
   \Tr e^{-\hK} 
   &= \int \mathcal{D} \phi \mathcal{D} \phi^\dag 
   \exp \left( \int_0^{\beta_0} d\tau 
   \int d^{d-1} x  \sqrt{h}
   \left[
   - \frac{1}{2}\phi^\dag \overleftrightarrow{\partial_\tau} \phi 
   + \beta_0^{-1} n_\mu
   \big( \Tcal^\mu_{~\nu} \beta^\nu + J_M^\mu \nu'_M \big)
   \right] \right) \\
   &= \int \mathcal{D} \phi \mathcal{D} \phi^\dag 
   \exp \Bigg( \int_0^{\beta_0} d\tau 
   \int d^{d-1} x e^\sigma \sqrt{h} \\
   &\quad \times \Bigg[
    \frac{i}{2} e^{-\sigma} 
   \phi^\dag \big( i \overlr{\partial_\tau} \big) \phi  
   - \frac{i}{2} u^\mu
   \phi^\dag \overlr{D_\perp}_{\mu} \phi 
   - \frac{1}{2m} h^{\mu\nu}  D_\mu \phi^\dag D_\nu \phi
   - \frac{1}{2} \lambda |\phi|^4
   + m \mu_M' \phi^\dag \phi  )
   \Bigg] \Bigg) \\
   &= \int \mathcal{D} \phi \mathcal{D} \phi^\dag 
   \exp \Bigg( \int_0^{\beta_0} d\tilx^d
   e^{\sigma} \sqrt{h}  \Bigg[
    \frac{i}{2} \tilv^{\mu}
   \phi^\dag \overlr{\tilD_{\mu}} \phi  
   -  \frac{1}{2m} \tilh^{\mu\nu} 
   \tilD_{\mu} \phi^\dag \tilD_{\nu} \phi
   - \frac{1}{2} \lambda |\phi|^4
   \Bigg] \Bigg) ,
  \end{split}
\end{equation}
where we used $\beta(x) \equiv \beta^\mu (x) n_\mu (x)$ together with
the following parametrization for local thermodynamic parameters
$\lambda^a (x) \equiv \{\beta^\mu(x), \nu'_M(x)\}$:
\begin{equation}
 \beta^\mu = \beta u^\mu 
  , \quad
  \nu'_M \equiv \beta \mu_M'  
  \with
  u^\mu n_\mu = 1. 
\end{equation}
We also defined $e^{\sigma(x)} = \beta(x)/\beta_0$ with 
an arbitrary reference temperature $\beta_0$. 
Here we introduced the background field in the emergent thermal spacetime
$\tilj \equiv \{\tiln_\mu, \tilv^\mu, \tilh_{\mu\nu}, \tilh^{\mu\nu}, 
\tila_\mu, \tilA_\mu\}$ and the covariant derivative 
\begin{equation}
 \tilD_\mu \phi 
  \equiv (\tildel_\mu + i m \tila_\mu + ie \tilA_\mu) \phi, 
  \quad 
 \tilD_\mu \phi^\dag 
 \equiv (\tildel_\mu - i m \tila_\mu - i e \tilA_\mu) \phi^\dag,
 \label{eq:tilD}
\end{equation}
with $\tildel_\mu \equiv \big(i\partial_\tau, \partial_{\ptc{i}} \big)$. 
The vital point here is that the effect of inhomogeneous 
temperature, fluid-velocity, and chemical potential is completely 
captured by the emergent thermal background field $\tilj (\tilx)$ 
in the manifestly covariant manner. 

Since we again have the Milne boost invariance in the emergent thermal 
spacetime, we need a kind of gauge fixing 
to write down explicit relations between the local thermodynamic 
parameters $\lambda^a(x)$ 
and the induced Newton-Cartan data $\tilj(x)$ in thermal spacetime: 
$\tilj = \tilj (\lambda,j)$.
If we choose a special gauge satisfying 
$\tilv^{\pmu} = (e^{-\sigma}, -u^{\ptc{i}})$, 
the Newton-Cartan data for thermal spacetime is given by  
\begin{equation}
 \begin{split}
 \tilde{n}_{\pmu} &\equiv 
  (e^\sigma, \bzero) , \quad
 \tilde{v}^{\pmu} \equiv 
 \begin{pmatrix}
  e^{-\sigma} \\
  - u^{\ptc{i}}
 \end{pmatrix}, \quad
 \tilh_{\pmu\pnu} 
  \equiv 
  \begin{pmatrix}
   e^{2\sigma} u^2 &  e^\sigma u_{\ptc{i}} \\
   e^\sigma u_{\ptc{j}} & h_{\ptc{i}\ptc{j}}
  \end{pmatrix} , \quad 
  \tilh^{\pmu\pnu} \equiv 
  \begin{pmatrix}
   0 & 0 \\
   0 & h^{\ptc{i}\ptc{j}}
  \end{pmatrix},  \\
  \tila_{\pmu} &\equiv 
  \big( - e^\sigma \mu_M' , \, a_{\ptc{i}} \big), \quad
  \tilA_{\pmu} \equiv 
  \big( 0 , \, A_{\ptc{i}} \big),
 \end{split} 
 \label{eq:Gauge1}
\end{equation}
where we defined $u_{\ptc{i}} \equiv h_{\ptc{i}\ptc{j}} u^{\ptc{j}}$ and 
$u^2 \equiv u_{\ptc{i}} u^{\ptc{i}} = h^{\mu\nu} u_\mu u_\nu$
with $u_{\mu} \equiv h_{\mu\nu} u^\nu$. 
From these relations, we can clearly see that the Newton-Cartan condition
\begin{equation}
  \tiln_{\pmu} \tilv^{\pmu} = 1, \quad 
   \tiln_{\pmu} \tilh^{\pmu\pnu}=0, \quad 
   \tilv^{\pmu} \tilh_{\pmu\pnu} = 0, \quad 
   \tilh^{\pmu\prho} \tilh_{\prho\pnu} 
   = \delta^{\pmu}_{\pnu} -\tilv^{\pmu} \tiln_{\pnu}
   \equiv \tilP^{\pmu}_{\pnu}
   = \begin{pmatrix}
      0 & 0 \\
      - e^{\sigma} u^{\ptc{i}} & \delta^{\ptc{i}}_{\ptc{j}}
     \end{pmatrix},
     \label{eq:thermalNCCond}
\end{equation}
is satisfied for the induced thermal Newton-Cartan data 
$\tilj \equiv \{\tiln_\mu, \tilv^\mu, \tilh_{\mu\nu}, \tilh^{\mu\nu}, 
\tila_\mu, \tilA_\mu\}$.
As is the same with our original spacetime, 
we can define non-degenerate ``metric'' $\tilgamma_{\pmu\pnu}$, 
and its inverse $\tilgamma^{\pmu\pnu}$ by
\begin{equation}
 \tilgamma_{\pmu\pnu} 
  \equiv \tilh_{\pmu\pnu} + \tiln_{\pmu} \tiln_{\pnu}
  = \begin{pmatrix}  
     e^{2\sigma}(u^2 + 1) &  e^\sigma u_{\ptc{i}}  \\
     e^\sigma u_{\ptc{j}} & h_{\ptc{i}\ptc{j}}
    \end{pmatrix}, \quad
  \tilgamma^{\pmu\pnu} 
  \equiv \tilh^{\pmu\pnu} + \tilv^{\pmu}  \tilv^{\pnu}
  = \begin{pmatrix}  
     e^{-2\sigma} &  - e^{-\sigma} u^{\ptc{i}} \\
     - e^{-\sigma} u^{\ptc{j}}  & h^{\ptc{i}\ptc{j}} + u^{\ptc{i}} u^{\ptc{j}}
    \end{pmatrix},
\end{equation}
whose determinant is given by 
$\tilgamma = \det \tilgamma_{\pmu\pnu} =  e^{2\sigma} h$. 
The above result thus shows that the path-integral formula for 
the Massieu-Planck functional is expressed 
in terms of the action in the thermally emergent Newton-Cartan background and 
gauge connection given by
\begin{equation}
 \begin{split}
  \tiln
  &= \tiln_{\pmu} d\tilx^{\pmu} = e^\sigma d\tilt , \\
  d\tilell^2 &= \tilh_{\pmu\pnu} d\tilx^{\pmu} \otimes d\tilx^{\pnu}  
  = e^{2\sigma} u^2 d\tilt \otimes d\tilt 
  + e^\sigma u_{\ptc{i}} 
  ( d\tilt \otimes d\px^{\ptc{i}} + d\px^{\ptc{i}} \otimes d\tilt )
  + h_{\ptc{i}\ptc{j}} d\px^{\ptc{i}} \otimes d\px^{\ptc{j}},  \\ 
  \tila &= \tila_{\pmu} d\tilx^{\pmu} 
  = - e^\sigma \mu_M' d\tilt + a_{\ptc{i}}  d\px^{\ptc{i}} \\
  \tilA &= \tilA_{\pmu} d\tilx^{\pmu} 
  =  A_{\ptc{i}}  d\px^{\ptc{i}},
 \end{split}
 \label{eq:Back1}
\end{equation}
where we defined $d\tilt \equiv - i d\tau$. 
Although this result looks fine for our discussion, 
as is discussed in the next subsection,
this gauge is not so useful from the viewpoint of the thermal
Milne boost invariance. 

As an alternative to the above gauge, we can choose another one in which 
$\tilv^{\pmu} = (e^{-\sigma}, \bzero)^t$ is satisfied. 
In this gauge, the term containing $u^\ptc{i}$ is installed into 
the thermal mass gauge field $\tila_{\pmu}$, and 
we obtain a different expression for the Newton-Cartan data,
\begin{equation}
 \begin{split}
  \tilde{n}_{\pmu} &\equiv  (e^\sigma, \bzero) ,  \quad 
  \tilde{v}^{\pmu} \equiv 
 \begin{pmatrix}
  e^{-\sigma} \\
   \bzero
 \end{pmatrix},  \quad 
 \tilh_{\pmu\pnu} 
  \equiv 
  \begin{pmatrix}
   0 & 0 \\
   0 & h_{\ptc{i}\ptc{j}}
  \end{pmatrix}
  , \quad
  \tilh^{\pmu\pnu} \equiv 
  \begin{pmatrix}
   0 & 0 \\
   0 & h^{\ptc{i}\ptc{j}}
  \end{pmatrix},  \\
  \tila_{\pmu} &\equiv 
  \big( - e^\sigma (\mu_M' + u^2/2 ) ,\, a_{\ptc{i}} - u_{\ptc{i}} \big), \quad 
  \tilA_{\pmu} \equiv 
  \big(0, \, A_{\ptc{i}}  \big),
 \end{split}
  \label{eq:Gauge2}
\end{equation}
which also satisfies the above 
Newton-Cartan condition \eqref{eq:thermalNCCond} except for 
the expression of $\tilP^{\mu}_{\pnu}$:
\begin{equation}
 \tilP^{\pmu}_{\pnu} = \delta^{\pmu}_{\pnu} - \tilv^{\pmu} \tiln_{\pnu} 
  = \begin{pmatrix}
     0 & 0 \\
     0 & \delta^{\ptc{i}}_{\ptc{j}}
    \end{pmatrix}.
\end{equation}
The non-degenerate ``metric'' also takes a different but simple form given by 
\begin{equation}
 \tilgamma_{\pmu\pnu} 
  \equiv \tilh_{\pmu\pnu} + \tiln_{\pmu} \tiln_{\pnu}
  = \begin{pmatrix}  
     e^{2\sigma} & 0  \\
     0 & h_{\ptc{i}\ptc{j}}
    \end{pmatrix}, \quad
  \tilgamma^{\pmu\pnu} 
  \equiv \tilh^{\pmu\pnu} + \tilv^{\pmu}  \tilv^{\pnu}
  = \begin{pmatrix}  
     e^{-2\sigma} & 0 \\
     0& h^{\ptc{i}\ptc{j}} 
    \end{pmatrix},
\end{equation}
which obviously gives the same determinant as before: 
$\tilgamma = \det \tilgamma_{\pmu\pnu} =  e^{2\sigma} h$.
Then, our resulting action is interpreted as the one in the 
emergent background given by 
\begin{equation}
 \begin{split}
  \tiln
  &= \tiln_{\pmu} d\tilx^{\pmu} = e^\sigma d\tilt , \\
  d\tilell^2 &= \tilh_{\pmu\pnu} d\tilx^{\pmu} \otimes d\tilx^{\pnu} 
  = h_{\ptc{i}\ptc{j}} d\px^{\ptc{i}} \otimes d\px^{\ptc{j}},  \\
  \tila &= \tila_{\pmu} d\tilx^{\pmu} 
  = - e^\sigma (\mu_M'+ u^2 /2 ) d\tilt 
  + (a_{\ptc{i}} -  u_{\ptc{i}}  ) d\px^{\ptc{i}} \\
  \tilA &= \tilA_{\pmu} d\tilx^{\pmu} 
  = A_{\ptc{i}} d\px^{\ptc{i}}.
 \end{split}
 \label{eq:Back2}
\end{equation}
From Eqs.~\eqref{eq:Back1} and \eqref{eq:Back2}, 
we now see that while $\tiln$ and $\tilA$ coinside with each other, 
$d\tilell^2$ and $\tila$ have different forms. 
As is mentioned above, this ambiguity is what we have already encountered 
in the original Newton-Cartan geometry due to the Milne boost redundancy
of our action.
In fact, these two gauges are connected with each other by the finite 
Milne boost transformation in thermal spacetime 
with the choice of a finite parameter 
$\tilLam_{\pmu} (x) = u_{\pmu} (x)$:
\begin{equation}
 \begin{cases}
  \delta_u \tilde{v}^{\pmu} 
  = \tilh^{\pmu\pnu} u_{\pnu} , \\
  \delta_u \tilh_{\pmu\pnu} 
  =  - ( \tiln_{\pmu} \tilP_{\pnu}^{\prho} 
  + \tiln_{\pnu} \tilP_{\pmu}^{\prho}  ) u_{\prho} 
  + \tiln_{\pmu} \tiln_{\pnu} \tilh^{\prho\psigma} u_{\prho} u_{\psigma}, \\
  \delta_u \tila_{\pmu} 
  = - \tilP^{\pnu}_{\pmu} u_{\pnu}  
  + \dfrac{1}{2} \tiln_{\pmu} \tilh^{\prho\psigma} u_{\prho} u_{\psigma}.
 \end{cases}
\end{equation}

\begin{figure}[t]
 \centering
 \includegraphics[width=0.93\linewidth]{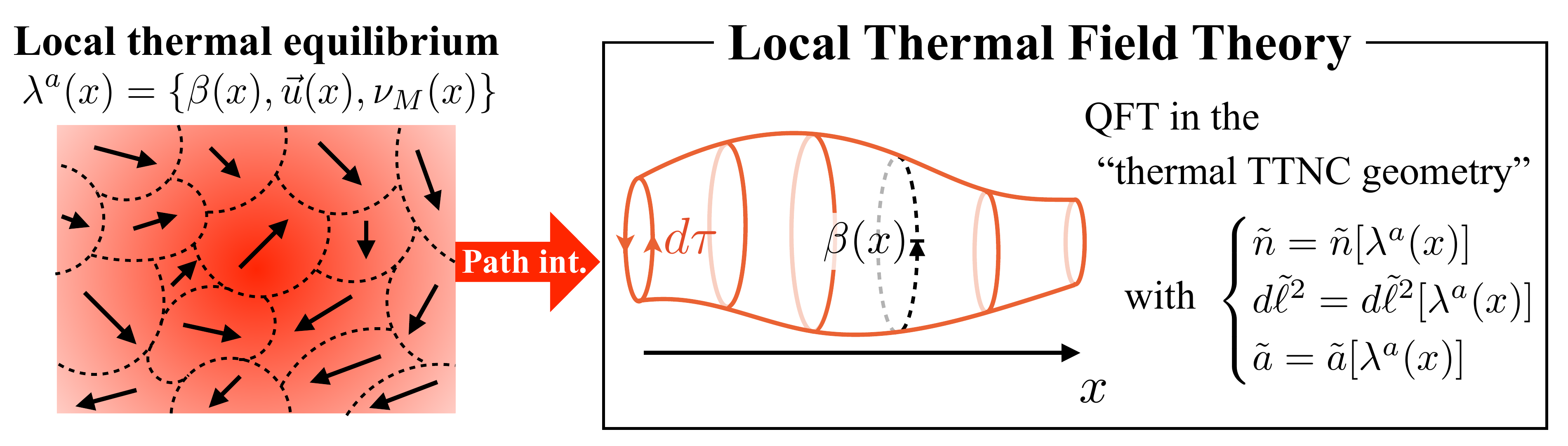}
 \caption{
 Schematic picture clarified in Sec.~\ref{sec:PathIntegral1}. 
 We develop the imaginary-time path-integral formalisms for local thermal 
 equilibrium, which brings about the emergence of the thermal 
 twistless torsional Newton-Cartan (TTNC) geometry. 
 }
 \label{Fig:path-integral}
\end{figure}

We here summarize our result clarified in this subsection.
Based on the local Gibbs distribution, we deal with the spinless nonlinear 
Schr\"odinger field as a concrete example and construct the path-integral 
formula for the Massieu-Planck funcional. 
The most notable result in this section is the following path-integral formula 
for the Massieu-Planck functional:
\begin{equation}
 \Psi [\pt;\lambda] 
  = \log \int \Dcal \phi \Dcal \phi^\dag \, e^{\Scal[\phi,\phi^\dag;\lambda,j]},
  \label{eq:PTforPsi}
\end{equation}
with the resulting action
\begin{equation}
 \begin{split}
  \Scal [\phi,\phi^\dag;\lambda,j] 
  &= \int_0^{\beta_0}  d\tau \int d^{d-1} \px \sqrt{\tilgamma} 
  \left[\frac{i}{2} \tilv^\mu \phi^\dag \overlr{\tilD_\mu} \phi
  - \frac{1}{2m} \tilh^{\mu\nu} \tilD_\mu \phi^\dag \tilD_\nu \phi  
  - \frac{1}{2} \lambda |\phi|^4
  \right]  \\
  &= \int_0^{\beta_0} d^d \tilx \sqrt{\tilgamma}  
  \tilLcal (\phi,\phi^\dag, \tilD_{\prho}\phi, \tilD_{\prho}\phi^\dag;\tilj). 
 \end{split}
\end{equation}
where we defined
$\displaystyle{\int_0^{\beta_0}d^d \tilx 
\equiv \int_0^{\beta_0} d\tau \int d^{d-1} \px }$ with an 
arbitrary reference temperatrue $\beta_0$. 
Here $\tilj(x) \equiv \{
\tiln_\mu(x),\tilv^\mu(x),\tilh_{\mu\nu}(x),\tilh^{\mu\nu}(x),\tila(x), 
\tilA_\mu(x)\}$ 
denotes a set of the background fields in emergent thermal spacetime, 
which is determined from configurations of hydrodynamic variables 
$\lambda^a(x)$ and original external fields $j(x)$: $\tilj = \tilj (\lambda,j)$.
We thus conclude that the effect of inhomogeneous temperature, fluid-velocity, 
and chemical potential naturally leads to 
the thermally emergent twistless torsional Newton-Cartan (TTNC) geometry.
This result is schematically shown in Fig.~\ref{Fig:path-integral}. 
It is worth emphasizing that the resulting action takes the same form 
as our original action, which means that we have the same symmetry properties 
elaborated in Sec.~\ref{sec:Matter}: diffeomorphism, and $U(1)$ gauge, 
and Milne boost invariance in the emergent thermal spacetime. 
Note that since we again have the Milne boost redundancy 
in emergent thermal spacetime, 
it requires a kind of gauge fixing to write down 
explicit relations between $\tilj (x)$ and $\{ \lambda^a(x),j(x) \}$.
For example, in some useful gauge, they are defined in Eqs.~\eqref{eq:Gauge1}, 
or \eqref{eq:Gauge2}. 
We have also introduced the covariant derivative 
in emergent thermal spacetime $\tilD_{\prho}$ in Eq.~\eqref{eq:tilD}.
Only difference with the original theory is
 that our external field $\tilj(x)$ does not have 
the imaginary time dependence. 
We will use these symmetry arguments to perform the derivative expansion 
of $\Psi[\pt;\lambda,j]$ in the following discussion.

\subsection{Symmetry and invariant of emergent thermal spacetime}
\label{sec:Symmetry}
As is obtained in the previous subsection, 
the path-integral formula for the Massieu-Planck functional is 
given by the covariant action in the thermally emergent background. 
As a consequence, we again encountered 
the Milne boost redundancy in thermal spacetime. 
This gauge ambiguity is not useful to construct 
the Massieu-Planck functional $\Psi[\pt;\lambda,j]$---Milne 
boost invariant quantity---in terms of $\tilj (x)$ 
since some members of $\tilj(x)$ are not Milne boost invariant.
We thus would like to describe our background 
in the Milne boost invariant manner. 
Nevertheless, we also need to pay attention to the $U(1)_M$ 
mass gauge invariance of $\Psi[\pt;\lambda,j]$ 
since there is a kind of tradeoff between 
the $U(1)_M$ gauge invariance and Milne boost invariance. 
In other words, if we are not careful, our Milne boost invariant quantities 
may not be $U(1)_M$ gauge invariant, or reversed case may occur.
We here explain how we can respect both of them and 
specify the Milne boost and $U(1)_M$ gauge invariant quantities
employed as basic building blocks for $\Psi[\pt;\lambda,j]$. 

\begin{figure}[htb]
 \centering
 \includegraphics[width=0.33\linewidth]{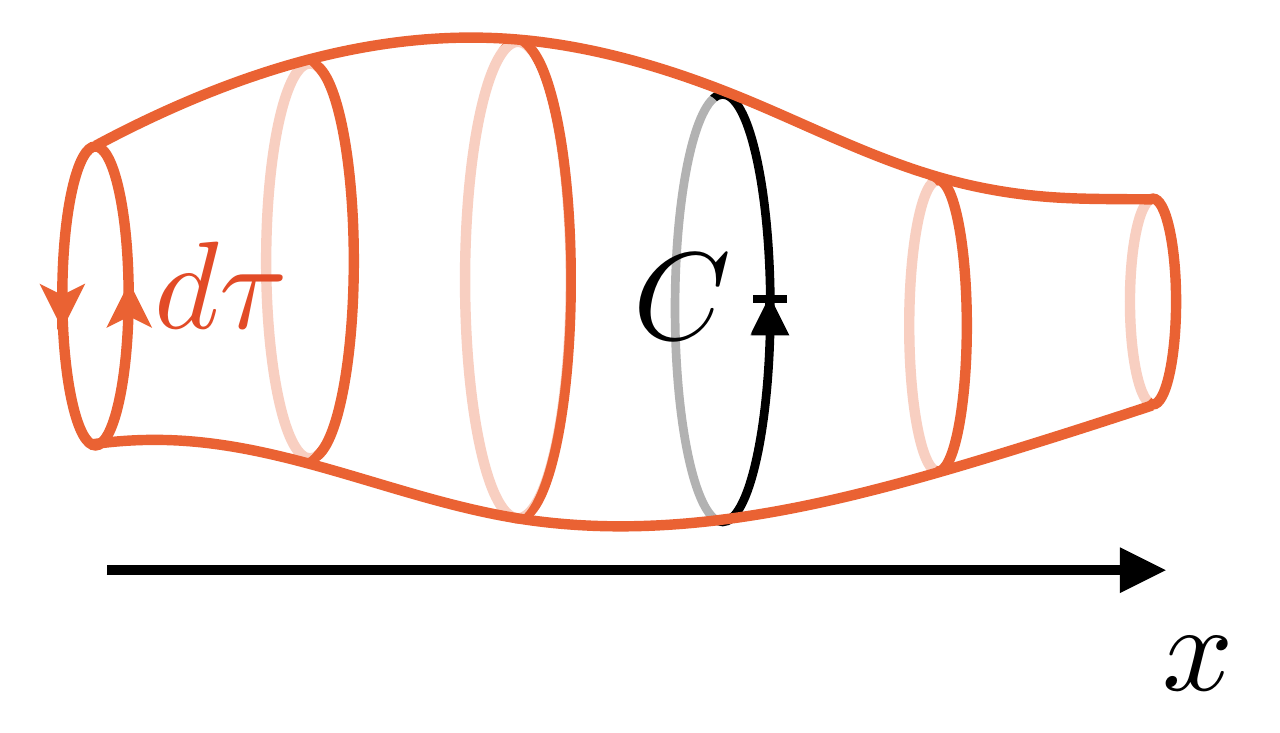}
 \caption{
 A contour of integral used in Eq.~\eqref{eq:contourInt}. 
 This contour results from the compactness 
 of thermal spacetime along the imaginary-time direction.
 }
 \label{Fig:contour}
\end{figure}

We first pay attention to the $U(1)_M$ gauge invariance. 
In the usual situation, we have the field strength tensor 
$F^a \equiv d\tila $ and $F = d A$ as gauge invariant building blocks. 
They contain one derivative, and thus, do not appear in the leading-order 
expression of $\Psi[\pt;\lambda,j]$
when we consider the derivative expansion with a usual power counting scheme 
as is employed in Sec.~\ref{sec:Derivation}. 
However, we have another gauge invariant quantity in our setup
due to the compactness of the imaginary-time direction.  
In fact, following contour integrals along the imaginary-time direction 
(See Fig.~\ref{Fig:contour}) provides us 
a diffeormophism and $U(1)_M$ gauge invariant quantities:
\begin{equation}
 \oint_C \tiln = 
  \oint_0^{\beta_0} d\tilt e^{\sigma} = \beta (x), \quad
  \oint_C \tila = 
  \oint_0^{\beta_0} d\tilt e^{\sigma} \mu'_M = \nu'_M (x) ,
  \label{eq:contourInt}
\end{equation}
where we employed the gauge \eqref{eq:Gauge1} for $\tila$. 
As is clearly seen, these do not contain any derivative, which can 
appear in the leading-order expression of $\Psi[\pt;\lambda,j]$. 
However, the second term $\nu'_M (x)$ is indeed not Milne boost invariant, 
and we have further restriction. 

We then discuss how the Milne boost invariance restricts the possible 
combination of our building blocks. 
Following Ref.~\cite{Jensen:2014ama}, let us first construct 
the Milne boost invariant ``line element''.
The important point clarified in the previous subsection is that 
spatial line element $d\tilell^2$ and $U(1)_M$ gauge connection $d\tila$ 
is not Milne boost invariant. 
However, utilizing these Milne boost covariance, 
we can construct the Milne boost invariant ``line element'' as follows: 
\begin{equation}
  \begin{split}
  d\tils^2 
   &\equiv d\tilell^2 - \tiln \otimes \tila - \tila \otimes n  \\
   &= \big( \tilh_{\pmu\pnu}
  - \tiln_{\pmu} \tila_{\pnu} - \tila_{\pmu} \tiln_{\pnu}\big)
  d\tilx^{\pmu} \otimes d\tilx^{\pnu} \\
   &= 2 e^{2\sigma} \Big( \mu_M' + \frac{1}{2}u^2 \Big) 
   d\tilt \otimes d\tilt
   - e^{\sigma} \big( a_{\ptc{i}} - u_{\ptc{i}} \big) 
   (d\tilt \otimes d\tilx^{\ptc{i}} + d\tilx^{\ptc{i}} \otimes d\tilt )
   + h_{\ptc{i}\ptc{j}} dx^{\ptc{i}} \otimes dx^{\ptc{j}},
  \end{split}
\end{equation}
from which we can read off the Milne boost invariant combinations: 
\begin{equation}
 \nu_M \equiv \nu_M' + \frac{1}{2} \beta u^2, \quad 
  a'_{\ptc{i}} \equiv a_{\ptc{i}} - u_{\ptc{i}}. 
\end{equation}
Note that while the Milne boost invariance is respected, 
the $U(1)_M$ gauge invariance is sacrificed; in other words,
$d\tils^2$ is Milne boost invariant but not $U(1)_M$ gauge invariant. 
However, we have already clarified the $U(1)_M$ gauge invariance 
of $\nu_M'$ and covariance of $a_{\ptc{i}}$.
Therefore, in order to construct the Massieu-Planck functional, 
paying attention to the diffeomorphism and gauge invariance, 
we only need to use $\beta(x)$ and 
$\nu_M (x) = \nu_M' (x) + \frac{1}{2} \beta u(x)^2$ 
as leading-order scalar quantities, 
$a'_{\ptc{i}} = a_{\ptc{i}} - u_{\ptc{i}}$ and $A_{\ptc{i}}$ as gauge fields, 
and $h_{\ptc{i}\ptc{j}}$ (or $h^{\ptc{i}\ptc{j}}$) as a spatial metric. 
This restriction on building blocks of $\Psi[\pt;\lambda,j]$ is 
a basic consequence resulting from the symmetry properties, 
which will be effectively utilized in Sec.~\ref{sec:Derivation}. 

Before closing this section, we put a comment on the relation between 
the gauge choice and Milne boost invariance. 
In this subsection, starting from one seemingly useful gauge \eqref{eq:Gauge1},
we discuss a way to implement the Milne boost invariance.
However, if we start from another gauge \eqref{eq:Gauge2}, 
we notice that all information on backgrounds in Eq.~\eqref{eq:Back2} 
is already Milne boost invariant. 
As is discussed in Ref.~\cite{Jensen:2014ama}, 
this comes from the fact that we have the Milne boost invariant vector 
$u^\mu$ which enables us to realize the Milne boost 
invariant gauge fixing. 
Therefore, from the viewpoint of the Milne boost invariance, 
the latter gauge \eqref{eq:Gauge2} is the most useful gauge choice 
(See e.g. \cite{Jensen:2014ama} for a detailed discussion).

\section{Fluctuation theorems and optimized perturbation theory for time evolution}
\label{sec:TimeEvo}
Let us consider systems in general nonequilibrium situations 
even far from local equilibrium in which 
innumerable microscopic degrees of freedom play an important role. 
As we emphasized in Sec.~\ref{sec:Intro},
hydrodynamics provides the macroscopic effective description of systems 
near local thermal equilibrium, and we do not know whether 
we can apply hydrodynamics to describe such really nonequilibrium situations.
Nevertheless, almost all microscopic degrees of freedom will go away 
a short while later, and only conserved charge densities remains 
since they cannot disappear due to the conservation laws%
\footnote{
If there exist other massless modes like 
the Nambu-Goldstone modes associated with spontaneous symmetry breaking,
we also have to consider them, which results in the superfluid hydrodynamics~
\cite{Landau:Fluid,Khalatnikov}.
}.
This brings about local thermalization, after which
hydrodynamics is expected to work.
In this section, based on the above expectation, 
assuming that our initial density operator is given by 
the local Gibbs distribution, we provide a way to derive an exact formula 
for the dissipative part of constitutive relations.
In Sec.~\ref{sec:Fluct}, 
we show two kinds of the so-called fluctuation theorems 
for local thermal equilibrium before discussing the constitutive relation.
In Sec.~\ref{sec:Dissipative}, we write down the exact formula for 
the dissipative part of constitutive relations based on 
the first fluctuation theorem.

\subsection{Fluctuation theorems for local thermal equilibrium}
\label{sec:Fluct}
We first put a most critical assumption that our initial density operator 
$\hrho_0$ takes a form of the local Gibbs distribution at initial time $\pt_0$: 
$\hrho_0 = \hrhoLG[\pt_0;\lambda]$.
Employing the Heisenberg picture, 
we express the average value of any Heisenberg operator $\hOcal(t)$ as
$\average{\hOcal(t)} \equiv \Tr \big(\hrho_0 \hOcal(t) \big)$.
Then, our problem is to derive the constitutive relations:
\begin{equation}
 \average{\hcurrent^\mu_{~a}(x)}
  = \current^\mu_{~a}[\average{\hcurrent^{\pzero}_{~a}}].
\end{equation}
Although we have fixed our initial density operator, 
it is inadequate to approximately evaluate the average value 
$\average{\hcurrent^\mu_{~a}(x)}
= \Tr \big(\hrhoLG [\pt_0;\lambda] \hcurrent^\mu_{~a}(x) \big)$ 
at later time $\pt~(>\pt_0)$. 
This is because we only have a set of local thermodynamic parameters
$\lambda^a (x) \big|_{\pt_0}$ at initial time $\pt_0$ while we expect 
$\average{\hcurrent^\mu_{~a}(x)}$ at later time $\pt$ is 
expressed by parameters at that time $\pt$.  
We thus introduce a new set of parameters $\lambda^a(x) \big|_{\pt}$ 
at $\pt$ and reconstruct the perturbative expansion 
on the top of the newly introduced local Gibbs distribution 
$\hrhoLG[\pt;\lambda]$, which is similar to 
the so-called optimized (or renormalized) perturbation theory. 
In other words, we decompose the initial density operator as 
\begin{equation}
 \begin{split}
  \hrho_0 
  &= \exp \left( - \hS[\pt_0;\lambda] \right)
  = \exp \left( - \hS[\pt;\lambda] 
  + \hS[\pt;\lambda] - \hS[\pt_0;\lambda]\right) \\
  &= \exp \left( - \hS[\pt;\lambda] \right)
  \hU[\pt,\pt_0;\lambda],
 \end{split}
 \label{eq:Decompose}
\end{equation}
where we introduced 
\begin{equation}
 \hU[\pt,\pt_0;\lambda] 
  = 
  T_\tau \exp \left( \int_0^1 d\tau \hSigma_\tau [\pt,\pt_0;\lambda] \right),
\end{equation}
with the entropy production operator
$\hSigma[\pt,\pt_0;\lambda] \equiv \hS[\pt,\lambda] - \hS[\pt_0;\lambda]$
and 
$\hOcal_\tau \equiv e^{\tau\hS[\pt,\lambda]} \hOcal e^{-\tau\hS[\pt,\lambda]}$.
Here $\hS[\pt;\lambda]$ represents the entropy functional operator 
defined in Eqs.~\eqref{eq:DefLG}-\eqref{eq:DefK} whose local 
thermodynamics parameters are given by new ones $\lambda^a(x)\big|_{\pt}$.

Since we do not put any condition to fix 
new parameters $\lambda^a(x)\big|_{\pt}$, they are arbitrary at this stage.
This means that if we are able to perform the exact calculation, 
the result does not depend on arbitrary parameters $\lambda^a(x) \big|_{\pt}$.
However, we cannot in general accomplish such an exact calculation,  
and rather perform the finite-order perturbative expansion  
on the top of a local Gibbs distribution with new parameters 
$\lambda^a(x) \big|_{\pt}$. 
This truncation leads to the result dependent on a way 
to define the new parameters $\lambda^a (x) \big|_{\pt}$.
In the hydrodynamic description of systems, 
we are interested in the spacetime evolution of the conserved charge 
densities $\average{\hcurrent^\ptc{0}_{~a}(x)}$.
We, therefore, employ a condition like the fastest apparent convergence (FAC) 
in the optimized perturbation theory~\cite{Stevenson:1981vj} 
for $\average{\hcurrent^\ptc{0}_{~a}(x)}$;
that is to say, the deviation of conserved charge density is minimized, 
or vanish in this case: 
$\averageLG{\hU \delta \hcurrent^{\ptc{0}}_{~a}(x)}_{\pt} = 0 $.
Since this condition is equivalent to 
\begin{equation}
 \average{ \hcurrent^{\ptc{0}}_{~a}(x)} \big|_{\pt}
  =  \averageLG{\hcurrent^{\ptc{0}}_{~a}(x)}_{\pt} , \label{eq:DefParameter}
\end{equation}
it means that the new parameters $\lambda^a(x)\big|_{\pt}$ is defined
so as to match with local thermodynamics for a given value of 
conserved charge densities $\average{ \hcurrent^{\ptc{0}}_{~a}(x)} \big|_{\pt}$.
With the help of the decomposition of the density operator 
\eqref{eq:Decompose}, 
we have a following exact identity to evaluate 
any Heisenberg operator $\hOcal(x)$:
\begin{equation}
 \average{\hOcal(x)} 
  = \averageLG{ \hU[\pt,\pt_0;\lambda] \hOcal(x)}_{\pt}.
  \label{eq:Fluctuation1}
\end{equation}
This identity belongs to a variant of the fluctuation theorem 
which will play a central role to evaluate $\average{\hcurrent^\mu_{~a}(x)}$ 
and perturbatively construct the constitutive relation in the next section. 

Unlike the classical systems~\cite{sasa:2013}, we cannot show 
the second law of thermodynamics directly from Eq.~\eqref{eq:Fluctuation1}
owing to the noncommutativity of quantum operators. 
However, we can derive a canonical quantum fluctuation theorem for 
local thermal equilibrium which contains the second law of thermodynamics 
as follows.
For that purpose, defining a time evolution operator from initial time $\pt_0$ 
to later time $\pt$ under the influence of the external field $j(x)$ 
as $\hUcal_j(\pt,\pt_0)$ which satisfies
$\hOcal(\pt) = \hUcal_j^\dag(\pt,\pt_0) \hOcal (\pt_0) \hUcal_j(\pt,\pt_0)$, 
we introduce the following quantity
\begin{equation}
 G_F (z;j] 
  \equiv 
  \Tr \left( 
       \hrho_0 \, \hUcal_j^\dag (\pt,\pt_0) 
       e^{iz \hS_0 [\pt;\lambda]} \hUcal_j (\pt,\pt_0)
       e^{-i z \hS[\pt_0;\lambda] } \right) 
  = \Tr 
  \left( 
   e^{-\hS [\pt_0;\lambda]}
   e^{iz \hS [\pt;\lambda]} 
   e^{-i z \hS[\pt_0;\lambda] } \right) ,
\end{equation}
where $\hS_0 [\pt;\lambda]$ denotes the entropy functional operator 
whose operator argument is not $\pt$ but $\pt_0$.
The notation $G(z;j]$ is used to clarify that $G(z;j]$ is 
a function of $z$ and functional of $j(x)$.
As is clear from the second expression, this quantity apparently gives  
a generating function for entropy production 
$\hSigma [\pt,\pt_0;\lambda] = \hS [\pt;\lambda] - \hS[\pt_0;\lambda]$ when 
we consider the forward time evolution in the presence of 
the external fields $j(x)$:
\begin{equation}
 \average{\hSigma [\pt,\pt_0;\lambda]} 
  = \frac{\partial}{i\partial z} G_F (z;j] \Big|_{z=0}, \quad 
  \average{(\hSigma [\pt,\pt_0;\lambda])^2} 
  = \left( \frac{\partial}{i\partial z} \right)^2 G_F (z;j] \Big|_{z=0}.
\end{equation}
While the above first- and second-order relations 
are certainly true, if we consider the third- or more order general terms,
this simple relation breaks down.
Nevertheless, considering the projection measurement of the entropy production,
we can regard $G_F (z;j]$ as the generating function 
for the entropy production being observed in an ensemble of the measurement.
In fact, introducing the probability to observe the entropy production 
being $\sigma$ by the Fourier transformation of $G_{F}(\sigma;j]$:
\begin{equation}
 P_{F} (\sigma;j] \equiv \int dz e^{-iz \sigma} G_{F}(z;j],
\end{equation}
we define the moments of $\sigma$ as
\begin{equation}
 \average{\sigma^n}_{P_F} 
  \equiv \int d\sigma P_F(\sigma;j] \sigma^n 
  = \left(\frac{\partial}{i\partial z}\right)^n G_F(z;j] \Big|_{z=0}.
\end{equation}
Then, as is mentioned above, we can directly relate the first two moments of 
$\sigma$ over the probability distribution $P_F(\sigma;j]$ 
with the expectation values of the entropy operators 
$\hSigma[\pt,\pt_0;\lambda]$ over the initial density operator 
$\hrho_0 = \hrhoLG[\pt_0;\lambda]$:
\begin{equation}
 \average{\hSigma[\pt,\pt_0;\lambda]} = \average{\sigma}_{P_F}, \quad
  \average{(\hSigma[\pt,\pt_0;\lambda])^2} = \average{\sigma^2}_{P_F}, \quad
  \average{(\hSigma[\pt,\pt_0;\lambda])^n} \neq \average{\sigma^n}_{P_F} 
  \quad (n = 3, 4,\cdots).
  \label{eq:Difference}
\end{equation}
In addition to $G_F(z;j]$, we also introduce 
\begin{equation}
 G_B (z;j] 
  \equiv 
  \Tr \left( 
       e^{-\Theta\hS[\pt_0;\lambda] \Theta^{-1}} \tilUcal_j^\dag 
       e^{iz \Theta \hS_0 [\pt;\lambda] \Theta^{-1}} \tilUcal_j 
       e^{-i z \Theta \hS[\pt_0;\lambda] \Theta^{-1}} \right) ,
\end{equation}
where we introduced $\Theta \equiv \Pcal \Tcal$ 
with $\Pcal$ and $\Tcal$ represent parity 
and time-reversal transformation, respectively.
Note that we insert $\PT$ transformation instead of the simple time-reversal 
transformation usually employed in the quantum fluctuation theorem%
\footnote{
Although we here introduced the parity transformation for the definition 
of $\Theta$, the following discussion is true with the choice of 
$\Theta = \Tcal$ as is the case for the quantum fluctuation theorem 
in global thermal equilibrium. 
The reason why we introduced the parity transformation is that it 
results in the cleaner form of the reversed density operator since it contains 
the time-reversal odd operator (momentum density operator $\hPcal_\mu$).
}.
Here $\tilUcal_j$ is defined as 
\begin{equation}
 \tilUcal_j \equiv \Theta\, \hUcal_j^\dag (\pt,\pt_0) \Theta^{-1},
  \label{eq:Backward}
\end{equation}
which represents the backward evolution, or the time-evolution with 
a spacetime-reversed protocol for external fields $j(x)$.
Moreover, in the same way as $P_F(\sigma;j]$, we introduce the 
probability distribution of the entropy production as
\begin{equation}
 P_B (\sigma;j] \equiv \int dz e^{-iz\sigma} G_B (z;j].
\end{equation}

Based on this setup, we can show the following simple identity
\begin{equation}
 G_F(z;j] = G_B (-z+i;j],
  \label{eq:FluctuationTheorem}
\end{equation}
which gives an extension of the canonical quantum fluctuation theorem 
in the case of local thermal equilibrium. 
The crucial assumptions to prove this identity is the 
local Gibbs form of the initial density operator.
We put a proof of this identity in Appendix \ref{sec:FTProof}.
Here, we demonstrate some consequences from this identity.
First of all, we can rewrite the quantum fluctuation theorem in an 
alternative form in terms of $P_{F,B}(\sigma;j]$ as follows:
\begin{equation}
 P_F (\sigma;j] = e^\sigma P_B (-\sigma;j].
\end{equation}
Using the above identity for $G_{F,R}(z;j]$, we can easily prove 
this as
\begin{equation}
 \begin{split}
 P_F (\sigma) 
  &= \int dz e^{-iz\sigma} G_F (z) 
  = \int dz e^{-iz\sigma} G_B (-z+i) \\
  &= e^{\sigma} \int dz e^{iz'\sigma} G_B (z') 
  = e^\sigma P_B(-\sigma) .
 \end{split}
\end{equation}
This identity immediately brings about the so-called 
the integral fluctuation theorem for local thermal equilibrium
\begin{equation}
 \average{e^{-\sigma}}_{P_F} \equiv 
  \int d\sigma P_F(\sigma;j] e^{-\sigma} 
  = \int d\sigma P_B(-\sigma;j] = 1,
\end{equation}
where we used the above identity for $P_{F,B}(\sigma;j]$.
It is worth to emphasize that 
$\average{f(\sigma)}_{P_F} \equiv \int d\sigma P_F(\sigma;j] f(\sigma)$ 
is not equal to $\average{f(\hSigma)} \equiv \Tr \big(\hrho_0 f(\hSigma) \big)$
except for the special case with the first-order or second-order terms of 
them due to Eq.~\eqref{eq:Difference}.
Then, taking into account Jensen's inequality ($e^{-x} \geq -x + 1$), 
this identity provides the following inequality
\begin{equation}
 1 = \average{e^{-\sigma}}_{P_F}  \geq 
  - \average{\sigma}_{P_F} + 1
  \quad 
  \Leftrightarrow
  \quad
 \average{\sigma}_{P_F} = 
  \average{\hSigma[\pt,\pt_0;\lambda]} \geq 0.
\end{equation}
This is precisely the second law of thermodynamics which we want to show%
\footnote{
Of course, we can directly show the second law of thermodynamics 
with the help of the Klein's inequality, 
or positivity of the relative entropy (see e.g. Ref.~\cite{RevModPhys.50.221})
\begin{equation}
 \Tr \hrho \log \hrho - \Tr \hrho \log \hrho' \geq 0, 
\end{equation}
by choosing $\hrho = \hrho_0 = \hrhoLG[\pt_0;\lambda]$ and 
$\hrho' = \hrhoLG[\pt;\lambda]$.
}. 
Moreover, expanding $e^{-\sigma}$ with respect to $\sigma$ and 
neglecting the $O(\sigma^3)$ terms, 
we can evaluate the left-hand-side of the integral fluctuation theorem as
\begin{equation}
 \begin{split}
  \average{e^{-\sigma}}_{P_F}
  &= 1 - \average{\sigma}_{P_F} + \frac{1}{2} \average{\sigma^2}_{P_F} 
  + O(\sigma^3) \\
  &= 1 - \average{\hSigma [\pt,\pt_0;\lambda]} 
  + \frac{1}{2} \average{(\hSigma[\pt,\pt_0;\lambda])^2}
  + O(\sigma^3) ,
 \end{split}
\end{equation}
where we used Eq.~\eqref{eq:Difference} to proceed the second line.
Therefore, we can also derive the following relation
\begin{equation}
 \average{\hSigma [\pt,\pt_0;\lambda]} = 
  \frac{1}{2} \average{(\hSigma [\pt,\pt_0;\lambda])^2} 
  + O(\sigma^3) ,
  \label{eq:Fluctuation2}
\end{equation}
which provides the relation between the dissipation 
(the left-hand-side) and fluctuation (the right-hand-side).
Although this relation does not play a central role 
in our derivation of hydrodynamic equations, 
we can understand this as a generalization of fluctuation-dissipation 
relations in our setup.

\subsection{Exact formula for dissipative constitutive relations}
\label{sec:Dissipative}
Based on the first fluctuation-like theorem obtained above,
let us derive the exact formula for the dissipative part of 
the constitutive relations. 
Although there are several differences, this procedure is  
accomplished in the same way as the relativistic case \cite{Hayata:2015lga}.
First of all, the identity \eqref{eq:Fluctuation1} enable us to decompose
$\average{\hcurrent^\mu_{~a}(x)}$ into two parts
\begin{equation}
 \average{\hcurrent^\mu_{~a}(x)} 
  = \averageLG{\hcurrent^\mu_{~a}(x)}_{\pt}
 +  \average{\delta \hcurrent^\mu_{~a}(x)} 
 \with 
  \average{\delta \hcurrent^\mu_{~a}(x)} 
  \equiv \averageLG{\hU \delta \hcurrent^\mu_{~a}(x)}_{\pt}, 
\end{equation}
where we introduced 
$\averageLG{\hOcal}_{\pt} \equiv \Tr ( \hrhoLG[\pt;\lambda] \hOcal )$ 
and $\delta \hOcal \equiv \hOcal - \averageLG{\hOcal}_{\pt}$.
The first term---the expectation values of conserved current operators 
over the local Gibbs distribution---can be evaluated from the 
Massieu-Planck functional as discussed in the previous section. 
We thus focus on the second term associated with the 
deviation from local thermal equilibrium.

In order to evaluate the second term, we first rewrite an expression 
of the entropy production operator $\hSigma[\pt,\pt_0;\lambda]$ 
staying in $\hU$ as
\begin{equation}
 \begin{split}
  \hSigma [\pt,\pt_0;\lambda] 
  &\equiv \hS[\pt;\lambda] -  \hS[\pt_0;\lambda] \\
  &= \int_{\pt_0}^{\pt} d\ps \partial_{\ps} 
  \left[
  - \int d\Sigma_{\ps\mu} \hcurrent^\mu_{~a} \lambda^a + \Psi [\ps;\lambda]  
  \right] \\
  &= - \int_{\pt_0}^{\pt} d\ps 
  \int d\Sigma_{\ps} N (\ps)
  \Big[
  \delta \hTcal^\mu_{~\nu} 
  \Big( 
  \nabla_\mu \beta^\nu - \beta^\rho 
  ( F^n_{\mu\rho}  v^\nu + n_\mu \nabla_\rho v^\nu ) 
  \Big)  \\
  &\hspace{93pt}
  + \delta \hJ_M^\mu \big( \nabla_\mu \nu'_M + \beta^\nu F^a_{\mu\nu} \big) 
  + \delta \hJ_Q^\mu \big( \nabla_\mu \nu'_Q + \beta^\nu F_{\mu\nu} \big)
  \Big].
 \end{split}
\end{equation}
Here we used the first line of Eq.~\eqref{eq:Divergence} 
together with the identity \eqref{eq:DtPsi} for 
$\partial_{\pt} \Psi [\pt;\lambda]$,
which leads to the subtraction of $\averageLG{\hcurrent^\mu_{~a}(x)}_{\pt}$.
Defining a shorthand notation 
$\nabla _\mu \Lambda^a~(a=0,1,\cdots,d-1,d,d+1)$ as
\begin{equation}
 \nabla_\mu \Lambda^\nu \equiv 
  \nabla_\mu \beta^\nu 
  - \beta^\rho ( F^n_{\mu\rho} v^\nu + n_\mu \nabla_\rho v^\nu ), 
  \quad
  \nabla_\mu \Lambda^d \equiv \nabla_\mu \nu'_M + \beta^\nu F^a_{\mu\nu} ,
  \quad
  \nabla_\mu \Lambda^{d+1} \equiv \nabla_\mu \nu'_Q + \beta^\nu F_{\mu\nu} ,
  \label{eq:AbbLam}
\end{equation}
we can express the entropy production operator in a compact form as
\begin{equation}
 \hSigma [\pt,\pt_0;\lambda]
  = - \int_{\pt_0}^{\pt} d \ps \int d\Sigma_{\ps}
  N  \delta \hcurrent^\mu_{~a} \big(\nabla_\mu \Lambda^a \big) . 
  \label{eq:EntropyProd1}
\end{equation}
This provides us the expression of the entropy production operator 
in terms of the 
local thermodynamic parameters $\lambda^a(x)$, external fields $j(x)$, and 
conserved current operators $\hcurrent^\mu_{~a}(x)$.
Nevertheless, this expression contains the time derivative of 
parameters $\lambda^a(x)$ whose time dependence is governed by 
the hydrodynamic equation. 
This means that we have the massless hydrodynamic mode which cause 
an undesirable behavior for correlation functions.
We then eliminate them in a self-consistent manner by 
formally rearranging hydrodynamic equations:
$(\nabla_\mu - \Gcal_\mu) \average{\hcurrent^\mu_{~a}} = \average{\hScal_a}$.
To accomplish this, taking into account the fact 
that $e^{-\hS[\pt;\lambda]}$ does not depend 
on spatial coordinate: $\nabla_{\perp \mu}\hS [\pt;\lambda]= 0$, 
we rewrite the local Gibbs part in hydrodynamic equations as 
\begin{equation}
 \begin{split}
  (\nabla_\mu - \Gcal_\mu) \averageLG{\hcurrent^\mu_{~a}(x)}_{\pt}
  &= (\nabla_\mu - \Gcal_\mu) 
  \Tr \left[ e^{-\hS[\pt;\lambda]} \hcurrent^\mu_{~a}(x)  \right] \\
  &= \Tr \left[ \frac{1}{N(x)} 
  (\partial_{\pt} e^{-\hS[\pt;\lambda]})  \hc_a (x) \right]
  + \averageLG{\hScal_a (x)}_{\pt}   \\
  &= -  \frac{1}{N(x)} 
  \int d\Sigma'_{\pt} N(x')
  \nabla_\mu' \Lambda^b(x')
  \ip{\delta \hc_a(x), \delta \hcurrent^\mu_{~b}(x')}_{\pt}
  + \averageLG{\hScal_a (x)}_{\pt},
 \end{split}
\end{equation}
where we decomposed the covariant derivative as 
$\nabla_\mu = n_\mu v^\nu \nabla_\nu + P^\nu_\mu \nabla_\nu 
=  n_\mu v^\nu \nabla_{\nu} + \nabla_{\perp\mu} $, and 
used conservation laws to proceed the second line. 
We also used the following result to evaluate the time derivative of 
$e^{-\hS[\pt;\lambda]}$:
\begin{equation}
 \begin{split}
  \partial_{\pt} e^{-\hS [\pt;\lambda] }
  &\equiv \lim_{\Delta\pt \to 0}
  \frac{e^{-\hS[\pt + \Delta\pt;\lambda]} - e^{-\hS[\pt;\lambda]}}{\Delta\pt} \\
  &= \lim_{\Delta\pt \to 0}
  \frac{e^{-\hS[\pt;\lambda]} (
  T_\tau e^{ -\int_0^1 d\tau \hSigma_\tau [\pt,\pt+ \Delta\pt;\lambda]} - 1)}{\Delta\pt}\\
  &= - e^{-\hS[\pt;\lambda]}
  T_\tau 
  \int_0^1 d\tau \int d\Sigma'_{\pt} N(x')
  \nabla_\mu' \Lambda^a(x')
  e^{\tau\hK[\pt;\lambda]} 
  \delta \hcurrent^\mu_{~a}(x') e^{-\tau\hK[\pt;\lambda]}.
 \end{split}
\end{equation}
Therefore, introducing the time-dependent local Gibbs 
version of the Kubo-Mori-Bogoliubov inner product 
$\ip{\hA, \hB}_{\pt}$ as 
\begin{equation}
 \ip{\hA,\hB}_{\pt} 
  \equiv \int_0^1 d\tau 
  \averageLG{e^{\hK [\lambda;\pt] \tau}\hA e^{-\hK [\lambda;\pt]\tau} \hB^\dag}_{\pt},
  \label{eq:ip}
\end{equation}
we can rewrite 
the full hydrodynamic equation 
$(\nabla_\mu  - \Gcal_\mu ) \average{\hcurrent^\mu_{~a}(x)} 
= (\nabla_\mu - \Gcal_\mu) \averageLG{\hcurrent^\mu_{~a} (x) }
  +  (\nabla_\mu - \Gcal_\mu) \average{\delta \hcurrent^\mu_{~a}} 
= \average{\hScal_a(x)} $ in the following form: 
\begin{equation}
 \begin{split}
  &\quad  \int d\Sigma'_{\pt} 
   \ip{\delta \hc_a(x), \delta \hc_b(x')}_{\pt}
  N(x') \nabla_{\pt}' \Lambda^b(x') \\
  &+  \int d\Sigma'_{\pt} 
   \ip{\delta \hc_a(x), \delta \hcurrent^\mu_{~b}(x')}_{\pt}
  N(x') \nabla_{\perp\mu}' \Lambda^b(x')
  = N(x)\left[  (\nabla_\mu - \Gcal_\mu) \average{\delta \hcurrent^\mu_{~b}} 
  - \average{\delta \hScal_b (x)} \right].
 \end{split}
\end{equation}
Noting that we have a generalized susceptibility 
$\chi_{ab}(x,x';\pt) \equiv \ip{\delta \hc_a(x), \delta \hc_b(x')}_{\pt}$ 
in front of the time derivatives in the first term, 
we multiply the its inverse
$\chi^{ab}(x,x';\pt) \equiv \ip{\delta \hc_a(x), \delta \hc_b(x')}_{\pt}^{-1}$ 
and integrate with respect to the spatical coordinate $x'$,
which results in 
\begin{equation}
\begin{split}
 N(x) \nabla_{\pt} \Lambda^a(x)
 &= - \int d\Sigma'_{\pt} \int d\Sigma''_{\pt} 
 \ip{\delta \hc_a(x), \delta \hc_b(x')}_{\pt}^{-1}
 \ip{\delta \hc_b(x'), \delta \hcurrent^\mu_{~c}(x'')}_{\pt}
 N(x'') \nabla_{\perp\mu}'' \Lambda^c(x'') \\
 &\quad + \int d\Sigma'_{\pt} 
 \ip{\delta \hc_a(x), \delta \hc_b(x')}_{\pt}^{-1} 
 N(x')\left[  (\nabla'_\mu - \Gcal_\mu) \average{\delta \hcurrent^\mu_{~b}(x')} 
 - \average{\delta \hScal_b (x')} \right].
\end{split}
\end{equation}
This equation enables us to eliminate the time derivative from 
the entropy production operator~\eqref{eq:EntropyProd1}.
It is then natural and convenient to introduce
the projection operator $\hPcal$ onto $\delta \hc_a$ 
used in Refs.~\cite{PhysRevA.8.2048,Hayata:2015lga,Esposito1994} by
\begin{equation}
 \hPcal \hOcal \equiv \int d\Sigma_{\pt} \int d\Sigma_{\pt}'
  \delta \hc_a (x)  \ip{\delta \hc_a(x), \delta \hc_b(x')}_{\pt}^{-1}
   \ip{\delta \hc_b(x'), \hOcal}_{\pt}.
   \label{eq:ProjectionOp}
\end{equation}
When our system is in global thermal equilibrium, this projection operator 
reduces to the so-called Mori's projection operator~\cite{Mori2}.
As is mentioned above, 
$\ip{\delta \hc_a(x), \delta \hc_b(x')}_{\pt}$ is the 
generalized susceptibility given by
\begin{equation}
 \chi_{ab}(x,x';\pt)  
  \equiv \ip{\delta \hc_a(x), \delta \hc_b(x')}_{\pt}
  = \frac{\delta c_b (x')}{\delta \lambda^a (x) }
  = \frac{\delta^2 \Psi[\pt;\lambda,j]}{\delta \lambda^a(x) \delta\lambda^b (x')},
  \label{eq:Sus}
\end{equation}
which brings about the following expression of the 
inverse generalized susceptibility:
\begin{equation}
 \chi^{ab}(x,x';\pt) \equiv \ip{\delta \hc_a(x), \delta \hc_b(x')}_{\pt}^{-1}
  = \frac{\delta \lambda^b(x')}{\delta c_a(x)} 
  = - \frac{\delta^2 S[\pt;c]}{\delta c_a(x) \delta c_b(x')},
  \label{eq:InvSus}
\end{equation}
where the second expressions of Eqs.~\eqref{eq:Sus}-\eqref{eq:InvSus}
are due to Eqs.~\eqref{eq:delPsi} and \eqref{eq:delS}.
This expression for the inverse susceptibility together with
\begin{equation}
 \ip{\delta \hc_b(x'), \hOcal}_{\pt} 
  = \frac{\delta}{\delta \lambda^b (x')} \averageLG{\hOcal}_{\pt},
\end{equation}
allows us to rewrite the projection operator in a more explicit form as
\begin{equation}
 \hPcal \hOcal 
  = \int d\Sigma_{\pt} \int d\Sigma_{\pt}'
  \delta \hc_a (x) \frac{\delta \lambda^b(x')}{\delta c_a(x)} 
  \frac{\delta}{\delta \lambda^b (x')} \averageLG{\hOcal}_{\pt}
  = \int d\Sigma_{\pt} 
  \delta \hc_a (x) \frac{\delta }{\delta c_a(x)} \averageLG{\hOcal}_{\pt},
\end{equation}
from which we can clearly see that $\hPcal$ gives 
the projection of an arbitrary operator $\hOcal$ onto $\delta c_a$.
As is clear from the definition~\eqref{eq:ProjectionOp}, 
$\delta \hc_a (x)$ is invariant under the 
operation of the projection: $\hPcal \delta \hc_a (x) = \delta \hc_a (x)$.
With the help of this projection operator, 
we can compactly express $\hSigma[\pt,\pt_0;\lambda]$ as
\begin{equation}
 \begin{split}
  \hSigma [\pt,\pt_0;\lambda] 
  &= - \int_{\pt_0}^{\pt} d\ps 
  \int d\Sigma_{\ps} N
  \left[
  (1-\hPcal)\delta  \hcurrent^\mu_{~a} \nabla_{\perp\mu} \Lambda^a 
  + \delta \hlambda^b
  \left(
  (\nabla_\mu - \Gcal_\mu) \average{\delta \hcurrent^\mu_{~b}} 
  - \average{\delta \hScal_b } \right) \right]  \\ 
  &= - \int_{\pt_0}^{\pt} d\ps 
  \int d\Sigma_{\ps} N
  \left[
  \tildelta  \hcurrent^\mu_{~a} \nabla_{\perp\mu} \Lambda^a 
  + \delta \hlambda^b
  \left(
  (\nabla_\mu - \Gcal_\mu) \average{\tildelta \hcurrent^\mu_{~b}} 
  - \average{\tildelta \hScal_b } \right) \right]  \\ 
  &= - \int_{\pt_0}^{\pt} d\ps 
  \int d\Sigma_{\ps} N
  \Big[
  \tildelta  \hTcal^\mu_{~\nu} 
  ( \nabla_{\perp\mu} \beta^\nu - \beta^\sigma 
  F^n_{\rho\sigma} P_\mu^\rho  v^\nu )
  + \tildelta \hJ_M^\mu  
  ( \nabla_{\perp\mu} \nu'_M + \beta^\sigma F^a_{\rho\sigma} P_\mu^\rho )
  \\
  &\hspace{80pt} 
  + \tildelta  \hJ_Q^\mu  
  ( \nabla_{\perp\mu} \nu'_Q + \beta^\sigma F_{\rho\sigma} P_\mu^\rho )
  + \delta \hlambda^b \left(
  (\nabla_\mu - \Gcal_\mu) \average{\tildelta \hcurrent^\mu_{~b}} 
  - \average{\tildelta \hScal_b } \right) \Big] .
 \end{split}
\end{equation}
where we defined 
$\tildelta \hcurrent^\mu_{~a} \equiv (1-\hPcal) \delta \hcurrent^\mu_{~a}$ 
and introduced the operator $\delta \hlambda^a(x)$ by 
\begin{equation}
 \delta \hlambda^a(x) 
  \equiv 
  \int d\Sigma_{\pt}'
  \delta \hc_b(x')  \ip{\delta \hc_b(x'), \delta \hc_a(x)}_{\pt}^{-1} 
  = \int d\Sigma_{\pt}'
  \delta \hc_b(x') \frac{\delta \lambda^a(x)}{\delta c_b (x')}.
\end{equation}
Recalling the consequence followed from 
the Milne boost invariance \eqref{eq:Milne},
the mass current $\hJ_M^\mu(x)$ is related to the conserved momentum density 
$\hPcal_\mu (x)$ by $\hPcal_\mu = h_{\mu\nu} \hJ_M^\nu$. 
Then, the trivial projection 
$\hPcal \delta \hc_a (x) = \delta \hc_a(x)$, or $\tildelta \hc_a (x) = 0$ 
results in a disappearance
of the mass current $\hJ_M^\mu(x)$ from the entropy production operator%
\footnote{
Furthermore, 
if our system composed of a single-component charged matter, the mass current 
and electric current gives the same current except for its 
unimportant coefficient. 
In that case, we only need to consider either of them as a conserved 
current, which results in a disappearance of the both current.
This case is discussed in Sec.~\ref{sec:Derivation} 
}: $\tildelta \hJ_M^\mu(x) = 0$.
Inserting the definition of the 
nonrelativistic energy-momentum tensor \eqref{eq:NonrelaEM} and
rearranging the integrand, we obtain
\begin{equation}
 \begin{split}
  \hSigma [\pt,\pt_0;\lambda] 
  &= - \int_{\pt_0}^{\pt} d\ps 
  \int d\Sigma_{\ps} N
  \Big[
   - \tildelta \hEcal^\mu 
  ( \nabla_{\perp\mu} \beta - \beta^\sigma 
  F^n_{\rho\sigma} P_\mu^\rho)
  + \tildelta \hT_{\mu\nu} \nabla_{\perp}^{\mu} \beta^\nu
  + \tildelta  \hJ_Q^\mu  
  ( \nabla_{\perp\mu} \nu'_Q + \beta^\sigma F_{\rho\sigma} P_\mu^\rho )
  \\ 
  &\hspace{80pt}
  + \delta \hlambda^b \left(
  (\nabla_\mu - \Gcal_\mu) \average{\tildelta \hcurrent^\mu_{~b}} 
  - \average{\tildelta \hScal_b } \right) \Big] .
 \end{split}
\end{equation}
As a last step, we perform the tensor decomposition of 
the stress tensor $\hT_{\mu\nu}$ as 
\begin{equation}
 \tildelta \hT_{\mu\nu} = h_{\mu\nu} \tildelta \hp + \tildelta \hpi_{\mu\nu}, 
  \with 
  \tildelta \hp 
  \equiv \frac{1}{d-1} h^{\mu\nu} \tildelta \hT_{\mu\nu}, \quad 
  \tildelta \hpi_{\mu\nu} 
  \equiv P_\mu^\rho P_\nu^\sigma \tildelta \hT_{\rho\sigma} 
   - \frac{h_{\mu\nu}}{d-1} h^{\rho\sigma} \tildelta \hT_{\rho\sigma},
\end{equation}
where $\tildelta \hp$ denotes the trace part and $\tildelta \hpi_{\mu\nu}$ 
the symmetric traceless part of the stress-tensor.
We eventually obtain the following expression for the 
entropy production operator
\begin{equation}
 \begin{split}
  \hSigma [\pt,\pt_0;\lambda] 
  &= \int_{\pt_0}^{\pt} d\ps 
  \int d\Sigma_{\pt} N
  \Big[
  \tildelta \hEcal^\mu 
  ( \nabla_{\perp\mu} \beta - \beta^\sigma 
  P_\mu^\rho  F^n_{\rho\sigma})
  - \tildelta \hp (h_{\mu\nu} \nabla_{\perp}^{\mu} \beta^\nu)
  - \tildelta \hpi_{\mu\nu} \nabla_{\perp}^{\langle\mu} \beta^{\nu\rangle}
  \\
  &\hspace{70pt} 
  - \tildelta  \hJ_Q^\mu  
  ( \nabla_{\perp\mu} \nu'_Q + \beta^\sigma P_\mu^\rho F_{\rho\sigma}  )
  - \delta \hlambda^a \left(
  (\nabla_\mu - \Gcal_\mu) \average{\tildelta \hcurrent^\mu_{~a}} 
  - \average{\tildelta \hScal_a } \right) \Big], 
  \label{eq:EntropyProd2}
 \end{split}
\end{equation}
where we defined the symmetric traceless projection of 
$\nabla^\mu \beta^\nu$ as
\begin{equation}
 \nabla^{\langle\mu} \beta^{\nu\rangle}
  \equiv 
  \frac{1}{2} P_\rho^\mu P_\sigma^\nu
  ( \nabla^{\rho} \beta^{\sigma} + \nabla^{\sigma} \beta^{\rho} )
  - \frac{h^{\mu\nu}}{d-1} h_{\rho\sigma} \nabla^\rho \beta^\sigma.
\end{equation}
Here we note that $\nabla_\mu \average{\tildelta \hcurrent^\mu_{~a}}$ does not 
contain the explicit time derivative because we have 
\begin{equation}
 \nabla_\mu \average{\tildelta \hcurrent^\mu_{~a}} 
  = (n_\mu v^\nu \nabla_{\nu} + \nabla_{\perp \mu})
  \average{\tildelta \hcurrent^\mu_{~a}} 
  = \big(-(v^\nu \nabla_\nu n_\mu)  + \nabla_{\perp \mu} \big)
  \average{\tildelta \hcurrent^\mu_{~a}} 
  =  \nabla_{\perp \mu} \average{\tildelta \hcurrent^\mu_{~a}},
\end{equation}
where we used the derivative of 
$n_\mu \average{\tildelta \hcurrent^\mu_{~a}} = 
  \average{\tildelta \hc_a} = 0 $ for the second equality and 
compatibility condition $\nabla_\nu n_\mu = 0 $ 
for the last equality. 
Therefore, the entropy production operator \eqref{eq:EntropyProd2} 
is written in terms of 
the external fields $j(x)$ and 
the spatial derivative of local thermodynamic parameters $\lambda^a(x)$. 
However, we also note that the time derivative of parameters may appear 
from the higher-order correction of $\average{\tildelta \hcurrent^\mu_{~a}}$.

Then, noting $\average{\delta \hcurrent^\mu_{~a}(x)} 
= \average{\tildelta \hcurrent^\mu_{~a}(x)}$ 
due to $\average{\delta \hc_a(x)} = 0$ 
followed from our condition to determine local thermodynamic parameters 
\eqref{eq:DefParameter}, we eventually obtain the final expression for 
$\average{\delta \hcurrent^\mu_{~a}}$ as 
\begin{equation}
 \average{ \delta \hcurrent^\mu_{~a} (x)}
  = \averageLG{
  T_\tau e^{\int_0^1 d\tau \hSigma_\tau[\pt,\pt_0;\lambda]}
  \tildelta \hcurrent^\mu_{~a} (x)}_{\pt} 
 \label{eq:dissipativeTJ}.
\end{equation}
This equation together with the expression of the 
entropy production operator \eqref{eq:EntropyProd2} provides an 
exact formula for the dissipative part of the constitutive relations. 
Since $\hSigma_\tau [\pt,\pt_0;\lambda]$ contains 
$\average{ \tildelta \hcurrent^\mu_{~a} (x)}$, 
the above equation gives a self-consistent equation to determine 
$\average{ \tildelta \hcurrent^\mu_{~a} (x)}$, which can be 
solved order-by-order with respect to the derivative expansion 
of $\lambda^a(x)$ as discussed in the next section.

\section{Derivation of hydrodynamic equations}
\label{sec:Derivation}
In this section, based on the exact formulae derived in the 
previous sections, we perform the derivative expansion and 
derive hydrodynamic equations order-by-order. 
We restrict ourselves to the simplest case---a single component 
parity-even fluid 
in the zeroth-order and first-order derivative expansion. 
As a consequence, 
we obtain the constitutive relations for the perfect fluid and 
Navier-Stokes fluid, respectively. 
After demonstrating our basic procedure, 
we give the leading-order (zeroth-order) result in Sec.~\ref{sec:Zeroth}.
In Sec.~\ref{sec:First}, we proceed to the first-order correction to 
the constitutive relation, which leads to the Navier-Stokes equation.

Before starting the discussion, 
we briefly summarize our starting point for the derivative expansion.
The result obtained so far is summarized as follows: 
We have decomposed the full average of conserved current operators 
into two parts:
\begin{equation}
 \average{\hcurrent^\mu_{~a}(x)} 
  = \averageLG{\hcurrent^\mu_{~a}(x)}_{\pt}
 +  \average{\delta \hcurrent^\mu_{~a}(x)}.
\end{equation}
Here the first term represents the nondissipative part appearing in 
local thermal equilibrium and the second terms does the dissipative part 
originated from the derivation from local thermal equilibrium.
We have derived the exact formulae for both of them as given in 
Eqs.~\eqref{eq:VarGeneral}, or \eqref{eq:VarHS} in the hydrostatic gauge, 
and Eq.~\eqref{eq:dissipativeTJ}.

Therefore, in order to evaluate the nondissipative part of 
constitutive relation order-by-order, we only need to 
perform the derivative expansion of the Massieu-Planck functional 
$\Psi[\pt;\lambda,j]$. 
For that purpose, we have to specify a power counting scheme for the 
parameters such as $\lambda^a(x)$, and external fields $j(x)$.
We employ the most standard choice in this paper where all parameters
are order $p^0$: $\lambda^a = j =  O(p^0)$, 
which allow us to apply the usual derivative expansion.
Here we use the momentum $p$ instead of the spatial derivative 
$\nabla_{\perp}$.
Nevertheless, note that this is not the unique choice 
since we should adopt other power counting scheme to describe systems 
e.g. in the presence of the strong magnetic field%
\footnote{
As will be discussed in the next papar \cite{Hongo},
we assume $A_i = O (p^{-1})$ in such a situation 
so that the magnetic field satisfies
$\vec{B} \equiv \vec{\nabla} \times \vec{A}  = O (p^0)$, 
which brings about the fact that the magnetic field can appear 
in the leading-order expansion.
}. 
Since we fix our power counting scheme, 
based on the symmetry arguments, we can perform the 
derivative expansion of the Massieu-Planck functional as
\begin{equation}
 \Psi[\pt;\lambda,j] 
  = \Psi^{(0)}[\pt;\lambda,j] + \Psi^{(1)}[\pt;\lambda,j] 
  + \Psi^{(2)}[\pt;\lambda,j] + \cdots,
\end{equation}
which provides nondissipative constitutive relations 
for $\averageLG{\hcurrent^\mu_{~a}(x)}_{\pt}$.
Here upper indices in the right-hand side of this equation
denotes the number of the spatial derivative (or momentum $p$). 
Note that we only have the spatial derivative 
due to the definition of the Massieu-Planck functional.

Furthermore, expanding $\hU$ in Eq.~\eqref{eq:dissipativeTJ} together with the 
entropy production operator \eqref{eq:EntropyProd2} provides us 
the dissipative part of constitutive relations in a self-consistent manner.
Since the entropy functional inescapably contain at least one spatial 
derivative of parameters, the expansion with respect to $\hSigma$ can be
regarded as the derivative expansion.
We then obtain 
\begin{equation}
 \begin{split}
  \average{ \tildelta \hcurrent^\mu_{~a} (x)}
  &= \averageLG{
  T_\tau e^{\int_0^1 d\tau \hSigma_\tau[\pt,\pt_0;\lambda]}
  \tildelta \hcurrent^\mu_{~a} (x)}_{\pt} \\
  &= \averageLG{\tildelta \hcurrent^\mu_{~a} (x)}_{\pt}  
  + \int_0^1 d\tau  \averageLG{T_\tau \hSigma_\tau[\pt,\pt_0;\lambda] 
  \tildelta \hcurrent^\mu_{~a} (x)}_{\pt}  \\
  &\hspace{65pt} 
  + \frac{1}{2} \int_0^1 d\tau \int_0^1 d\tau'
  \averageLG{T_\tau \hSigma_\tau[\pt,\pt_0;\lambda] 
  \hSigma_{\tau'} [\pt,\pt_0;\lambda] 
  \tildelta \hcurrent^\mu_{~a} (x)}_{\pt}
  + \cdots .
 \end{split}
 \label{eq:DissipCons}
\end{equation}
Here the first term in the second line vanish by definition:
$\averageLG{\tildelta \hcurrent^\mu_{~a} (x)}_{\pt}  
=\averageLG{\delta \hcurrent^\mu_{~a} (x)}_{\pt} = 0 $.
Then, the leading-order dissipative correction appears with 
at least one spatial derivative. 
Although we do not discuss the next-leading-order dissipative correction, 
we note that second-order corrections also arises from 
the second term with the single $\hSigma_\tau$ in addition to 
contributions from the third term.

\subsection{Zeroth-order result: Perfect fluid}
\label{sec:Zeroth}
As is clarified above, 
dissipative corrections to constitutive  relations are inevitably 
accompanied by at least on spatial derivative of local 
thermodynamic parameters $\lambda^a(x)$.
We, therefore, do not have the zeroth-order dissipative corrections, 
and we only need to evaluate the Massieu-Planck functional in 
the leading-order derivative expansion.

As is elaborated in Sec.~\ref{sec:Symmetry}, we can only use 
$\beta(x)$ and $\nu_M$ as basic building blocks
 of the leading-order Massieu-Planck functional $\Psi^{(0)}[\pt;\lambda,j]$. 
Then, the most general form of $\Psi^{(0)}[\pt;\lambda,j]$ 
respecting diffeomorphism and $U(1)_M$ gauge invariance in emergent 
thermal spacetime is given by
\begin{equation}
 \Psi^{(0)} [\pt;\lambda]
  = \int_0^{\beta_0} d^d \tilx \sqrt{\tilgamma} p (\beta, \nu_M) 
  = \int d^{d-1} \px \sqrt{h} \beta p (\beta, \nu_M) , 
  \label{eq:0thPsi}
\end{equation}
where we used $ \sqrt{\tilgamma} = e^{\sigma} \sqrt{h}$ with 
$e^{\sigma(x)} = \beta(x) / \beta_0$ and 
performed the integration with respect to $\tilt$ to derive the right-hand side
of this equation. 
Here $p(\beta,\nu_M)$ represent a certain function 
dependent on $\beta$ and $\nu_M$ which satisfies a following relation,
\begin{equation}
 d (\beta p) = c_a d \lambda^a = p_\mu d\beta^\mu + n_M d \nu_M', 
\end{equation}
due to the thermodynamic properties of 
the Massieu-Planck functional \eqref{eq:delPsi}.
From this equation, we can read off its relations to 
the conserved charge densities as 
\begin{equation}
 \frac{\partial p}{\partial \beta} 
  = - \frac{1}{\beta} 
  \left( n_\mu \Ecal^\mu + p + \frac{1}{2} n_M u^2  \right), \quad
  \frac{\partial p}{\partial \nu_M} 
  = \frac{n}{\beta}, 
  \label{eq:delp}
\end{equation}
whese we used 
$\beta = \beta^\mu n_\mu$, $\nu_M  = \nu_M'  + \dfrac{1}{2} \beta u^2$, 
and $p_\mu v^\mu = - n_\mu \Ecal^\mu$ 
with $\Ecal^\mu \equiv \averageLG{\hEcal^\mu}_{\pt}$. 

Since we have obtained the explicit form of the leading-order 
Massieu-Planck functional $\Psi^{(0)}[\pt;\lambda]$, 
the variational formulae \eqref{eq:VarGeneral} 
enables us to obtain the corresponding leading-order constitutive relations. 
However, we further simplify the problem by employing the hydrostatic gauge 
developed in Sec.~\ref{sec:Variation2}, and use 
the corresponding variational formulae \eqref{eq:VarHS}.
In the hydrostatic gauge, recalling 
the gauge fixing condition \eqref{eq:HydrostaticGauge}, we have
$ N (x) \big|_{\mathrm{hs}} \equiv n_\mu (x) t^\mu (x) \big|_{\mathrm{hs}} 
= n_\mu (x) \beta^\mu (x) / \beta_0 = e^\sigma $, which leads to 
$\sqrt{\gamma} \big|_{\mathrm{hs}} = \sqrt{\tilgamma}$. 
Thus, we can simply express $\Psi^{(0)}[\pt;\lambda]$ 
by the use of the original background field $j(x) \big|_{\mathrm{hs}}$: 
\begin{equation}
 \Psi^{(0)} [\pt;\lambda] \big|_{\mathrm{hs}}
  = \int_0^{\beta_0} d^d \tilx \sqrt{\gamma} p (\beta, \nu_M) 
  = \beta_0 \int d^{d-1} \px \sqrt{\gamma} p (\beta, \nu_M) \big|_{\mathrm{hs}}.
  \label{eq:0thPsiHS}
\end{equation}
Let us then take the variation of 
$\Psi^{(0)} [\pt;\lambda] \big|_{\mathrm{hs}}$ with respect to 
the independent background fields 
$\ptc{j}(x) \equiv \{ n_\mu, \ptc{v}^\mu, \ptc{h}^{\mu\nu},v^\mu, a_\mu \}$. 
For that purpose, we use a following variational formulae 
\begin{equation}
 \begin{split}
  \delta \beta
  &= \beta^\mu \delta n_\mu ,
  \\
  \delta \sqrt{\gamma} 
  &= \sqrt{\gamma} 
  \Big( v^\mu \delta n_\mu 
  - \frac{1}{2} h_{\mu\nu} \delta \ptc{h}^{\mu\nu} \Big) , 
  \\
  \delta \nu_M \big|_{\mathrm{hs}}
  &= - \beta^\mu \delta a_\mu + \frac{1}{2} u^2 \beta^\mu \delta n_\mu
  - \beta u_\mu \delta \ptc{v}^\mu 
  - \frac{1}{2} \beta u_\mu u_\nu \delta \ptc{h}^{\mu\nu}, 
 \end{split}
\end{equation}
where we used $\nu_M \big|_{\mathrm{hs}} 
= - \beta^\mu a_\mu + \dfrac{1}{2} \beta u^2 $. 
By using these relations, we can calculate the variation of 
the leading-order Massieu-Planck functional as follows:
\begin{equation}
 \begin{split}
  \delta \Psi^{(0)} [\pt;\lambda] \big|_{\mathrm{hs}}
  &= \beta_0 \int d^{d-1} \px 
  \left[ p \delta \sqrt{\gamma} 
  + \frac{\partial p}{\partial \beta} \delta \beta 
  + \frac{\partial p}{\partial \nu_M} \delta \nu_M \big|_{\mathrm{hs}}
  \right] \\
  &= - \beta_0 \int d^{d-1} \px \sqrt{\gamma}  
  \Bigg[ 
  \Big(  (\Ecal \cdot n) u^\mu +  p P^\mu_\nu u^\mu  
  \Big) \delta n_\mu
  \\
  &\hspace{90pt}
  + n_M u_\mu \delta \ptc{v}^\mu
  + \frac{1}{2} 
  \Big( n_M u_\mu u_\nu  + p h_{\mu\nu} \Big)  \delta \ptc{h}^{\mu\nu}
  + n_M u^\mu \delta a_\mu 
   \Bigg] .
 \end{split}
\end{equation}
Recalling the variational formulae in the hydrostatic gauge \eqref{eq:VarHS}, 
we eventually obtain 
\begin{align}
 \averageLG{\hEcal^\mu(x)}_{\pt}  
 &= - \dfrac{1}{\beta_0 \sqrt{\gamma}}
 \dfrac{\delta \Psi[\pt;j]}{\delta n_\mu(x)} \Bigg|_{\mathrm{hs}} 
 = \big( \Ecal \cdot n \big) u^\mu + p P^\mu_\nu u^\nu , 
 \label{eq:0thEcal}
 \\
 \averageLG{\hPcal_\mu(x)}_{\pt}  
 &= - \dfrac{1}{\beta_0 \sqrt{\gamma}}
  \dfrac{\delta \Psi[\pt;j]}{\delta \ptc{v}^\mu(x)} \Bigg|_{\mathrm{hs}}
 = n_M u_\mu 
 \label{eq:0thPcal}  ,
 \\
 \averageLG{\hT_{\mu\nu}(x)}_{\pt}  
 &= - \dfrac{2}{\beta_0 \sqrt{\gamma}}
  \dfrac{\delta \Psi[\pt;j]}{\delta \ptc{h}^{\mu\nu}(x)} \Bigg|_{\mathrm{hs}}
 = n_M u_\mu u_\nu + p h_{\mu\nu} ,
 \label{eq:0thT}  \\
 \averageLG{\hJ_M^\mu(x)}_{\pt}  
 &= - \dfrac{1}{\beta_0 \sqrt{\gamma}}
  \frac{\delta \Psi[\pt;j]}{\delta a_\mu(x)} \Bigg|_{\mathrm{hs}}
 =  n_M u^\mu 
 \label{eq:0thJM}. 
\end{align}
These equation provide the leading-order constitutive relation
which correctly reproduces the equation of motion for a perfect fluid 
in conjunction with the conservation laws. 
From these equations, we see that $p$ and $u^\mu$ is simply 
regarded as a pressure and velocity of the fluid, respectively. 
Combination of Eqs.~\eqref{eq:0thEcal}-\eqref{eq:0thT} gives 
the leading-order expression for the nonrelativistic energy-momentum tensor as
\begin{equation}
 \begin{split}
  \averageLG{\hTcal^\mu_{~\nu}(x)}_{\pt} 
  &= - \Big( \big( \Ecal \cdot n \big) u^\mu + p P^\mu_\rho u^\rho \Big) n_\nu
  + n_M v^\mu u_\nu + n_M P^\mu_\rho u^\rho u_\nu + p P^\mu_\nu \\
  &= - \big( \Ecal \cdot n \big) u^\mu  n_\nu 
  + n_M u^\mu u_\nu
  + p (\delta^\mu_\nu - u^\mu n_\nu).
 \end{split}
\label{eq:0thTcal}
\end{equation}
Note that the fluid pressure in Eq.~\eqref{eq:0thPsi} 
can be, in principle, calculable from the microscopic quantum theory 
by evaluating the path-integral formula \eqref{eq:PTforPsi}.
Therefore, we have derived a universal form of the leading-order 
constitutive relations together with a way to calculate its all contents.
This provides the leading-order answer to our question to derive
nonrelativistic hydrodynamics raised in Sec.~\ref{sec:Intro}.

\subsection{First-order result: Navier-Stokes fluid}
\label{sec:First}
Let us proceed the first-order derivative expansion and derive 
the constitutive relation for the Navier-Stokes fluid.
First of all, it is important to notice that we do not have the first-order
nondissipative correction
since we are now considering systems with parity symmetry, 
which leads to $\Psi^{(1)}[\pt;\lambda,j] = 0$.
Thus, the nondissipative part is the same as the leading-order results 
\eqref{eq:0thEcal}-\eqref{eq:0thJM}, 
and we only need to take into account the leading-order dissipative corrections
in the first-order derivative expansion.

In order to derive the first-order dissipative correction, 
we first rewrite the second term in Eq.~\eqref{eq:DissipCons} 
by redefining the integration variable as $\tau \to \tau'= 1-\tau$ 
and using the cyclic property of traces 
: $\Tr(AB) = \Tr (BA)$, which leads to 
\begin{equation}
 \begin{split}
  \average{\delta \hcurrent^\mu_{~a}(x)}
  &= \int_0^1 d\tau
  \averageLG{e^{\hK \tau} \hSigma[\pt,\pt_0;\lambda] e^{-\hK\tau}
  \tildelta \hcurrent^\mu_{~a} (x)}_{\pt} + O (\nabla^2)  \\
  &= \int_0^1 d\tau
  \averageLG{e^{\hK\tau} \tildelta \hcurrent^\mu_{~a} (x)e^{-\hK \tau}
  \hSigma[\pt,\pt_0;\lambda] }_{\pt}  + O (\nabla^2) \\
  &= \ip{\tildelta \hcurrent^\mu_{~a} (x), \hSigma[\pt,\pt_0;\lambda]}_{\pt} 
   + O (\nabla^2) .
  \label{eq:1stDiss}
 \end{split}
\end{equation}
Here note that $\average{\delta \hcurrent^\mu_{~a}(x)} $ contains at least 
one spatial derivative.
This allows us to neglect the term proportional to 
$\delta \hlambda^a$ in the entropy production operator 
$\hSigma[\pt,\pt_0;\lambda]$ since it contains two derivatives.
Furthermore, we are considering the single-component fluid, and thus, 
the mass current and electric current (if charged) gives the same current 
except for the unessential coefficient. 
In this case, the electric current also disappears from 
the entropy production operator due to 
the consequence of the Milne boost invariance: 
$\tildelta \hJ_Q^\mu \propto \tildelta \hJ_M^\mu = 0$. 
We thus need to evaluate the above equation with the following reduced form of 
$\hSigma[\pt,\pt_0;\lambda]$:
\begin{equation}
 \begin{split}
  \hSigma [\pt,\pt_0;\lambda] 
  &= \int_{\pt_0}^{\pt} d\ps 
  \int d\Sigma_{\ps} N
  \Big[
  \tildelta \hEcal^\mu 
  ( \nabla_{\perp\mu} \beta - \beta^\sigma P_\mu^\rho F^n_{\rho\sigma} )
  - \tildelta \hp (h_{\mu\nu} \nabla_{\perp}^{\mu} \beta^\nu)
  - \tildelta \hpi_{\mu\nu} \nabla_{\perp}^{\langle\mu} \beta^{\nu\rangle} 
  \Big] .
 \end{split}
\end{equation}
In Eq.~\eqref{eq:1stDiss}, we still have higher-order contributions 
coming from the expansion of the correlation function. 
We then assume that our correlation functions behaves in a moderate manner 
showing the exponential damping with respect to spacetime differences%
\footnote{
As is well-known, this assumption breaks down in 
low dimensional systems due to the hydrodynamic fluctuations.
Considerations of hydrodynamic fluctuations will be gien elsewhere.
}.
This assumption enables us to construct the local (Markovian) 
constitutive relations with transport coefficients.
To see this, recalling that all the dissipative term is 
perpendicular to $n_\mu$ and $v^\mu$, 
we perform the tensor decomposition of Eq.~\eqref{eq:1stDiss} 
only by the use of $h^{\mu\nu}$ (or $h_{\mu\nu}$). 
Then, the expectation value of e.g. the energy current   
$\average{\tildelta \hEcal^\mu (x)}$ in the first-order 
derivative expansion can be evaluated as
\begin{equation}
 \begin{split}
  \average{\tildelta \hEcal^\mu (x)}
  &= \ip{\tildelta \hEcal^\mu (x),\hSigma[\pt,\pt_0;\lambda]}_{\pt} 
  + O(\nabla^2) \\
  &= \int_{\pt_0}^{\pt} d\pt' \int d\Sigma_{\pt'} N'
  \ip{\tildelta \hEcal^\mu (x),\tildelta \hEcal^\nu (x')}_{\pt} 
  \big( \nabla_{\perp\nu}' \beta(x') - \beta^\sigma (x') 
  P_\nu^\rho (x') F^n_{\rho\sigma} (x') \big)  
  + O(\nabla^2) \\
  &= \int_{\pt_0}^{\pt} d\pt' \int d\Sigma_{\pt'} N'
  \ip{\tildelta \hEcal^\mu (x),\tildelta \hEcal^\nu (x')}_{\pt} 
  \big( \nabla_{\perp\nu} \beta(x) - \beta^\sigma (x) 
  P_\nu^\rho (x) F^n_{\rho\sigma} (x) \big)  
  + O(\nabla^2) .
 \end{split}
\end{equation}
To proceed the second line, 
we used the fact that a possible term for non-vanishing correlation 
functions is only 
$\ip{\tildelta \hEcal^\nu (x),\tildelta \hEcal^\mu (x')}_{\pt} $ due to 
the number of the tensor indices. 
We also used the above assumption on the correlation function 
to derive the last line.
Then, regarding the integral part as a transport coefficient, 
this equation gives the local constitutive relations for the 
energy current. 
Similar analysis also works for the stress-tensor 
of the trace part $\average{\tildelta \hp(x)}$ and traceless symmetric 
part $\average{\tildelta \hpi_{\mu\nu}(x)}$.
Then, recalling that 
$\tildelta \hPcal_\mu (x)= \tildelta \hJ_M^\mu (x) = \tildelta \hJ_Q(x) = 0$, 
we obtain the first-order derivative corrections to constitutive relations
as follows:
\begin{equation}
 \begin{split}
 \average{\tildelta \hEcal^\mu (x)}
  &= \frac{\kappa}{\beta} h^{\mu\nu}
  (\nabla_{\perp\nu} \beta + \beta^\sigma F^n_{\sigma\rho} P^\rho_\nu ) 
  + O (\nabla^2) , \\
  \average{\tildelta \hT_{\mu\nu}(x)}
  &= - h_{\mu\nu}
  \frac{\zeta}{\beta} h_{\rho\sigma} \nabla_\perp^{\rho} \beta^\sigma
  - \frac{2\eta}{\beta} h_{\mu\rho} h_{\nu\sigma} 
  \nabla_{\perp}^{\langle\rho} \beta^{\sigma\rangle}
  + O (\nabla^2) , \\
  \average{\tildelta \hPcal_ \mu (x)}  &= 
  \average{\tildelta \hJ_M^\mu (x)} = 
  \average{\tildelta \hJ_Q^\mu (x)}  =0 ,
 \end{split}
\end{equation}
where the transport coefficients $L_i =\{\zeta, \eta, \kappa\}$---%
the bulk viscosity $\zeta$, shear viscosity $\eta$, 
and thermal conductivity $\kappa$---are given by
\begin{align}
 \zeta 
 &= \beta(x) \int_{-\infty}^{\pt} d\pt' \int d\Sigma_{\pt}' N' 
 \ip{\tildelta \hp(x), \tildelta \hp(x')}_{\pt} 
 \label{eq:zeta} \\
 \eta 
 &= \frac{\beta(x)}{(d+1)(d-2)} 
 \int_{-\infty}^{\pt} d\pt' \int d\Sigma_{\pt}' N' 
 \ip{\tildelta \hpi_{\mu\nu}(x), \tildelta \hpi_{\rho\sigma}(x')}_{\pt} 
 h^{\mu\rho} h^{\nu\sigma}   
 \label{eq:eta}\\
 \kappa
 &= \frac{\beta(x)}{d-1} 
 \int_{-\infty}^{\pt} d\pt' \int d\Sigma_{\pt}' N' 
 \ip{\tildelta \hEcal^\mu(x), \tildelta \hEcal^\nu(x')}_{\pt} h_{\mu\nu} 
 \label{eq:kappa}.
\end{align}
These are the so-called Green-Kubo formulae for transport coefficients~
\cite{Green,Nakano,Kubo}.
Combining these with the result for nondissipative part, we finally obtain 
the following constitutive relations in the first-order derivative 
expansion:
\begin{align}
 \average{\hEcal^\mu(x)}
 &= \big( \Ecal \cdot n \big) u^\mu + p P^\mu_\nu u^\nu 
 + \frac{\kappa}{\beta} h^{\mu\nu}
  (\nabla_{\perp\nu} \beta + \beta^\sigma F^n_{\sigma\rho} P^\rho_\nu ) , 
 \label{eq:1stEcal}
 \\
 \average{\hPcal_\mu(x)}
 & = n_M u_\mu 
 \label{eq:1stPcal} ,
 \\
 \average{\hT_{\mu\nu}(x)}
 &= n_M u_\mu u_\nu + p h_{\mu\nu} 
 - h_{\mu\nu}
  \frac{\zeta}{\beta} h_{\rho\sigma} \nabla_\perp^{\rho} \beta^\sigma
  - \frac{2\eta}{\beta} h_{\mu\rho} h_{\nu\sigma} 
  \nabla_{\perp}^{\langle\rho} \beta^{\sigma\rangle},
 \label{eq:1stthT}  \\
 \average{\hJ_M^\mu(x)}
 &=  n_M u^\mu 
 \label{eq:1stthJM}. 
\end{align}
Then, recalling the definition of the nonrelativistic energy-momentum tensor, 
we obtain 
\begin{equation}
 \begin{split}
  \average{\hTcal^\mu_{~\nu}(x)} 
  &=  - \big( \Ecal \cdot n \big) u^\mu  n_\nu 
  + n_M u^\mu u_\nu
  + p (\delta^\mu_\nu - u^\mu n_\nu)  
  \\ &\quad 
  + \frac{\kappa}{\beta} h^{\mu\lambda}
  (\nabla_{\perp\lambda} \beta
  + \beta^\sigma P^\rho_\lambda F^n_{\sigma\rho}  ) n_\nu 
  - P^\mu_\nu
  \frac{\zeta}{\beta} h_{\rho\sigma} \nabla_\perp^{\rho} \beta^\sigma
  - \frac{2\eta}{\beta} P^\mu_\rho h_{\nu\sigma} 
  \nabla_{\perp}^{\langle\rho} \beta^{\sigma\rangle} .
 \end{split}
\end{equation}
These are our final results for the derivation of hydrodynamic equation 
in the first-order derivative expansion.
We emphasize that 
this form of the constitutive relation is universal and independent of 
microscopic ingredients/interactions of systems 
as long as symmetry properties given in Sec.~\ref{sec:Matter} are satisfied.
On the other hand, 
the functional form of the equation of state $p = p(\beta,\nu_M)$
and the transport coefficients $L_i = L_i(\lambda^a)$ depend on 
the microscopic details of systems.
The crucial point here is that 
once we determine the microscopic system under consideration,
we can, in principle, calculate all of them 
based on the path-integral formula for the Massieu-Planck functional 
\eqref{eq:PTforPsi} with the leading-order form \eqref{eq:0thPsi} and
the Green-Kubo formula~\eqref{eq:zeta}-\eqref{eq:kappa}
for given $\beta^\mu$ and $\nu_M'$ which 
have one-to-one correspondences to the conserved charge densities $c_a$.
Therefore, all quantities appearing in the constitutive relations are 
now calculable for given values of conserved charge densities, 
which provides a next-leading-order complete answer to the problem of 
the derivation of the nonrelativistic hydrodynamic equation 
raised in Sec.~\ref{sec:Intro}.

\section{Discussion}
\label{sec:Discussion}
In this paper, we only consider the parity-even normal fluid 
composed of the spinless Schr\"odinger field. 
There are several prospects which should be clarified based on our approach.
One is a generalization to systems with spin degrees of freedom, e.g. 
systems composed of the \textit{spinful} Schro\"odinger field. 
The reason why we do not consider the spinful case in this paper 
is that it requires another elaborate preliminary
in order to deal with spinful fields in the curved geometry. 
For example, we need to introduce the vielbein formalism 
with the spin connection in the covariant derivative of spinor fields 
which comes from invariance under the local spatial rotation. 
As a result, we have to show whether the path-integral formula for local 
thermal equilibrium---Eq.~\eqref{eq:PTforPsi} in this paper---contains
the appropriate spin connection in emergent thermal spacetime or not
(See Ref.~\cite{Hongo:2016mqm} for the discussion 
on the Dirac field in the relativistic setup). 
In the companion paper~\cite{Hongo}, we will deal with 
the spinful Schr\"odinger field and clarify the derivation of 
hydrodynamic equations with spin degrees of freedom.

It is also interesting to consider transport phenomena 
which does not take place in the normal fluid---one typical example is 
a hall transport in the parity-odd fluid, 
and another is a transport in the superfluid. 
The former is relatively easy to take into account 
since it is classified into the nondissipative transport
captured by the Massieu-Planck functional $\Psi[\pt;\lambda,j]$. 
Based on the symmetry argument discussed in Sec.~\ref{sec:PathIntegral}, 
we can write down the possible form of derivative corrections of 
$\Psi[\pt;\lambda,j]$. 
Furthermore, with the help of the field-theoretical technique 
like the diagrammatic calculation, we can evaluate their explicit form, 
which includes electric hall conductivity, thermal hall conductivity, 
and hall viscosity and so on. 
Compared to this, the derivation of the superfluid hydrodynamics 
is a little bit complicated since we have to consider the new massless mode 
known as the Nambu-Goldstone mode~
\cite{Nambu:1961tp,Goldstone:1961eq,Goldstone:1962es} from the starting point. 
Because of this new massless degree of freedom, we have both nondissipative 
and dissipative corrections to the hydrodynamic equations, which may 
lead to the famous two-fluid hydrodynamic equation 
(See Refs.~\cite{Landau:Fluid,Khalatnikov}). 
Extending this direction enables us to justify a unified 
hydrodynamic treatment of superfluid, liquid crystal and 
crystal~\cite{PhysRevA.6.2401} which is regarded as hydrodynamics
with spontaneous symmetry breaking of internal, rotational, and 
translational symmetry. 
Consideration of these is left for future works.

We lastly point out an unsettled point not captured by our approach. 
Our derivation is for the conventional hydrodynamic equation without  
thermal fluctuation. 
In other words, our hydrodynamic equation is regarded as one obtained
after integrating out the nonlinear hydrodynamic fluctuation.
In the usual setup (like a normal fluid in $d=3+1$ dimension), 
the hydrodynamic fluctuation does not cause 
serious problem, and what we only need to do is to use the renormalized 
transport coefficients for our hydrodynamic equation.
However, in some situations---e.g. low dimensional systems---the 
hydrodynamic fluctuation breaks our assumption on 
the moderate behaviour of correlation functions, and we cannot 
construct the local (Markovian) constitutive relations, 
which means the breakdown of the conventional (non-fluctuating) hydrodynamics. 
We thus need to construct a systematic way to take into account the effect of
the nonlinear hydrodynamic fluctuation. 
Here we only point out the possibility that recent developments on 
the effective field theoretical approach to 
relativistic dissipative hydrodynamics~(See \cite{Crossley:2015evo,Jensen:2017kzi,Glorioso:2017fpd}) may help us to consider this.

\begin{acknowledgements}
The author thanks K. Fujii, Y. Hidaka, K. Jensen, Y. Kikuchi, M. Roberts, 
K. Saito, S-i. Sasa, H. Taya, and T. Tsuboi for useful discussions.
M.H. was supported by the Special Postdoctoral Researchers Program
at RIKEN. This work was partially supported by the RIKEN iTHES/iTHEMS Project 
and iTHEMS STAMP working group.
\end{acknowledgements}

\appendix
\section{Proof of the quantum fluctuation theorem \eqref{eq:FluctuationTheorem}}
\label{sec:FTProof}
Here we give a proof of the quantum fluctuation theorem 
for local thermal equilibrium \eqref{eq:FluctuationTheorem}.
In order to prove it, we insert $1 = \Theta^{-1} \Theta$ 
with $\Theta \equiv \PT$ denotes a shorthand notation 
for the combined $\PT$ transformation. 
Note that $\Theta$ is an anti-unitary operator because of $\Tcal$.
Then, from the definition of $G_F(z;j]$, we obtain 
\begin{equation}
 \begin{split}
  G_F (z;j]
  &=  \Tr \left( 
  \hrho_0 \, \hUcal_j^\dag 
  e^{iz \hS_0 [\pt;\lambda]} \hUcal_j
  e^{-i z \hS[\pt_0;\lambda] } \right)  \\
  &=  \Tr \left( 
  \Theta^{-1} \Theta
  e^{-\hS [\pt_0;\lambda]} \Theta^{-1} \Theta
  \hUcal_j^\dag \Theta^{-1} \Theta
  e^{iz \hS_0 [\pt;\lambda]} 
  \Theta^{-1} \Theta \hUcal_j \Theta^{-1} \Theta
  e^{-i z \hS[\pt_0;\lambda] } \Theta^{-1} \Theta \right)  \\
  &=  \Tr \left( 
  \Theta^{-1} 
  e^{- \Theta \hS [\pt_0;\lambda]\Theta^{-1}  } 
  \tilUcal_j  e^{-iz \Theta \hS_0 [\pt;\lambda] \Theta^{-1}} 
  \tilUcal_j^\dag  e^{i z \Theta \hS[\pt_0;\lambda] \Theta^{-1} }
  \Theta \right) ,
 \end{split}
\end{equation}
where we used the definition of the backward evolution operator 
\eqref{eq:Backward} and the anti-unitarity of $\Theta$. 
In order to eliminate $\Theta$ at both ends, 
we again utilize the anti-unitarity of $\Theta$, 
which brings about
\begin{equation}
 \bra{\Psi_1} \Theta^{-1} \Theta \ket{\Psi_2}
  = \braket{\Theta \Psi_1 | \Psi_2}^*.
\end{equation}
As a consequence, assuming that integral for $\varphi$ 
is invariant under the combined $\PT$ transformation, we obtain
\begin{equation}
 \Tr \big( \Theta^{-1} \hOcal \Theta \big)
  = \int d\varphi \bra{\Theta \varphi} \Ocal \ket{\Theta \varphi}^*
  = \int d\varphi \bra{\Theta \varphi} \Ocal^\dag \ket{\Theta \varphi}
  = \Tr \big( \hOcal^\dag \big).
\end{equation}
where $\varphi$ denote all dynamical fields under consideration.
With the help of this, we further rewrite $G_F(z;j]$ as 
\begin{equation}
 \begin{split}
  G_F (z;j]
  &=  \Tr \left( 
  e^{- \Theta \hS [\pt_0;\lambda]\Theta^{-1}  } 
  \tilUcal_j  e^{-iz \Theta \hS_0 [\pt;\lambda] \Theta^{-1}} 
  \tilUcal_j^\dag  e^{i z \Theta \hS[\pt_0;\lambda] \Theta^{-1} }
  \right)^\dag  \\
  &=  \Tr \left( 
  e^{-i z \Theta \hS[\pt_0;\lambda] \Theta^{-1} }
  \tilUcal_j 
  e^{iz \Theta \hS_0 [\pt;\lambda] \Theta^{-1}} 
  \tilUcal_j^\dag  
  e^{- \Theta \hS [\pt_0;\lambda]\Theta^{-1}  } 
  \right)  \\
  &=  \Tr \left( 
  e^{-i z \Theta \hS[\pt_0;\lambda] \Theta^{-1} }
  \tilUcal_j 
  e^{- i(i-z) \Theta \hS_0 [\pt;\lambda] \Theta^{-1}} 
  e^{- \Theta \hS [\pt_0;\lambda] \Theta^{-1}} 
  \tilUcal_j^\dag  
  e^{- \Theta \hS [\pt_0;\lambda]\Theta^{-1}  } 
  \right)  ,
 \end{split}
\end{equation}
where we inserted $1 =  e^{\Theta \hS [\pt_0;\lambda] \Theta^{-1}}  
e^{- \Theta \hS [\pt_0;\lambda] \Theta^{-1}}$ just before 
$\tilUcal_j^\dag$ to obtain the third line. 
Then, using the cyclic property of traces: $\Tr(AB) = \Tr (BA)$,
we eventually obtain
\begin{equation}
 \begin{split}
  G_F (z;j]
  &=  \Tr \left( 
  e^{-i z \Theta \hS[\pt_0;\lambda] \Theta^{-1} }
  \tilUcal_j 
  e^{- i(i-z) \Theta \hS_0 [\pt;\lambda] \Theta^{-1}} 
  e^{- \Theta \hS [\pt_0;\lambda] \Theta^{-1}} 
  \tilUcal_j^\dag  
  e^{- \Theta \hS [\pt_0;\lambda]\Theta^{-1}  } 
  \right)  \\
  &=  \Tr \left( 
  e^{- \Theta \hS [\pt_0;\lambda] \Theta^{-1}} 
  \tilUcal_j^\dag  
  e^{i (i -z) \Theta \hS[\pt_0;\lambda] \Theta^{-1} }
  \tilUcal_j 
  e^{- i(i-z) \Theta \hS_0 [\pt;\lambda] \Theta^{-1}} 
  \right)  \\
  &= G_B (i-z;j]. 
 \end{split}
\end{equation}
This is what we want to prove. \qed

\bibliographystyle{spmpsci}
\bibliography{non-rela-hydro-quantum}
\end{document}